\documentclass[aps,prr,superscriptaddress]{revtex4-1}

\usepackage{graphicx}
\usepackage[utf8]{inputenc}
\usepackage[english]{babel}
\usepackage{amsmath,graphicx,enumerate}
\usepackage{epsfig}
\usepackage{hyperref}
\usepackage{amssymb}

\usepackage{color}

\newcommand{\be}{\begin{equation}}
\newcommand{\ee}{\end{equation}}
\newcommand{\bea}{\begin{eqnarray}}
\newcommand{\eea}{\end{eqnarray}}
\newcommand{\bse}{\begin{subequations}}
\newcommand{\ese}{\end{subequations}}
\newcommand{\comment}[1]{}

\begin{document}



\title{Statistical Mechanics of Systems with Negative Temperature}


\author{Marco Baldovin}
\affiliation{Dipartimento di Fisica, Universit\`a Sapienza, 
P.le A. Moro 2, I-00185 Roma, Italy}
\author{Stefano Iubini}
\affiliation{Consiglio Nazionale delle Ricerche, Istituto dei Sistemi Complessi, via Madonna del Piano 10, I-50019 Sesto Fiorentino, Italy}
\affiliation{Istituto Nazionale di Fisica Nucleare, Sezione di Firenze, 
via G. Sansone 1 I-50019 Sesto Fiorentino, Italy}
\author{Roberto Livi}
\affiliation{Consiglio Nazionale delle Ricerche, Istituto dei Sistemi Complessi, via Madonna del Piano 10, I-50019 Sesto Fiorentino, Italy}
\affiliation{Istituto Nazionale di Fisica Nucleare, Sezione di Firenze, 
via G. Sansone 1 I-50019 Sesto Fiorentino, Italy}
\affiliation{Dipartimento di Fisica, Universit\`a di Firenze, 
Via Sansone 1, Sesto Fiorentino, I-50019, Italy}
\author{Angelo Vulpiani}
\affiliation{Dipartimento di Fisica, Universit\`a Sapienza, 
P.le A. Moro 2, I-00185 Roma, Italy}
\affiliation{Complexity Science Hub Vienna, Josefst\"adter Str. 39, 1080 
Vienna, Austria}

\begin{abstract}
Do negative absolute temperatures matter physics and specifically Statistical 
Physics?  We provide evidence that we can certainly answer positively to this 
\emph{vexata quaestio}. The great majority 
of models investigated by statistical mechanics over almost one century and a 
half exhibit positive absolute temperature, because their entropy is a 
nondecreasing function of energy. Since more than half a century ago it has 
been realized that this may not be the case for some physical systems as
incompressible fluids, nuclear magnetic chains, lasers,
cold atoms and optical waveguides. We review these examples and
discuss their peculiar thermodynamic properties, which
have been associated to the presence of thermodynamic regimes,
characterized by negative absolute temperatures. As reported in this review,
the ambiguity inherent the definition of entropy has recurrently raised a harsh debate about
the possibility of considering negative temperature states as genuine
thermodynamic equilibrium ones. 
Here we show that negative absolute temperatures are consistent with equilibrium
as well as with non-equilibrium thermodynamics. In particular,  
thermometry, thermodynamics of cyclic transformations, ensemble equivalence, 
fluctuation-dissipation relations, response theory and even transport 
processes can be reformulated to include them, thus dissipating any prejudice 
about their exceptionality, typically presumed as a manifestation of
transient metastable effects.
\end{abstract}

\maketitle


\newpage
\tableofcontents

\newpage
\section{Introduction}

\subsection{Introductory remarks and plan of the paper}
Among all physical quantities in thermodynamics and statistical mechanics, temperature has for sure a privileged role.
 It is the first observable which is introduced in any course on thermodynamics, and it rules the
 macroscopic behaviour of systems in and out of equilibrium; one of the main aims of statistical mechanics
 is precisely to establish exact relations between the temperature of a system and its underlying microscopic dynamics.
Despite its importance, at a first glance the notion of temperature may appear to be relatively simple, expecially
to beginners: the mantra {\it temperature is a quantity proportional to the mean kinetic energy of a particle} is
often believed to be exhaustive about the whole subject. However, although the statement is correct for most physical systems, one should always keep in mind that the concept of temperature is much more sophisticated than that~\cite{puglisi17}.

The richness of the notion of temperature and its relation with
many important aspects of the statistical description of macroscopic objects are often underestimated.
For instance,  even in good books  of  philosophy of science,  one can find  the  naive  idea
 that  the relation  between temperature and  average kinetic energy is  the  {\it  bridge law} 
 between mechanics and thermodynamics~\cite{nagel79}.
This is wrong:  the expression of temperature  in terms of  mean 
kinetic energy  holds only for a special class of phenomena, although very important.
A more general bridge law is represented by the celebrated relation, engraved on the Boltzmann's  tombstone, between the entropy $S$ and the number of states $W$ which are accessible to the considered system: $S = k_B \log W$, where $k_B$ is the Boltzmann constant.

An accurate analysis of the concept of temperature, and in particular of its relationship  with energy an entropy, shows the possibility, for suitable systems, to attain equilibrium states at negative absolute temperature (NAT).
The existence of this class of models is not a mere theoretical curiosity; on the contrary, many important systems, from several branches of physics, belong to this category: vortices in two-dimensional hydrodynamics, magnetic spins and cold atoms in optical lattices are just some examples. For such systems, NAT states have been observed in real laboratory experiments.
 
In the last decades, much work was devoted to the theoretical understanding of such unusual thermal states. Starting from the pioneering works by Onsager~\cite{onsager49} and Ramsey~\cite{ramsey56}, who originally introduced the physical idea and the first theoretical analyses, many authors have discussed the consequences of including negative temperature in the building of
thermodynamics and statistical mechanics. Equilibrium and out-of-equilibrium situations have been analysed and compared to the familiar cases with positive temperature, leading to a large body of work on the subject. Some authors have also criticised the concept of negative temperature, for several reasons, raising an interesting debate on the very foundations of statistical mechanics.

In this review we aim at providing a picture of the state of the art on this wide subject, trying to clarify the fundamental 
theoretical aspects and the open controversies, discussing results from classical works on the subject as well as the most recent findings.
In the remaining part of this Section we give a brief  overview on the concept of temperature, 
for a generic Hamiltonian system, focusing  on  the relation between energy and entropy and the conditions to observe
negative absolute temperatures.
In Section~\ref{sec:examples} we give an account of the most important examples of physical systems which can be found in NAT states; some relevant experimental aspects are briefly mentioned and discussed. Section~\ref{sec:alternative} reports the main points of the interesting debate on the consistency of negative tempertaure: alternative proposals for the statistical interpretation of entropy and temperature, which would rule out the possibility of NAT equilibrium states, are analysed in some detail; the problem of thermodynamic cycles at negative tempertaure is also discussed. In Section~\ref{sec:equilibrium} we describe the main aspects of equilibrium states at negative temperature, and we introduce in this context some mechanical models which are paradygmatic for this class of systems; particular attention is given to the problem of measuring temperature and to the equivalence between statistical ensembles (and its violations). Section~\ref{sec:fdr} is devoted to the analysis of the fluctuation-dissipation relation when NAT states are allowed; Langevin equation and response theory are revised, and the possibility to design thermal baths at negative temperature is discussed. Section~\ref{sec:dnls} accounts for the large body of works on the Discrete Non-Linear Schr\"odinger Equation, an important physical model which shows a rich and nontrivial phenomenology, which also includes states at NAT. Finally in Section~\ref{sec:fourier} we discuss some recent results and open questions on the problem of heat transport in one-dimensional systems, when also negative temperatures are present.

\subsection{Entropy and temperature}
\label{sec:entropy}
When pondering about the foundations of thermodynamics and statistical 
mechanics, one is often left with the feeling that the very constitutional 
concepts of those theories, as entropy and temperature, are not easy at all to 
master, even in equilibrium conditions. As clearly expressed by 
Truesdell~\cite{truesdell13}:
\begin{quotation} \noindent Entropy, like 
force, is an undefined object, and if you try to define it, you will suffer the 
same fate as the force definers of the seventeenth and eighteenth centuries. 
Either you will get something too special or you will run around in a circle. 
\end{quotation}
One may be tempted to say that, since thermodynamics is a phenomenological
theory, it is somehow unavoidable that also its 
primitive concepts only refer to practical experience; in this light, it seems 
natural to introduce all fundamental quantities via proper operative 
definitions, so that, for instance, temperature should be only seen as the 
result of a measuring protocol involving some kind of thermometer. While this 
point of view is, without any doubt, largely satisfactory at a pragmatic level 
(e.g. in most engineering applications), from a more conceptual perspective it 
can actually raise some concern.

Indeed, we know that macroscopic systems are made of particles (atoms, 
molecules) ruled 
by microscopic mechanical laws. In the eighteenth century, physicists started 
to 
understand the necessity of a clear link between the mechanical world (i.e. 
Newtons's laws, in the classical description) on the one hand, and the 
phenomena 
described by thermodynamics on the other. As far as we know, the first attempt 
in this direction is due to Daniel Bernoulli. In his seminal book 
\textit{Hydrodynamica}, he considered ideal gases constituted by point-like 
particles of mass $m$:  the pressure resulted then as a consequence of the many 
collisions of the molecules with the walls of the container~\cite{hall03}. 
Using  this model for gaseous matter, Clausius was able to show that 
temperature 
$T$ is proportional to the mean kinetic energy of the system:
\begin{equation}
m \langle v^2 \rangle= k_B T
\end{equation}
being $v$ any component of the velocity, and $k_B$ the Boltzmann constant (in 
modern terminology).

The decisive step to understand the relation between mechanics and 
thermodynamics was made by Ludwig Boltzmann by introducing ergodicity, i.e. by 
assuming that an isolated mechanical systems will eventually explore, in a 
homogeneous way, the whole fixed-energy hypersurface accessible to it in the
space of its configurations. This assumption, which turns out 
to be true for almost all practical purposes, allowed him to make a precise 
connection between the microscopic world and the macroscopic, observable 
quantities of thermodynamics; this relation is summarized by the celebrated 
equation:
\begin{equation}
\label{eq:entropy}
S(E)= k_B \log W(E)\,,
\end{equation}
 engraved on Boltzmann's tombstone. Here $S(E)$ denotes the entropy of the 
macroscopic body
and  $W(E)$ is the number of microscopic states accessible to the system, once a 
proper discretization of the phase-space has been introduced.
Since $S$ is a thermodynamic observable, while $W$ is clearly a mechanical-like 
quantity, in the 
terminology of the philosophy of science Eq.~\eqref{eq:entropy} is a 
\textit{bridge law} between mechanics and thermodynamics.

Given a system with Hamiltonian $\mathcal{H}(\mathbf{Q}, \mathbf{P})$, to 
understand its macroscopic behaviour we must 
 first express $W(E)$ as a function of the energy $E$ and of the other 
parameters of the system, and then compute the entropy through 
Eq.~\eqref{eq:entropy}.
$W(E)$ is obviously proportional to the density of states
\begin{equation}
\label{eq:omega}
\omega(E)=\int \mathcal{D}\mathbf{Q}\mathcal{D}\mathbf{P}\, 
\delta(E-\mathcal{H}(\mathbf{Q}, \mathbf{P}))\,,
\end{equation} 
so that Eq.~\eqref{eq:entropy} is usually written as
\begin{equation}
\label{eq:boltentropy}
S(E)= k_B \log \omega(E)\,.
\end{equation}
Even if this designation is not historically precise~\cite{mehra01}, the above 
quantity is sometimes referred to as ``Boltzmann entropy''.
In the following we shall adopt this convention.
Once entropy is known, the thermodynamic relations can be invoked to find all 
interesting thermodynamic quantities. In particular, the well known relation
\begin{equation}
\label{eq:temperature}
{1 \over T}= {\partial S \over \partial E}
\end{equation} 
can be used to obtain a mechanical definition of temperature which is usually found in                 Statistical mechanics textbooks~\cite{ma85,huang88} and which we shall refer to as ``Boltzmann temperature''.

The deep connection between Boltzmann entropy 
(and temperature) and the corresponding thermodynamic counterpart can be 
understood by recalling the following, straightforward argument. Let us 
consider two independent systems $\mathcal{A}$ and $\mathcal{B}$, at energy 
$E_{\mathcal{A}}$ and $E_{\mathcal{B}}$ respectively. It is a trivial 
consequence of Eq.~\eqref{eq:entropy} that the compound system $\mathcal{A} 
\cup 
\mathcal{B}$ has a total entropy
\begin{equation}
 S_{A\cup B}=S_{\mathcal{A}}(E_{\mathcal{A}})+S_{\mathcal{B}}(E_{\mathcal{B}})\,.
\end{equation} 
Let us now assume that the two subsystems  $\mathcal{A}$ and $\mathcal{B}$ are 
now put into thermal contact, i.e. they are allowed to exchange energy with 
each 
other, still being isolated from the environment. We also assume that 
contributions to the total energy $E=E_{\mathcal{A}}+E_{\mathcal{B}}$ coming 
from the interaction among the two systems can be neglected, a condition 
typically verified when only short-range interactions are involved. The 
equilibrium condition is reached when $S_{\cup}$ is maximized with respect to 
$E_{\mathcal{A}}$ (and $E_{\mathcal{B}}=E-E_{\mathcal{A}}$), i.e when 
\begin{equation}
\frac{\partial S_{\mathcal{A}}}{\partial 
E_{\mathcal{A}}}=\frac{\partial S_{\mathcal{B}}}{\partial E_{\mathcal{B}}}\,. 
\end{equation}
It is therefore clear that $\partial S/\partial E$ has a 
straightforward interpretation as (an invertible function of) the temperature of 
a system at equilibrium, if 
$S$ is the Boltzmann entropy.

Let us anticipate that in Section~\ref{gibbstemperature} we shall examine a 
possible alternative definition of entropy, and temperature, attributed to Josiah
Willard Gibbs. Even though, as we will discuss in the following, this 
alternative description has several interesting properties, it fails, in our 
opinion, to give an interpretation of equilibrium temperature as straightforward 
as that exposed above: this is the main reason why, in this review, we shall 
mainly refer to the standard Boltzmann formalism.
We need to specify this point 
because, for the physical systems discussed in the following, the two 
``entropies'' (and the two ``temperatures'') show completely different 
behaviors, even in the thermodynamic limit: in particular, as we will discuss, 
the Gibbs temperature is always positive, by definition; the Boltzmann 
temperature, on the other hand, can assume negative values in the high-energy 
regime,
for some classes of physical systems. These ``negative-temperature'' states are 
the main subject of
the present review.

\subsection{Negative absolute temperature }
\label{sec:nat}

From the discussion of the above section it is clear that the Boltzmann 
temperature assumes negative values whenever the density of states $\omega(E)$, 
defined by Eq.~\eqref{eq:omega}, shows a negative slope. If this is the case, 
in 
that interval also entropy is a decreasing function of energy, and 
Eq.~\eqref{eq:temperature} assures that the absolute temperature is negative. 
This possibility may sound quite unusual or even paradoxical, for sure very far 
from everyday experience. In Section~\ref{sec:examples} we shall illustrate some 
relevant examples of physical systems in which thermodynamic states at NAT appear;
for now, let us just discuss why 
they are not as common as positive-temperature states.

Consider an isolated system described by the generic Hamiltonian
\begin{equation}
 \mathcal{H}(\mathbf{Q},\mathbf{P})=\sum_{i=1}^N \frac{p_i^2}{2m}+ 
\mathcal{U}(\mathbf{Q})\,,
\end{equation}
where $\mathcal{U}$ is the potential contribution to the Hamiltonian, a regular 
function of (in general) all positions and bounded from below (stability 
condition); without any lack of generality, we can assume that 
$\min_{\mathbf{Q}}\mathcal{U}(\mathbf{Q})=0$. We also assume that 
the only quantity conserved by dynamics is the total energy $E$. The density of 
states of this system can be written as
\begin{equation}
\begin{split}
 \omega(E)&=\int \mathcal{D} \mathbf{Q}  \mathcal{D} \mathbf{P}\, 
\delta(E-\mathcal{H}(\mathbf{Q},\mathbf{P}))\\
 &=\int \mathcal{D} \mathbf{Q}  \mathcal{D} \mathbf{P}\, \int_0^E d u\, 
\delta(u-\mathcal{U}(\mathbf{Q}))\delta\left(E-u-\sum_{i=1}^N 
\frac{p_i^2}{2m}\right)\,.
 \end{split}
\end{equation}
Due to the quadratic form of the kinetic energy, we can explicitly compute the 
integral over the momenta; as a result, we obtain
\begin{equation}
\label{eq:omega_newton}
  \omega(E)=\int_0^E d u\, \omega_{\mathbf{Q}}(u) c_N(E-u)^{N/2-1}\,,
\end{equation} 
where $c_N=\frac{N}{2}(2 \pi)^{N/2}/\Gamma(1+N/2)$ is a constant (being 
$\Gamma(x)$ the Euler Gamma function). In the above equation we have introduced
a ``configurational'' $\omega_{\mathbf{Q}}$ density of states only depending on 
positions, i.e.
\begin{equation}
 \omega_{\mathbf{Q}}(u)=\int \mathcal{D} \mathbf{Q} 
\,\delta(u-\mathcal{U}(\mathbf{Q}))\,.
\end{equation}
By deriving Eq.~\eqref{eq:omega_newton} with respect to the energy, we obtain, for 
$N>4$,
\begin{equation}
 \omega'(E)=\frac{(N-2)c_N}{2}\int_0^E d u\, \omega_{\mathbf{Q}}(u) 
(E-u)^{N/2-2} \ge 0\,,
\end{equation} 
the last inequality coming from the positivity of $\omega_{\mathbf{Q}}(u)$.

The above argument should clarify why in most physical systems only 
positive-temperature states can be observed: as far as the dynamics of the 
system can be modeled by a Hamiltonian with quadratic kinetic terms, with 
potential energy not depending on momenta and no additional conserved 
quantities, the density of states is always an increasing function of the 
energy. 
As a consequence, entropy is also monotonic and temperature can never assume 
negative sign.

\begin{figure}[h!]
\centering
\includegraphics[width=.47\linewidth]{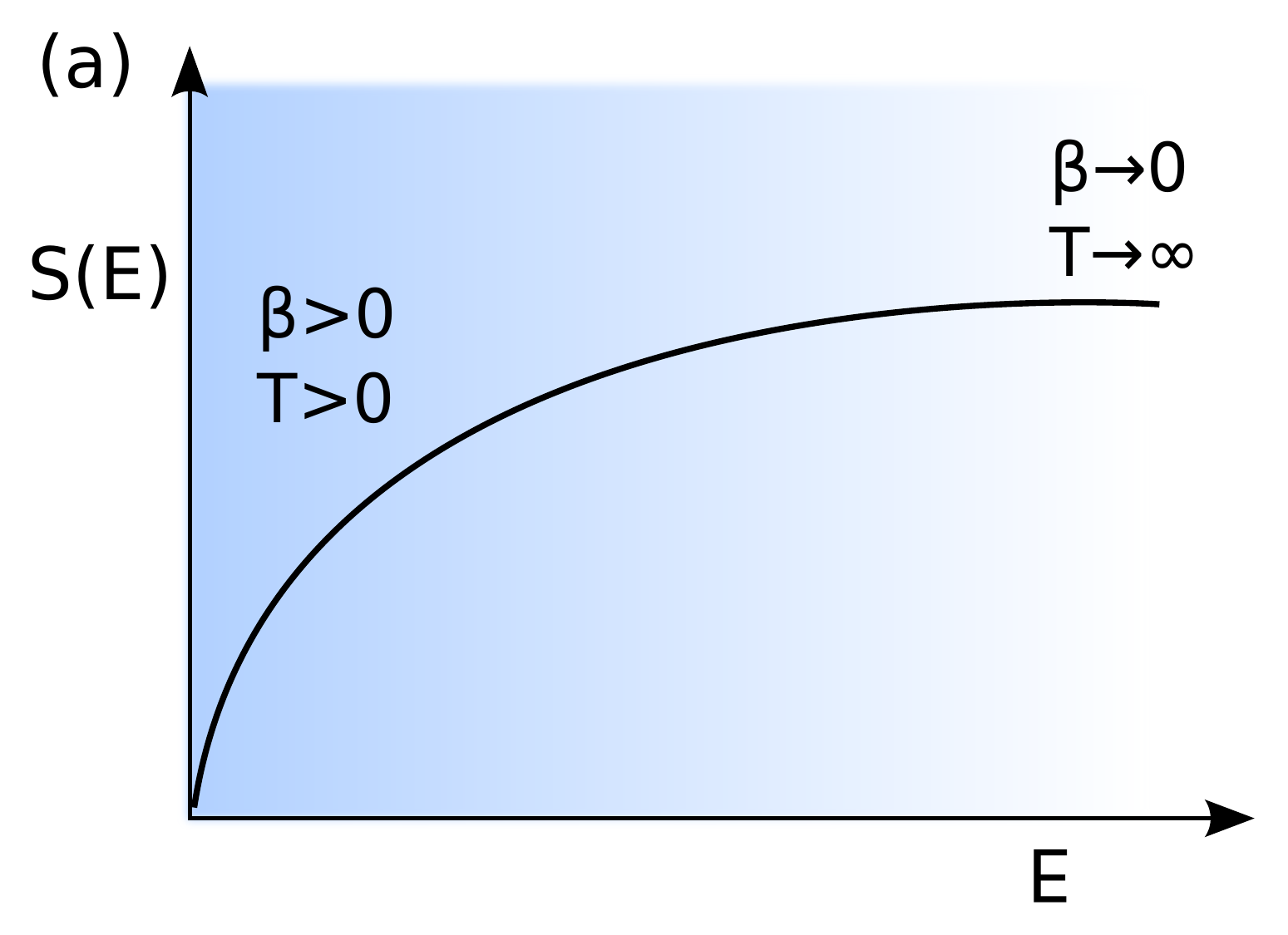}
\quad
\includegraphics[width=.47\linewidth]{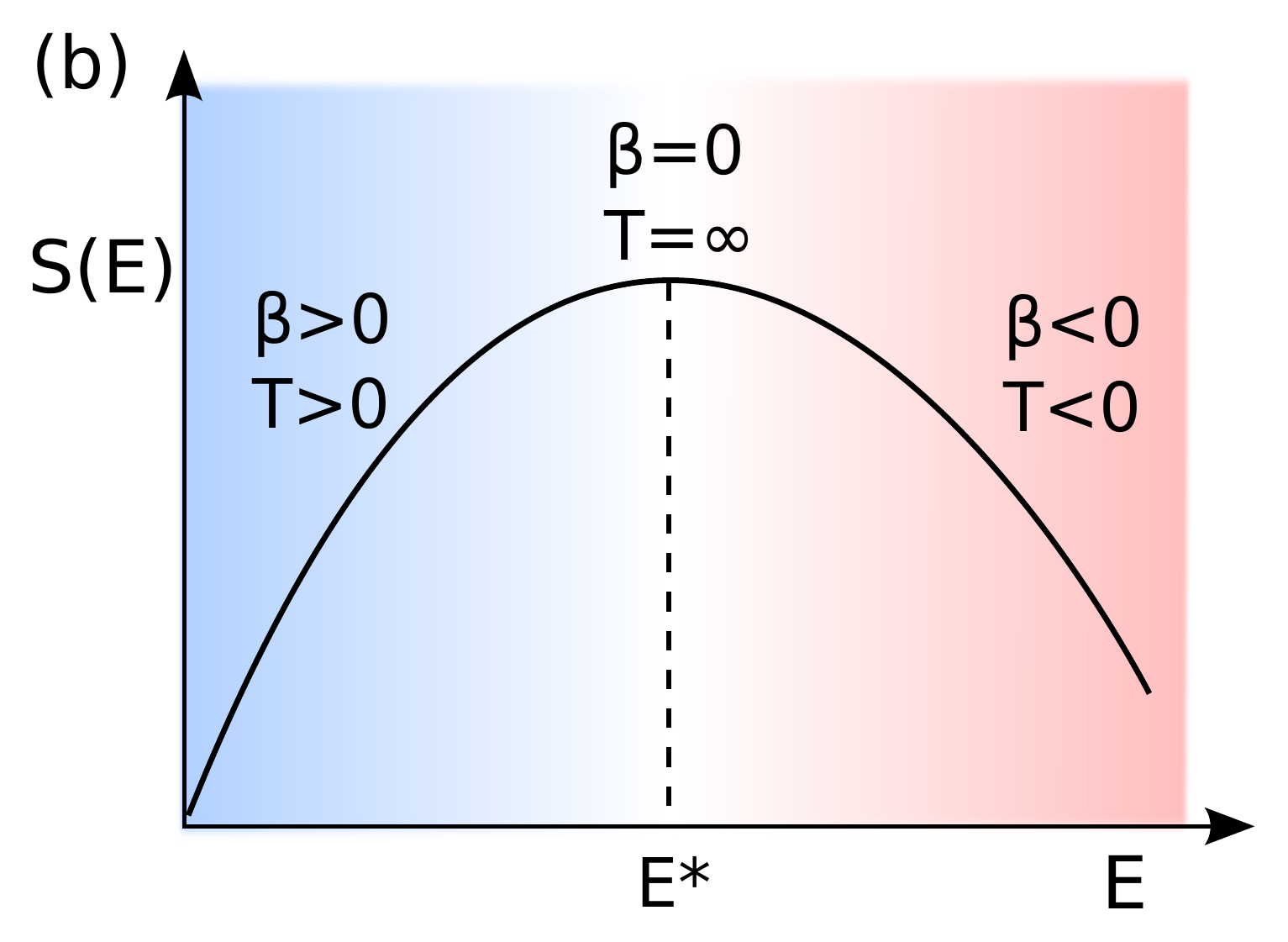}
\caption{\label{fig:entropy} Qualitative sketch of the entropy vs energy curve for systems
with unbounded phase space (panel (a)) and for typical cases with bounded phase space (panel (b)).}
\end{figure}

Negative temperatures are instead typically observed in systems with bounded 
phase spaces, some of which will be discussed in the next Section. In most 
cases, two regimes can be identified: at low energy, $\omega(E)$ is usually an 
increasing function, since additional energy allows the system to explore wider 
regions of the phase space; but since the phase-space volume is finite, the 
number of states cannot increase indefinitely, and it must exist some energy 
threshold $E^*$ such that, for $E>E^*$, $\omega(E)$ starts decreasing. 
Introducing the inverse temperature
\begin{equation}
\label{eq:beta}
 \beta=\frac{1}{k_B}\frac{\partial S}{\partial E}
\end{equation}
we can summarize this typical scenario as follows:
\begin{equation}
 \begin{cases}
  E<E^*\,,\quad\beta>0\quad \longrightarrow \quad T>0\\
  E=E^*\,,\quad\beta=0\quad \longrightarrow \quad T\to\pm \infty\\
  E<E^*\,,\quad\beta<0\quad \longrightarrow \quad T<0\,.
 \end{cases}
\end{equation}
The situation is qualitatively sketched in Fig.~\ref{fig:entropy}, where the
typical entropy vs energy curve of usual systems with quadratic kinetic energy is compared
to that of systems with bounded phase space.  As we are going to show in
what follows, this qualitative scenario concerns also models other than those 
with bounded phase space, like the Discrete Nonlinear Schr\"odinger Equation 
(See Sections~\ref{sec:esdnls} and~\ref{sec:remarks}).

Some confusion may arise from the fact that, according to the above discussion, 
any system with $T<0$ is, to the purpose of establishing the energy flux, 
``hotter'' than any other system with $T>0$.
To clarify this point, let us consider again two systems ${\cal A}$  and ${\cal 
B}$,
composed by  $N_{{\cal A}}$ and  $N_{{\cal B}}$   particles and described by the 
Hamiltonians
$\mathcal{H}_{{\cal A}} ({\bf Q}_{{\cal A}}, {\bf P}_{{\cal A}})$ and 
$\mathcal{H}_{{\cal B}} ({\bf Q}_{{\cal B}}, {\bf P}_{{\cal B}})$, respectively.
Here we limit ourselves to the case in which $\mathcal{H}_{{\cal A}}$ and 
$\mathcal{H}_{{\cal B}}$ have the same functional dependencies on the canonical 
variables (i.e. they correspond to systems with the same microscopic dynamics, 
with possibly different
 sizes $N_{{\cal A}}$  and  $N_{{\cal B}}$), and only short-range interactions 
are involved.
Using the additivity of entropy, in the thermodynamic limit we can introduce the 
entropy per particle $s(e)$
\begin{equation}
s(e)= \frac{S(E,N)}{N}\,,
\end{equation} 
where $e$ is the specific energy. With our assumptions, $s(e)$ has the same
functional form in the two   systems. 

Let us now suppose that systems ${\cal A}$ and ${\cal B}$ have energy $E_{{\cal 
A}}=N_{{\cal A}} e_{{\cal A}}$ and $E_{{\cal B}}=N_{{\cal B}} e_{{\cal B}}$ 
respectively, and that the corresponding inverse temperatures are  $\beta_{\cal 
A}$
and $\beta_{\cal B}$. When the two systems are put in contact via some kind 
of small, short-range interaction, a new system is realized, composed by 
$N=N_{{\cal A}}+ N_{{\cal B}}$ particles. If we denote by   $a=N_{{\cal A}}/N$ 
the fraction of particles from the system ${\cal A}$, the final energy that the 
total system achieves at equilibrium is clearly
\begin{equation}
E_f=N e_f\,, \quad \text{with} \quad  e_f= a e_{{\cal A}}+(1-a)e_{{\cal B}}\,.
\end{equation} 
Since, according to ergodic hypothesis, the system will spend most time in the 
macroscopic state which corresponds to the largest number of microscopic 
configurations, the final entropy reached by the system cannot be smaller than 
the initial one:
\begin{equation}
N s(e_f)\ge N_{{\cal A}} s(e_{{\cal A}}) +N_{{\cal B}} s(e_{{\cal B}})=
N [a s(e_{{\cal A}}) +(1-a) s(e_{{\cal B}})] \,;
\end{equation}
this is consistent with the second principle of thermodynamics. Note that the 
specific entropy of the global system is the same as those of the subsystems, 
due to our assumption that the interacting potentials are short-range. The 
previous inequality is nothing but a way to express the concavity  of  $s(e)$:
\begin{equation}
s(a e_{{\cal A}} +(1-a) e_{{\cal B}})\ge a s(e_{{\cal A}}) +(1-a) s(e_{{\cal 
B}})\,.
\end{equation}
As a consequence, the inverse temperature $\beta$ is a monotonically decreasing 
function of the energy. The final inverse temperature  $\beta_{f}$ is 
intermediate between   $\beta_{\cal A}$ and $\beta_{\cal B}$, i.e.  if 
$e_{{\cal 
B}} > e_{{\cal A}}$,  that is  $\beta_{\cal A} > \beta_{\cal B}$, then
\begin{equation}
  \beta_{\cal B} < \beta_{f}<\beta_{\cal A} \,\, .
 \end{equation} 
The energy flux obviously goes from smaller  to larger $\beta$, i.e. from hot 
to cold, irrespectively of the sign of $\beta$. The absolute temperature $T=(k_B 
\beta)^{-1}$ fails to verify an analogous ordering; and indeed the privileged 
role of $\beta$ over $T$ in statistical mechanics is also quite clear if one 
notes that $\beta$ is the variable actually associated to energy, e.g. when 
passing from microcanonical to canonical ensemble. The fact that we use $T$, 
instead of $\beta$, as a measure of thermal equilibrium may be seen as an 
historical heritage of the first phenomenological studies on thermodynamics, 
but 
it has no fundamental reason.

Let us now conclude this section by briefly discussing the case in which two 
different systems $\mathcal{A}$ and $\mathcal{B}$ are put at contact, assuming 
that $\mathcal{A}$ can admit negative temperature, while $\mathcal{B}$ cannot. 
It is quite easy to understand that the coupling of the system $\mathcal{A}$, 
initially prepared in a negative temperature state, with the system 
$\mathcal{B}$, always at positive temperature, will result in a system with 
final positive temperature.  Indeed, by repeating the argument discussed in the 
previous section, we find that at equilibrium the relation
\begin{equation}
\beta_\mathcal{A}={\partial S_\mathcal{A} \over \partial 
E_\mathcal{A}}={\partial S_\mathcal{B} \over \partial 
E_\mathcal{B}}=\beta_\mathcal{B}
\end{equation}
must hold. Since $\beta_\mathcal{B}$ is positive for every value of 
$E_\mathcal{B}$, the final common temperature must also be positive. The above 
result helps to understand 
why it is not common to observe negative temperature: even if one deals with 
systems allowed to achieve such kind of states, it is necessary to completely 
isolate them from the external environment, made by ordinary matter which can 
only assume positive temperature.  In Section~\ref{sec:termometro} we shall 
consider again this aspect from a theoretical point of view. This difficulty is 
crucial in experimental contexts, where one has the practical need to isolate 
the system from the environment, at least for a time long enough to allow for 
internal equilibration, as we will briefly see in the next section.

\newpage
\section{Examples and phenomenology }
 \label{sec:examples}
\subsection{Onsager's vortices}
\label{sec:vortices}

One of the first, and most important, systems showing  negative 
temperature was studied by Onsager in a seminal paper at the origin
of the modern statistical hydrodynamics~\cite{onsager49}.
Due to the historical and technical relevance of that work, it is worth to 
briefly summarize its 
main results.

 Consider a two-dimensional incompressible ideal flow  in a domain 
$\mathcal{A}$.
The time evolution of the flow ${\bf u}$ is ruled  by Euler equation
 \begin{equation}
 \label{eq:EU0}
 \begin{cases}
 \partial_t {\bf u}+({\bf u} \cdot \nabla) {\bf u}= -\frac{ \nabla p}{\rho_0}\\
\nabla \cdot {\bf u}=0 \,,
 \end{cases}
 \end{equation}
 being  $\rho_0$ and $p$ the constant density and the pressure of the fluid, 
respectively.
Let us introduce the vorticity $w$, defined by
\begin{equation}
\nabla \times {\bf u}=w \hat{\bf z}\,,
\end{equation} 
 where $\hat{\bf z}$ is the unitary vector perpendicular to the plane of 
the flow. From Eq.~\eqref{eq:EU0} one has that $w$ evolves according to
\begin{equation}
\partial_t w +({\bf u} \cdot \nabla) w =0 \,\, .
\label{EU1}
\end{equation}
The previous equation is nothing but
the conservation of vorticity along fluid-element paths~\cite{kraichnan80}.
The incompressibility condition allows us to introduce
 the stream function $\psi$: 
\begin{equation}
 {\bf u}=\nabla^{\perp}\psi=(\partial_y, -\partial_x)\psi\,,\quad \quad w=- 
\Delta \psi\,.
 \end{equation}
 Therefore, the velocity can be expressed in terms of $w$ as
\begin{equation}
{\bf u}({\bf x}, t)=-\nabla^{\perp}\int d {\bf x}' {\cal G}({\bf x}, {\bf 
x}')w({\bf x}', t) 
\end{equation} 
where  ${\cal G}({\bf r}, {\bf r}')$  is the Green function 
of the Laplacian operator  $\Delta$, which depends on the shape of
the domain $\mathcal{A}$.

 Consider now an initial condition at $t=0$ such that the vorticity is localized 
on $N$
point-vortices  
 \begin{equation}
w({\bf r}, 0)=\sum_{i=1}^N \Gamma_i \delta({\bf r} -{\bf r}_i(0)) \,\, , 
 \end{equation} 
where $\Gamma_i$  is the circulation of the $i-$th vortex.
Using the Kelvin theorem~\cite{truesdell18} one realizes  
that the vorticity  must remain localized at any time:  
 \begin{equation}
w({\bf r}, t)=\sum_{i=1}^N \Gamma_i \delta({\bf r} -{\bf r}_i(t)) \,. 
 \end{equation} 
Plugging the above formula in Eq.~\eqref{EU1}  it is possible to derive
the evolution law for  the vortex positions  ${\bf r}_i=(x_i, y_i)$:
\begin{equation}
\label{eq:onsager}
\frac{ d x_i}{dt}=
\frac{1} {\Gamma_i} \frac{\partial \mathcal{H}} {\partial y_i} \,\,\, , \,\,\,
\frac{ d y_i}{dt}=
-\frac{1} {\Gamma_i} \frac{\partial \mathcal{H}} {\partial x_i} \,\,\, , 
\end{equation} 
with
\begin{equation}
\mathcal{H}(\mathbf{r}_1,\dots,\mathbf{r}_N)=\sum_{i \neq j}  \Gamma_i \Gamma_j {\cal 
G}({\bf r}_i, {\bf r}_j) \,.
\end{equation} 
So the  $N$ point vortices constitute a $N$ degree of freedom
Hamiltonian system~\cite{newton10} with canonical coordinates 
\begin{equation}
q_i=\sqrt{|\Gamma_i|} x_i\, , \quad
p_i= \text{sign}(\Gamma_i)\sqrt{|\Gamma_i|}  y_i \,\,.
\end{equation}
and the corresponding Hamiltonian 
$\mathcal{H}(\mathbf{Q},\mathbf{P})=H[\mathbf{r}_1(q_1,p_1),\dots,\mathbf{r}
_N(q_N,p_N)]$

Consider now  $N$ point vortices confined in a bounded domain $\mathcal{A}$ of 
area $A$.
 Since for each point vortex ${\bf r}_i \in \mathcal{A}$  one has that the 
phase-space volume
 enclosed by the constant-energy hypersurface 
$\mathcal{H}(\mathbf{X},\mathbf{Y})=E$ verifies
\begin{equation}
 \Omega(E)=\int_{\mathcal{H}<E} d q_1 \cdots d q_N d p_1 \cdots d p_N\le C_N A^N 
\,, \quad C_N=\prod_{i=1}^N |\Gamma_i| \,,
 \end{equation} 
 it follows that the density of states $\omega(E)=d \Omega(E)/d E$  must 
approach to zero for 
$E \to \pm \infty$, and it must therefore attain its maximum at a certain 
finite value $E_M$.
This implies   that for 
$E > E_M$ the entropy $S(E)=k_B \ln  \omega(E)$ is a decreasing function and 
hence 
$T(E)=(dS/dE)^{-1}$ is negative.

It is easy to understand that both the low- and high-energy regimes, 
corresponding to $\beta \gg 1$ and $\beta \ll -1$, 
are characterized by spatially ordered configurations. Indeed, in any finite 
domain $\mathcal{A}$, for ${\bf r} \sim  {\bf r}'$ we have
 ${\cal G}({\bf r}, {\bf r}')\simeq-1/(2 \pi) \ln |{\bf r} -{\bf r}'|$, and the 
energy contribution of a pair of vortices located in ${\bf r}$ and $ {\bf r}'$
 is maximized, in modulus, if they are very close to each other. The sign of the 
energy contribution is, of course, determined by the sign of the vorticities.
As a result, when energy is small, $E \ll E_M$, the system tends to organize 
itself into pairs of very close vortices with opposite signs of $w$;
at large energy $E \gg E_M$, on the other hand, the typical states are those in 
which the vortices 
are crowded in two separate clusters, depending on the sign of their 
vorticity~\cite{onsager49, kraichnan80, newton10}.
 Let us notice that in the negative-temperature scenario energy tends to be 
transferred from small-scale structures (wandering pairs of opposite-sign 
vortices)
to a large-scale configuration (two clusters containing almost all vortices).
This mechanism is somehow opposite to what usually happens in three-dimensional 
turbulence,
where energy is carried from large to small spatial scales: this phenomenon
is therefore called ``inverse energy cascade''.

The model proposed by Onsager has been studied in many works, both analytically 
and numerically.
It has been shown, for instance, that it is possible to derive a 
canonical distribution for a small portion of the system even at negative 
temperature; in the same context, an attempt to solve the BBGKY hierarchy for 
the distribution of the vorticity was discussed, using the Vlasov 
approximation~\cite{montgomery74}. Similar studies were done in the context of 
guiding-center plasma, that can be modeled with the same equations of point 
vortices in two-dimensional hydrodynamics~\cite{smith89,smith90}.

In the 70's, early numerical simulations of Eq.~\eqref{eq:onsager} supported 
the 
picture that negative temperature would signal the emergence of a new ordered 
phase at high energies~\cite{montgomery74}. Nowadays modern computational tools 
allow for much more 
extensive simulations of the vortices dynamics, and the underlying physical 
mechanisms can be efficiently observed and 
studied~\cite{yatsuyanagi05,yatsuyanagi15}.
 As predicted by Onsager, in the $\beta<0$ 
regime, negative and positive vortices separate into two large clusters.

In more recent years, the above theory has been adopted and studied in the 
context
of quantum superfluids~\cite{simula14,groszek17,valani18}. In particular it has 
been shown by numerical simulations
that isolated Bose-Einstein condensates, under suitable conditions, relax 
towards
the ordered phase at negative temperature described by Onsager. The
mechanism behind the formation of large clusters of equal sign vortices is the 
so-called
``evaporative heating''. From time to time, pairs of opposite-sign vortices 
happen to melt together and 
disappear, while their energy is transferred to the other vortices through the 
produced sound waves.
During such process the energy of the system is conserved, but the entropy
of the vortices decreases, due to the disappearing of a pair~\cite{simula14}. 

Experimental evidences of inverse energy cascade and of hydrodynamics states at 
negative temperature
have been reported in two very recent works~\cite{gauthier19,johnstone19}. The 
two research groups
have realized, independently of each other and using different methodologies, 
experimental setups
that allow for the observation of vortices in superfluid Bose-Einstein 
condensates. In both cases,
steady large-scale clusters of vortices with equal sign are observed, certifying 
the validity
of Onsager's interpretation in the context of quantum superfluids.

Let us close this subsection with a remark. The Hamiltonian of the 
point-vortices system contains  long-range interactions, and one may be led to 
the conclusion that the presence of NAT states is due to this peculiarity.
This guess is wrong, and the reason can be understood as follows: if ${\cal 
G}({\bf r}_i, {\bf r}_j)$ is replaced by any bounded function with its maximum 
at  ${\bf r}_i= {\bf r}_j$ and  fast decreasing to zero for large $|{\bf 
r}_i- 
{\bf r}_j|$, a situation corresponding to short-range interacting systems, the 
argument about the possibility of negative temperature used in the case of 
point-vortices system can be repeated exactly in the same way.

\subsection{Magnetic systems}\label{sec:mag_sys}

In parallel with the discoveries on statistical hydrodynamics, the physics of 
nuclear spin systems 
represented another independent field where the concept of negative absolute 
temperature
naturally emerged. In this context, novel experimental techniques on nuclear 
magnetic resonance
allowed to probe the nuclear magnetization of materials subject to external 
magnetic fields.

\begin{figure}[h!]
\centering
\includegraphics[width=.6\linewidth]{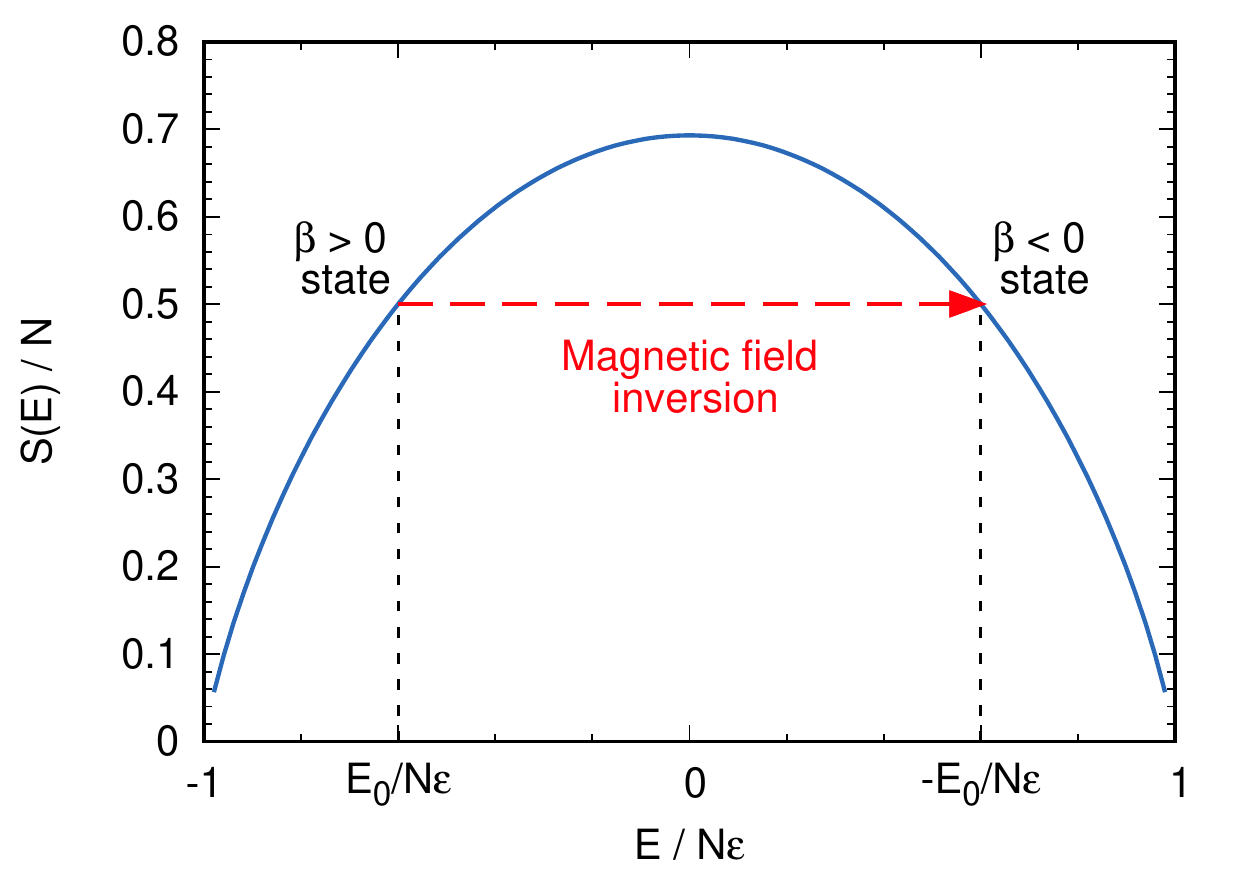}
\caption{Entropy vs energy for the spin system~\eqref{eq:hammag} in the case of $1/2$ spins (two levels). Here $\varepsilon=\hbar \gamma|\mathbf{B}|/2$. The effect of a sudden inversion of the external magnetic field, as that performed in the experiments discussed in the main text, is qualitatively represented by the red arrow.}
\label{fig:spin}
\end{figure}

As a preliminary remark, it is important to clarify the physical meaning of 
``spin temperature" 
in such systems, see~\cite{abragam58,oja97,vladimirova18} and references 
therein. In a first series of experiments, Pound~\cite{pound51} discovered that 
sufficiently pure
crystals of lithium fluoride (LiF) display nuclear spin relaxation times of the 
order of
minutes at room temperature.
Such relaxation process is produced by the interaction between nuclear
spins and the crystal lattice and it results to be much longer than the typical 
time scale 
of spin interactions, which is of order $10^{-5}$~s~\cite{pound51}. Given this 
large separation
of time scales, a spin temperature can be introduced to characterize the thermal 
state of
the nuclear spin subsystem, which is thereby assumed to be isolated from the  
external environment.
In a second experiment, Purcell and Pound~\cite{purcell51} used this property of 
LiF crystals to reverse
the magnetization of a sample initially in equilibrium in a strong magnetic 
field.
This operation was achieved by reversing the direction of the magnetic field 
very rapidly with respect
to the nuclear spin  time scales.  As a result, the final state displayed a 
magnetization opposed to
the new field direction. Before decaying again to the original state (over 
several minutes), the spin
system was therefore in equilibrium in a peculiar  high-energy state which gains 
entropy as it loses internal
energy. As stressed by the authors themselves, this state can be properly 
described by a negative  spin temperature.
In particular they wrote ``{\it Statistically, the most
probable distribution of systems over a \emph{finite} number of equally
spaced energy levels, holding the total energy constant, is the
Boltzmann distribution with either positive or negative temperature
determined by whether the average energy per system is
smaller or larger, respectively, than the mid-energy of the available levels.
 The sudden reversal of the magnetic field produces
the latter situation.}''~\cite{purcell51}. 
In practice, the effect of the field inversion  can be understood by considering 
a simple model of paramagnetic
system~\cite{ramsey56}. The Hamiltonian of a system of $N$ non interacting 
nuclear spins $\{\mathbf{I}_j\}$ in a magnetic field $\mathbf{B}$ takes the form
\begin{equation}
\label{eq:hammag}
\mathcal{H}=-\hbar \gamma\mathbf{B}\cdot \sum_{j=1}^N  \mathbf{I}_j 
\end{equation}
where $\gamma$ is the gyromagnetic ratio. The behavior of the total entropy $S$ 
of the system as a function of its internal energy
$E$  is shown qualitatively in Fig.~\ref{fig:spin}, 
see~\cite{ramsey56,buonsante16} and Section~\ref{sec:fourierspin} of the present review
for further details. A region with negative 
temperature states is present for positive values of the internal energy, which 
are determined by a magnetization vector opposed to
the field direction. 
If the reversal of the magnetic field is much faster than the typical spin-spin 
interaction
times, the spin system can perform no redistribution during the field variation. 
Therefore,
one can assume that the spin configuration  $\{\mathbf{I}_j\}$ is not modified 
by the reversal. 
With these assumptions, given the initial equilibrium configuration with 
temperature $T_0$ and  internal energy $E=E_0$, with $E_0$ proportional to $B$, 
the field flip moves the system to a new state with energy $E'=-E_0$ and 
temperature $T'=-T_0$ (see  Fig.~\ref{fig:spin}).
Notice that if the field inversion is slower than the intrinsic time 
scales of the spins, the latter are able to follow adiabatically the
field variation and no NAT states are produced~\cite{oja97}.
 
These concepts were  discussed theoretically 
 in the pioneering work by Ramsey~\cite{ramsey56},
  which represents
the first systematic attempt to generalize equilibrium thermodynamics  to 
negative
temperatures states.
The essential requirements identified by Ramsey and further justified by 
Klein~\cite{klein56}
for the existence of negative-temperature states are:
\begin{enumerate}
\item 
the system must be in thermodynamical equilibrium so that a proper temperature 
can be
introduced to characterize its thermodynamic state;
\item	the possible energy of the allowed states of the system must be bounded;
\item the system must be thermally isolated from all other systems which do not
satisfy conditions (1) and (2).
\end{enumerate}
Ramsey also mentions that in order to consider negative temperature, Kelvin-Planck formulation of the second law of thermodynamics needs to be modified in the following way: 
\begin{quotation}
 \noindent It is impossible to construct an engine that will operate in a closed cycle and produce no effect other than (1) the extraction of heat from a positive temperature reservoir with the performance of an equivalent amount of work or (2) the rejection of heat into a negative-temperature reservoir with the corresponding work being done on the engine.
\end{quotation}
The reason of the above formulation can be easily understood by looking at Fig.~\ref{fig:spin}. As long as the system is in the positive-temperature regime, extracting heat from it results in a decrease of its entropy; if such energy was completely converted into work, the total entropy would decrease too, leading to a paradox. Similarly, in the negative-temperature regime adding energy to the system by mean of mechanical work would result in a global decrease of entropy, and therefore this process is forbidden. This topic will be further discussed in Section~\ref{sec:cycles} of the present review.

It should be stressed that the emergence of negative temperatures in spin 
systems is not limited to  paramagnetic systems.
 More recent experiments have demonstrated the existence of negative temperature 
states also
in the presence of dominating ferromagnetic/antiferromagnetic interactions.  
 An important example is the
case of silver in cryogenic conditions~\cite{hakonen92},
whose nuclear Hamiltonian can be approximated by
 \begin{equation}
\mathcal{H}=-\sum_{\langle j,k \rangle} J_{jk}\mathbf{I}_j \cdot \mathbf{I}_k 
-\hbar \gamma\mathbf{B}\cdot \sum_{j=1}^N  \mathbf{I}_j 
\end{equation}
where the (negative) coupling constants $J_{jk}$ identify  a nearest-neighbor 
antiferromagnetic interaction.
Here, equilibrium negative temperatures are signaled by the change of the sign 
of the interaction energy. Indeed, susceptibility
measurements confirm the presence of ferromagnetic domains, which would be 
unstable at positive temperatures~\cite{hakonen92}. The emergence of 
macroscopic 
ferromagnetic order  from microscopic antiferromagnetic interactions is a 
relevant example of the physical significance of negative
temperatures in spin systems.

\subsection{Laser systems} 

Population inversions play a fundamental role in the physics of laser 
systems~\cite{schawlow58}. The simplest example of a two-level
laser discussed by Machlup~\cite{machlup75} clarifies that this condition, 
applied to open systems, implies negative power absorption, i.e. power
emission.
 Let indeed consider such a system where the quantum state ``0'' is the ground 
state with energy $E_0$ and the state
  ``1'' is the excited state with energy $E_1$. Let $h \nu=(E_1-E_0)$ be the 
(positive) energy difference between 
the two levels, where $h$ is the Planck constant.  Here we focus on a steady 
nonequilibrium process where the system is coupled to an external 
electromagnetic
field. The rate of absorption $\mathcal{P}$ of energy from the field reads
\begin{equation}
\mathcal{P}=h \nu (w_{0\rightarrow 1}N_0 - w_{1\rightarrow 0}N_1)
\end{equation}
where $w_{0\rightarrow 1}$ and $w_{1\rightarrow 0}$ are the induced transition 
rates between the two levels while $N_0$ and $N_1$ are the
number of atoms in the lower and higher state, respectively.  Time-reversal 
invariance
implies that induced transition probabilities are symmetrical, i.e. 
$w_{0\rightarrow 1}= w_{1\rightarrow 0}=w$~\cite{yariv89}. Therefore, the 
absorption rate
simplifies to $\mathcal{P}=h \nu w(N_0 -N_1)$. Accordingly, when a population 
inversion is steadily sustained in the system, i.e. when 
$N_1>N_0$, a power emission occurs.

The concept of negative absorption is central in the history of the development 
of laser systems  
and dates back to the results by Kramers~\cite{kramers24} and the early 
experiments by Ladenburg and coworkers on an electrically excited Neon 
gas~\cite{kopfermann28}.
It is important to note that the occurrence of population inversions in steady 
nonequilibrium regimes can not be related to a global negative temperature in 
the sense of equilibrium thermodynamics (see first point of Ramsey's 
requirements in Section~\ref{sec:mag_sys})~\cite{machlup75}. If a {\it local 
equilibrium hypothesis} is verified, a negative temperature 
can be defined to characterize population inversions in a small but still 
macroscopic portion of the system~\cite{lepri03} (see also Section~\ref{sec:fourier}). 
Nevertheless, population inversions may occur even in the absence
of local equilibrium, when no proper temperature is well defined.

Several techniques to obtain population inversions and negative-temperature 
states were proposed and tested.
Among the main mechanisms, it was discovered that in a three-level system with 
unequally spaced energy levels $E_3>E_2>E_1$, a large saturating field at 
frequency 
$(E_3-E_1)/h$ can induce a population inversion  in the pairs of levels $(3,2)$ 
or $(2,1)$~\cite{bloembergen56}.
Other examples included  the use of  electric 
discharges~\cite{sanders59,javan59} or rapid gas expansions~\cite{basov63}.
In the case of semiconductor lasers~\cite{chow99}, 
Basov discusses in his Nobel lecture~\cite{basov65} three different methods, 
namely:{\it(i)} optical pumping by means of an exciting field
shined one the sample, {\it(ii)} creation of electron-hole pairs through beams 
of fast electrons, {\it(iii)} injection of electrons and holes
through p-n junctions.  

\subsection{Cold atoms in optical lattices}
\label{sec:coldatoms}

In the last decades, the physics of cold atoms in optical lattices has become a 
prominent topic of investigation,
with important applications in quantum control~\cite{haycock00} and quantum 
computation~\cite{brennen99,pachos03}. They also represent the ideal set-up for 
the study of quantum phase transitions~\cite{greiner02,simon11}  and many-body 
localization~\cite{schreiber15,kaufman16}, as well as of the interplay between 
nonlinearity,
discreteness and disorder in low-dimensional quantum 
systems~\cite{fallani07,franzosi11}. 
Optical lattices~\cite{jessen96} are obtained by  interference patterns of 
counter-propagating laser beams which produce
a stable and spatially periodic external potential for neutral atoms. This 
technique can be used to confine cold
atoms in the wells of the optical potential and to make them experience a 
quantum tunneling effect between neighboring wells,  with a resulting frequency 
spectrum characterized by bands. In particular, it can be shown that the 
quantum dynamics of a system of bosonic atoms in an optical lattice is described 
by a Bose-Hubbard model~\cite{jaksch98,franzosi11}.

In this context, Mosk observed that the combination of a band gap and of a 
negative effective band curvature (aka negative band mass) could satisfy the 
requirements for thermalization of the gas at negative 
temperatures~\cite{mosk05}.
In analogy with early experiments on magnetic systems (see 
Section~\ref{sec:mag_sys}), this peculiar thermalization process requires that 
the typical time scales for energy recombination are much faster than energy 
losses.  
More precisely, it is necessary that the selected  band  with negative effective 
mass displays a reduced rate
of interband scattering processes, which
provide an effective mechanism of energy dissipation towards different 
bands~\cite{mosk05}. 

Let us analyze more in detail this mechanism of thermalization at negative 
temperature in the framework of 
the Bose-Hubbard model~\cite{gersch63} for an optical lattice with $N$ sites. 
The quantum many-body Hamiltonian operator reads
\begin{equation}
\hat{\mathcal{H}} = \frac{1}{2} U \sum_{j=1}^N \hat{n}_j(\hat{n}_j-1) 
-J\sum_{\langle j k \rangle} \left( \hat{a}_j^\dagger \hat{a}_k + \mbox{h.c.} 
\right) + \sum_{j=1}^N V(\mathbf{x}_j) \hat{n}_j
\end{equation}
 where $\hat{a}_j^\dagger$ and $\hat{a}_j$ are bosonic creation and annihilation 
operators for an atom on lattice 
 site $j$, satisfying standard commutation relations $[\hat{a}_j, 
\hat{a}_k^\dagger] = \delta_{j,k}$, and 
 $\hat{n}_j=\hat{a}_j^\dagger \hat{a}_j$ is the number operator for site $j$.
 The first sum in $\hat{\mathcal{H}}$ accounts effectively for the  interactions 
of atoms in the same lattice well
 and is modulated by the parameter $U$, whose sign is determined by the 
attractive or repulsive nature of atom 
 interactions. The second contribution, proportional to $J$, represents the 
hopping energy between nearest neighbor sites $\langle j k \rangle$. Finally, 
the
 last sum describes the contribution of an external trapping potential 
$V(\mathbf{x}_j)$ (a magnetic trap), in  the form of shifts of local site 
energies. In the following, we shall specialize $V(\mathbf{x})$  to the case of 
an harmonic 
 trapping potential 	$V(\mathbf{x})=V_0 |\mathbf{x}|^2$.

Let us assume that  the parameters  $U$ and $V_0$ are tunable, in such a way 
that they can assume both positive and negative values. The procedure  proposed 
by Mosk~\cite{mosk05} and later improved by Rapp, Mandt and Rosch~\cite{rapp10} 
essentially consists of preparing  an ultracold atomic gas with repulsive 
interactions ($U>0$)  in an optical lattice with confinement $V_0>0$. In this 
configuration, the atomic system can be assumed in equilibrium at positive 
temperature.  The height of the optical lattice potential is then increased  
such that $U/J\gg 1$.
The system enters a deep Mott-insulator phase, which inhibits any transition 
between different lattice sites and ``freezes'' the gas in its initial state. 
The signs of $V_0$ and $U$ are then rapidly changed with no entropy variation 
and finally the intensity of the laser beams is lowered again as to bring $U/J$ 
to its original value. Basically, this protocol inverts the sign of the 
effective masses of the occupied states, operating a sort of controlled 
``population inversion'': in the equilibrium state that is reached, the new 
temperature is negative, the inter-particle potential is attractive ($U<0$) and 
the cloud of cold atoms is confined by the harmonic potential with 
$V_0<0$~\cite{rapp12,rapp13,mandt13}. The change of sign of the interaction term 
$U$ can be realized by tuning the magnetic bias near a Feshbach 
resonance~\cite{feshbach62,inouye98}.

The above described NAT state was actually realized in a famous experiment on 
cold atoms by Braun et al.~\cite{braun13} in 2013. The authors used a 
Bose-Einstein condensate of $^{39}$K atoms in a 3-dimensional simple-cubic 
optical lattice.
At the end of the experimental protocol, the momentum distribution of the atomic 
cloud was probed  by mean of time-of-flight imaging~\cite{kastberg95}, i.e. 
measuring the spatial spreading of the cloud during a small time interval (7 
ms); in this way it was possible to verify that the distribution was peaked at 
the borders of the Brillouin zone, corresponding to maximal kinetic energy, as 
expected for a NAT state. The density of states was also in optimal agreement 
with the expected negative-temperature Bose-Einstein distribution. The 
equilibrium state, for optimized choices of the experimental parameters, was 
found to last for times of the order of hundreds of milliseconds (before decaying 
due to energy and atom losses), i.e. much longer 
than the typical lattice tunneling time. 
These experimental results are therefore a convincing evidence of the 
possibility to realize equilibrium thermodynamic states at negative temperature 
also for motional degrees of freedom.

Unlike point vortices and spin systems, whose energy  does not include any 
contribution from motional degrees of freedom, a negative effective mass 
provides  an efficient mechanism for bounding kinetic energies from above.
Otherwise, the presence of a standard positive-definite kinetic term would 
forbid negative temperatures, since the phase-space region accessible to 
motional degrees of freedom would increase with the total energy, due to the 
quadratic dependence on momenta, see Sec.~\ref{sec:nat}. 
From a more general point of view, we point out that the experimental protocol 
here discussed amounts essentially  to an inversion of the sign the total 
Hamiltonian operator of the trapped gas. This operation is  analogous to the one 
performed in paramagnetic systems in the early experiments by Purcell and 
Pound~\cite{purcell51} by means of a magnetic field inversion. 
We conclude this subsection by noting that also genuinely dynamical effects, 
usually related to nonlinearities, can contribute to restrict the accessible 
phase-space and produce NAT states. We will discuss this point in  Sec.~\ref{sec_6.1}
in the context of the Discrete Nonlinear Schr\"odinger Equation.

\subsection{Discrete Nonlinear Schr\"odinger Equation} 
\label{sec:esdnls}
 
The Discrete Nonlinear Schr\"odinger (DNLS) equation describes a simple model of 
a nonlinear lattice of coupled 
oscillators~\cite{Eilbeck85,Kevrekidis09}. For a one-dimensional chain with 
nearest neighbor interactions, it can be written as
\be
\label{eq:Intro_DNLS}
i \dot z_j = -\Lambda |z_j|^2 z_j -z_{j+1} - z_{j-1} \,,\quad j=1,\cdots,N
\ee
where $z_j$ are complex-valued amplitudes and $\Lambda$ is a nonlinear 
coefficient. Unless otherwise specified, throughout this paper $\Lambda$ will be
assumed positive without any loss of generality.

Since the pioneering investigations of the 50's, the DNLS equation has been 
widely studied in several
domains of physics. It was firstly derived by Holstein~\cite{holstein59} within 
a tight-binding approximation for
the motion of polarons in molecular crystals and 
later introduced in the context of the Davydov's theory to describe energy 
transfer mechanisms
in proteins~\cite{davydov73}. Its novelty was immediately noticed  as, unlike 
the {\it continuous} nonlinear Schr\"odinger
and the Ablowitz-Ladik equation~\cite{ablowitz76,scott03}, it is a nonintegrable 
model. 
Rapidly, it became clear that the DNLS equation and its generalizations 
including high-order~\cite{weinstein99}
or saturable~\cite{khare05} nonlinearities or nonlocal 
interactions~\cite{kevrekidis03} deserved great 
attention for the peculiar properties of its chaotic trajectories and periodic 
orbits~\cite{Eilbeck03}. Indeed, this model is of
particular interest for the study of {\it discrete breather} solutions, i.e. 
nonlinear and spatially localized
excitations characterized by periodic oscillations in 
time~\cite{flach98,flach08} and it has become a prototype model
of nonlinear lattice dynamics~\cite{Kevrekidis09}. 

In addition to its  theoretical interest, the DNLS equation is also 
representative model of nonlinear wave transport in a 
broad range of experimentally accessible setups~\cite{Kevrekidis09}. These 
include nonlinear optical waveguide 
arrays~\cite{christodoulides88,eisenberg98,morandotti99}, cold atoms in optical 
lattices~\cite{smerzi97,trombettoni01,cataliotti01,cataliotti03}, electric 
transmission
lines~\cite{sato07} and nanomagnetic systems~\cite{borlenghi14,borlenghi15}.

\begin{figure}[h!]
\centering
\includegraphics[width=.6\linewidth]{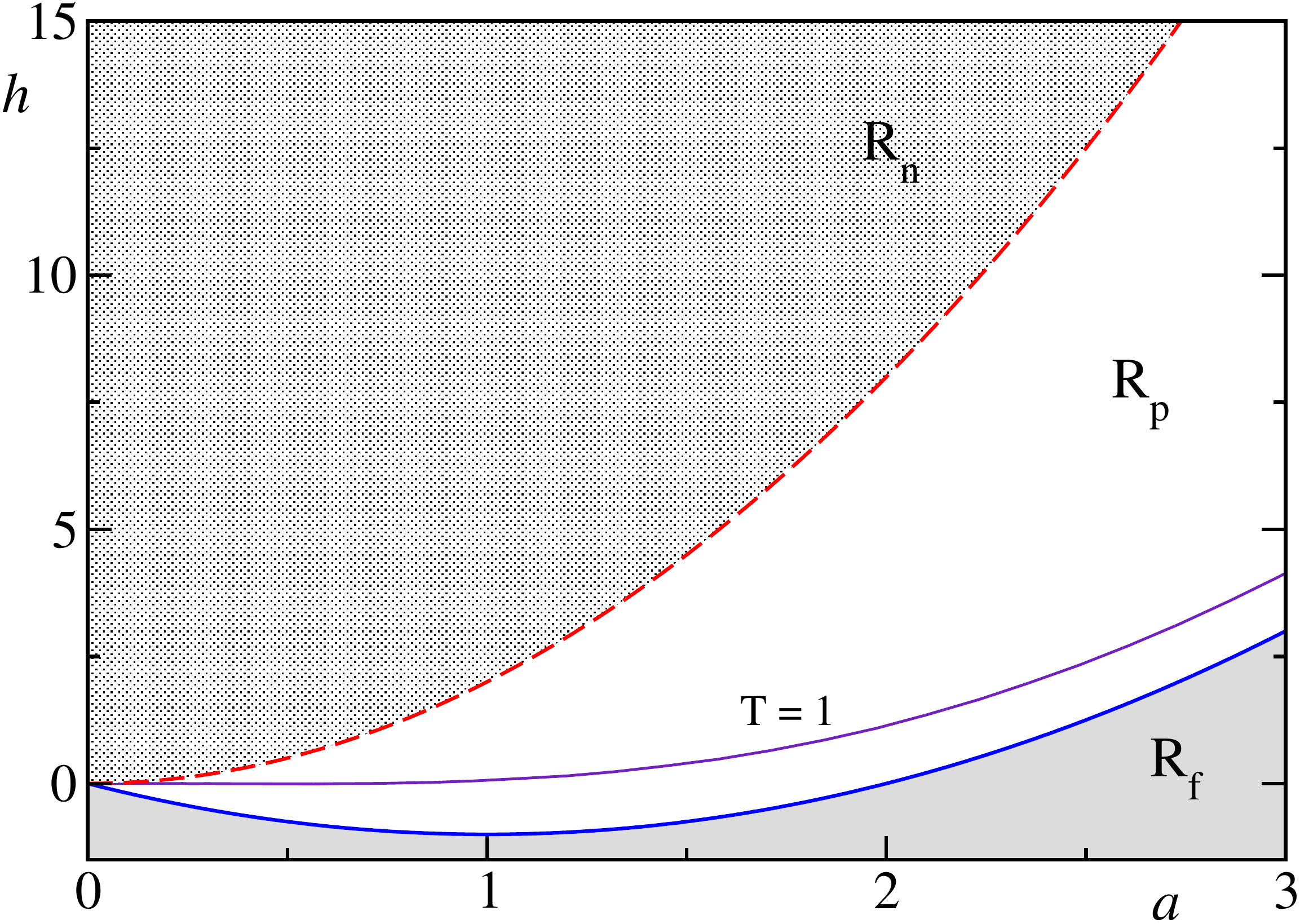}
\caption{Equilibrium phase diagram $(a,h)$ of the DNLS equation for $\Lambda=2$. We show the 
isothermal lines $T=0$ (solid blue), 
$T=1$ (solid purple) and $\beta=0$ (dashed red). }
\label{fig:phase_diagDNLS}
\end{figure}

The statistical mechanics of the DNLS model was discussed in 2000 by Rasmussen 
{et al.} within the grand-canonical 
ensemble~\cite{rasmussen00}.  A Gibbsian measure was introduced which takes into 
account the two conserved quantities of the model, namely the total energy
\footnote{It can be verified that Eq.~(\ref{eq:Intro_DNLS}) is obtained from the 
Hamilton equations of motion $\dot z_j=-\partial H / \partial iz_j^*$,
where $(z_j,iz_j^*)$ are canonical variables.}
\be
\label{eq:Hdnsle}
H=\sum_{j=1}^N \frac{\Lambda}{2}|z_j|^4 +  \left(z_j^* z_{j+1} + c.c.\right)\ee   
and the total norm 
\be
\label{eq:Adnsle}
A=\sum_{j=1}^N |z_j|^2 \,.
\ee  	
Accordingly, the resulting equilibrium {\it phase diagram} is two-dimensional 
and it can be represented in the space of
densities $(a,h)$, where $a=A/N$ is the average norm density and $h=H/N$ is the 
average energy density, see Fig.~\ref{fig:phase_diagDNLS}. 
In this diagram, the ground state of the model ($T=0^+$) is identified by the 
condition $h=\Lambda a^2/2-2a$. States below this 
line are not accessible and belong to the forbidden region $\mathrm{R_f}$. 
Moreover, the limit of diverging temperatures ($\beta=0$) defines the  line 
$\mathcal C_\infty$, $h=\Lambda a^2$. Accordingly, positive-temperature states 
are confined in the region 
 $\mathrm{R_p}$  between $\mathcal C_0$ and  $\mathcal C_\infty$.  
Negative-temperature states  were conjectured to exist in the region 
$\mathrm{R_n}$ above the infinite temperature line,
although a direct treatment was impossible because of the ill definiteness of 
the grand-canonical measure in $\mathrm{R_n}$~\cite{rasmussen00}.	
  
In successive papers, it was shown that negative-temperature states in the DNLS 
equation are metastable and
that in the thermodynamic limit the system eventually relaxes   to  an 
equilibrium
state characterized by a single breather excitation superposed to an 
infinite-temperature background~\cite{RN,R1,R2,R3,R4}.

In parallel, the nature of the negative-temperature region has been extensively 
studied numerically. In particular, it was shown that
the relaxation process can be extremely slow  and that the system evolves 
towards a stationary state with a finite density
of breathers and negative microcanonical 
temperature~\cite{JR,Iubini13}. This apparent contradiction with the 
theoretical predictions  was
 clarified in~\cite{Iubini14,Iubini17,ICOPP} by pointing out
two different sources of slowness.
On the one hand localization occurs via a slow coarsening of breathers. On the 
other hand, the large oscillation
frequency of a breather state tends to decouple it from the rest of the system, 
so that the observed timescales for 
breather-background norm transfer  are exponentially long in the breather 
norm~\cite{ICOPP}. 
As a result the asymptotic convergence to the single-breather 
equilibrium state can be practically unattainable. 
In addition to the above phenomena, the equilibrium properties of the model in 
the $\mathrm{R_n}$  region  have been recently reconsidered in the framework of 
large deviation theory~\cite{GILM1}, finding that negative temperature states do 
exist as genuine equilibrium states for finite but large system sizes.

The DNLS equation has recently used as a prototypical model to study the role of 
negative-temperature
states and localization in out-of-equilibrium contexts~\cite{iubini17entropy}. 
In particular, it was shown that a DNLS chain in contact with a reservoir at 
positive temperature and
a pure norm dissipator reaches a partially localized nonequilibrium steady state 
where negative temperatures
are spontaneously  created.  This phenomenon
brings to the fore several aspects of the role of negative-temperature states in 
non-equilibrium statistical mechanics.
Some of these points will be discussed in Sections~\ref{sec:dnls},~\ref{sec:fourier}.

We conclude this section by observing that in the limit of vanishing 
nonlinearity, $\Lambda\rightarrow 0$, the DNLS equation
reduces to a standard Schr\"odinger equation on a lattice. The thermodynamics of 
this model resembles the behavior of the class
of paramagnetic systems discussed by Ramsey and implies the existence of a 
genuine negative-temperature region where no condensation
occurs~\cite{iubini19b, buonsante16} (see also Section~\ref{ssec1}).


\newpage

\section{Alternative interpretations }
\label{sec:alternative}

All physical systems described in Section~\ref{sec:examples} admit
long-lasting states at high energy, whose statistical properties are
conveniently portrayed by an equilibrium description at negative absolute temperatures.
While no doubts are usually raised about the aforementioned experimental
observations, their theoretical interpretation as
thermal equilibrium states at negative temperature has raised some concern and criticisms,
for different reasons. First, the very definition of temperature
given in Section~\ref{sec:entropy} has been questioned by some
authors~\cite{berdichevsky91, hilbert14, sokolov14, hanggi15, campisi15},
claiming that the definition of absolute temperature usually adopted in statistical mechanics
does not properly reproduce the basic principles of thermodynamics, and should be
replaced by a different one, which does not admit negative values. Other authors criticize
the description of the physical states observed in experiments as genuine equilibrium states,
and point out the occurrence of paradoxical results for thermodynamic cycles~\cite{romero13,struchtrup18}.
In this Section we first discuss the alternative statistical description proposed by
the authors who oppose to the traditional one; then we review the main contributions
to the (still ongoing) debate about negative temperatures. 

\subsection{Two definitions of entropy (and temperature)}
\label{gibbstemperature}

As discussed in Section~\ref{sec:entropy}, the link between mechanics and 
thermodynamics
is given by the formula
\begin{equation}
\label{eq:entropy2}
 S(E)=k_B \ln W\,,
\end{equation}
where, in Boltzmann's picture, $W$ is the number of different, equiprobable 
states that the system is allowed to assume along its dynamics at fixed energy 
$E$,
given a proper discretization of the phase-space. A precise mathematical 
expression of $W$ is available for any system of $N$ particles described by a 
Hamiltonian  $\mathcal{H}({\bf Q}, {\bf P})$,
being ${\bf Q}\in \mathbb{R}^{dN}$ and ${\bf P}\in \mathbb{R}^{dN}$ the 
canonical positions and momenta in a $d$-dimensional space: one just exploits
the proportionality between $W(E)$ and the density of states $\omega(E)$,
\begin{equation}
 W(E)= \epsilon \omega(E)\,
\end{equation}
leading to Eq.~\eqref{eq:entropy2}. It has been pointed out~\cite{hilbert14} that 
Eq.~\eqref{eq:entropy} is -- strictly speaking -- not consistent from a 
dimensional point of view, since the argument of the logarithm should be 
dimensionless. For this reason, some authors prefer to keep the multiplicative 
factor $\epsilon$ inside the logarithm, with the physical meaning of the 
uncertainty associated with the measurement of $E$ (which is non-vanishing, due 
to Heisenberg's uncertainty principle, in every physical system). Of course the 
particular choice of $\epsilon$ does not affect any measurable quantity, since 
it results in an unessential additive correction to the entropy.

An alternative definition of entropy can be found in a seminal work by J. W. 
Gibbs~\cite{gibbs02}, later adopted also by Hertz~\cite{hertz10}. The idea is to define $W(E)$ as the number of states in the 
phase-space volume \textit{enclosed} by  the hypersurface at constant energy 
$E$, instead of those \textit{on} the hypersurface. In other words, one can 
define the quantity
\begin{equation}
\label{eq:sigma}
 \Sigma(E)= \int \mathcal{D}\mathbf{Q}\mathcal{D}\mathbf{P}\, 
\Theta\left(E-\mathcal{H}(\mathbf{Q},\mathbf{P})\right)\,,
\end{equation}
$\Theta$ being the Heaviside function,
and the corresponding \textit{Gibbs entropy}:
\begin{equation}
S_G(E,N)=k_B \ln \Sigma(E,N)\,.
\end{equation} 
Note that $\Sigma(E)$ and the density of states verify the relation
\begin{equation}
 \omega(E)=\frac{\partial \Sigma}{\partial E}\,.
\end{equation} 
Sometimes Boltzmann entropy, $S_B$, is also called \textit{surface entropy}, 
while Gibbs' one is referred to as \textit{volume entropy}.
The above volume entropy can be used to define, in a natural 
way, the so-called ``Gibbs temperature'' $T_G$:
\begin{equation}
\frac{1}{T_G}=\frac{\partial S_G(E,N)}{\partial E}\, .
\end{equation} 
In Section~\ref{sec:tgibbs} we shall recall that this quantity is characterized 
by very interesting properties.

It is important to recognize that in most physical situations $S_B$ and $S_G$ 
are equivalent, as soon as the thermodynamic limit $N\gg1$ is considered~\cite{huang88} . For 
instance, it is easy to verify that for a ideal gas of $N$ particle with mass 
$m$ in a $d-$dimensional space,
\begin{equation}
 \Sigma(E)=c(2mE)^{Nd/2}
\end{equation}
and
\begin{equation}
 \omega(E)=c m Nd (2mE)^{Nd/2-1}
\end{equation}
where $c=\pi^{Nd/2}/\Gamma(Nd/2+1)$, so that the difference between Gibbs and 
Boltzmann entropy,
\begin{equation}
 S^{id}_G(E)-S^{id}_B(E)=\ln \left( \frac{2E}{Nd} \right)
\end{equation} 
is subextensive, and it can be neglected in the thermodynamic limit.

\subsection{Main properties of Gibbs' formalism}
\label{sec:tgibbs}
Due to the equivalence between $T_B$ and $T_G$ for most physical systems, 
usually it is not so important to specify whether the temperature appearing in 
a 
formula comes from the Boltzmann's formalism or from the Gibbs' one: the 
relative
difference is of order $N^{-1}$, i.e. it is negligible to any practical 
purpose. 
Conversely, when dealing with systems with bounded spectrum, as those discussed 
in Section~\ref{sec:examples}, one has to bear in mind that $T_B$ and $T_G$ can 
assume
very different values: indeed, for such systems 
it is possible
to find some energy regime in which $\omega(E)$ is a decreasing function of $E$,
resulting in negative values of $T_B$; on the other hand, $\Sigma(E)$ is 
monotonically increasing,
by definition, and therefore $T_G$ is always positive. It is then crucial to 
distinguish between
the two alternatives.

As we will widely discuss in the next Sections, most results of Statistical 
Mechanics
are expressed in terms of $S_B$ and $T_B$; these relations still hold true when 
$T_B$ assumes
negative values, leading sometimes to counterintuitive effects, which can be 
actually observed in
experiments or numerical simulations.
There are, however, also some classical results
which are derived in the Gibbs' formalism, so that temperature and entropy 
appearing in the formulae
have to be intended as $T_G$ and $S_G$ in these cases. This is a serious 
limitation if Gibbs' and Boltzmann's
formalisms are not equivalent, as it happens for the kind of systems presented 
in Section~\ref{sec:examples}:
as we shall discuss in Section~\ref{sec:equilibrium}, $T_G$ and $S_G$ cannot be easily measured in experiments, and the mentioned 
results have therefore little
practical relevance. In the following we recall the main results involving $S_G$ 
and $T_G$. 
 
 \subsubsection{Equipartition theorem}
 \label{sec:equipartition}
 
 A useful result of statistical mechanics is the so-called ``Equipartition 
theorem'', which states that,
 given a Hamiltonian system $\mathcal{H}(\mathbf{Q},\mathbf{P})$, the following 
relation holds:
 \begin{equation}
 \label{eq:intro_equip}
  \left\langle x_i \frac{\partial \mathcal{H}}{\partial x_i} \right\rangle = k_B 
T\,,
 \end{equation}
where $x_i$ is a certain canonical coordinate and $\langle \cdot \rangle$ 
denotes a microcanonical average. If $\mathcal{H}(\mathbf{Q},\mathbf{P})$ 
depends quadratically on the momenta,
Eq.~\eqref{eq:intro_equip} implies that energy is equally distributed, on 
average, among the different kinetic degrees of freedom,
hence the name of the theorem.

In deriving such result (see e.g. Ref.~\cite{huang88}) one makes use of Gibbs' 
definition of entropy; consequently, Eq.~\eqref{eq:intro_equip} is exactly 
verified only if one interprets $T$ as $T_G$. See also Ref.~\cite{dunkel14} on 
this point.

 \subsubsection{Exact validity of the Thermodynamic Relations}

 In thermodynamics, energy conservation is expressed by the First Law, which 
describes the energetic balance of a physical system subjected to external 
forces and/or thermal coupling with the environment, during an infinitesimal 
thermodynamic transformation. First Law can be expressed as
 \begin{equation}
  dE=\delta Q+dW\,,
 \end{equation}
 where $dE$ is the total amount of energy acquired by the system during the 
transformation, given by the sum of the absorbed heat $\delta Q$ and of the work 
$dW$ which the system is subjected to. This infinitesimal work is due to the 
action of external thermodynamic forces, e.g. pressure or magnetic field, 
resulting in infinitesimal variations of macroscopic parameters of the system, 
such as volume or magnetization. In the following, we shall indicate the $n$ 
external forces acting on the system as $a_1, a_2, ..., a_n$, and we shall refer 
to the corresponding macroscopic parameters as $A_1, A_2, ..., A_n$. In formulae 
we have, by definition,
\begin{equation}
\label{eq:workmacro}
 dW= -\sum_{j=1}^n a_j d A_j\,.
\end{equation}
It is a phenomenological result of thermodynamics
that a state function $S(E, A_1, ..., A_n)$, the \textit{thermodynamic} entropy, 
exists such that
\begin{equation}
 \delta Q = T d S\,,
\end{equation} 
as far as reversible transformations are considered. The First Law can then be 
expressed as
\begin{equation}
\label{eq:first2}
 dE=  -\sum_{j=1}^n a_j d A_j + T d S\,.
\end{equation}
As a consequence of the existence of the thermodynamic entropy we have that
\begin{equation}
\label{eq:thermrel1}
 \frac{1}{T}=\frac{\partial S}{\partial E}\,,
\end{equation}
as it can be deduced from Eq.~\eqref{eq:first2} in the case of a reversible 
transformation with fixed $\{A_j\}$. Similarly, if we consider an infinitesimal 
transformation in which only $A_j$ is varied, and such that energy remains constant, 
we have
\begin{equation}
\label{eq:thermrel2}
 \frac{a_j}{T}=\frac{\partial S}{\partial A_j}\,.
\end{equation}
Equations~\eqref{eq:thermrel1}~and~\eqref{eq:thermrel2} are often referred to as \textit{Thermodynamic Relations}.

In order to give a statistical mechanics interpretation of the above experimental 
findings, one usually identifies the thermodynamic potentials with 
microcanonical averages of suitable mechanical observables. Let us denote by 
$\mathcal{H}(\mathbf{Q},\mathbf{P}; A_1,...A_n)$ the Hamiltonian of the system, 
so that we can express work as
\begin{equation}
\label{eq:workmicro}
 dW= \sum_j \left\langle \frac{\partial \mathcal{H}}{\partial A_j}\right\rangle 
d A_j\,,
\end{equation}
where $\langle \cdot \rangle$, under the assumption of ergodicity, is an average 
on the microcanonical ensemble. Let us just mention that, according to the 
definition introduced by Jarzynski in Ref.~\cite{jarzynski07}, we are 
considering here the \textit{inclusive} definition of work. By comparing 
Eq.~\eqref{eq:workmacro} and  Eq.~\eqref{eq:workmicro}, one deduces
\begin{equation}
\label{eq:aj}
 a_j= -\left\langle \frac{\partial \mathcal{H}}{\partial A_j}\right\rangle\, 
\quad \quad \forall j.
\end{equation} 

It can be shown~\cite{campisi15} that Eq.~\eqref{eq:thermrel1} 
and~\eqref{eq:thermrel2} are verified exactly, also for a small number of 
particles $N$, only if the entropy $S$ is a function of $\Sigma(E)$ (the 
phase-space volume defined by Eq.~\eqref{eq:sigma}), i.e. only if
\begin{equation}
\label{eq:campisi}
 S(E)=g[\Sigma(E)]\,,
\end{equation}
for some, sufficiently regular, $g: \mathbb{R} \mapsto \mathbb{R}$. Indeed, if 
this is the case, Eq.~\eqref{eq:thermrel2} can be written as
\begin{equation}
 \frac{a_j}{T}=g'(\Sigma)\frac{\partial \Sigma}{\partial A_j}\,,
\end{equation} 
which, by taking into account Eq.~\eqref{eq:aj}, leads to
\begin{equation}
\begin{aligned}
  T g'(\Sigma) \frac{\partial \Sigma}{\partial A_j}&=-\frac{1}{\omega}\int 
\mathcal{D}\mathbf{Q}\,\mathcal{D}\mathbf{P}\,\delta(E-\mathcal{H}) 
\frac{\partial \mathcal{H}}{\partial A_j}\\
  &=\frac{1}{\omega}\frac{\partial}{\partial A_j}\int 
\mathcal{D}\mathbf{Q}\,\mathcal{D}\mathbf{P}\,\Theta(E-\mathcal{H})\\
  &=\frac{1}{\omega}\frac{\partial \Sigma}{\partial A_j}\,.
\end{aligned}
\end{equation} 
This result is clearly consistent with Eq.~\eqref{eq:thermrel1}:
\begin{equation}
 \frac{1}{T}=g'(\Sigma)\omega=\frac{\partial S}{\partial E}\,.
 \end{equation} 
The simultaneous validity of Eqs.~\eqref{eq:thermrel1} and~\eqref{eq:thermrel2} 
is only guaranteed if $S(E)$ has the form~\eqref{eq:campisi}.
 The particular choice $g(\Sigma)=\Sigma$ corresponds to Gibbs' definition of 
entropy. It can be deduced, for instance,
 from the specific case of the ideal gas~\cite{campisi15}.

 \subsubsection{Adiabatic invariance of the entropy}
 
Making reference to the notation introduced in the previous paragraph, suppose 
now that parameters $\{A_j\}$ are time-dependent, and that the protocol which 
rules their evolution has a typical time $\tau$. The total energy $E$ is, in 
general, not conserved. If $\tau$ is very large, i.e. the transformation of the 
macroscopic parameters $\{A_j\}$ is very slow, the process is called 
\textit{adiabatic} in the \textit{mechanical} sense. A function $I(E(t); 
\{A_j(t)\})$, depending on the total energy of the system and on the macroscopic 
parameters, is called \textit{adiabatic invariant} if 
\begin{equation}
\lim_{\tau \to \infty}  I(E(t); \{A_j(t)\})-I(E(0); \{A_j(0)\}) = 0\quad 
\quad\forall t :\quad 0\le t<\tau
\end{equation} 
i.e. if its evolution can be approximated by a constant behavior in the interval 
$[0,\tau)$ when the characteristic time $\tau$ of the evolution is sufficiently 
large.

One of the assumptions of thermodynamics is that a very slow change of the 
parameters $\{A_j\}$ will result in a thermodynamic transformation in which no 
heat is exchanged between the system and the environment; in other words, it is 
generally believed that an adiabatic process in the mechanical sense should be 
also adiabatic in the \textit{thermodynamic} sense~\cite{campisi05}. This 
condition can be implemented by imposing that
\begin{equation}
 \lim_{\tau \to \infty}  \frac{d E}{dt} + \sum_{j=1}^n a_j \frac{d 
A_j}{dt}=0\quad \quad\forall t :\quad 0\le t<\tau\,,
\end{equation} 
meaning that in the infinite $\tau$ limit the energy variation is only due to 
the work done on the system by the external forces.
This is only true if $S$ verifies
\begin{equation}
\lim_{\tau \to \infty}  \frac{d S}{dt} =0\quad \quad\forall t :\quad 0\le 
t<\tau\,,
\end{equation} 
i.e. if $S$ is an adiabatic invariant in the mechanical sense~\cite{campisi05, 
dunkel14}. 
 It can be proved that $S_G$ possesses the above property, even for systems with 
a small number of degrees of freedom~\cite{hertz10, kasuga61, berdichevsky97}. 
Conversely, in general $S_B$ does not verify this condition.
 
 \subsubsection{Helmholtz's theorem}
\label{sec:helmotz}

The last remarkable properties of Gibbs entropy (and temperature) that we 
discuss in this Section
is maybe a little out of the scope of this review, since it only applies to 
systems with quadratic kinetic energy;
it is the so-called ``Helmholtz's theorem'', a classical result providing a link  
between ergodicity 
and thermodynamics; this theorem had a crucial role in the
development of  Boltzmann's ideas on statistical mechanics. This topic has been 
recently discussed by Gallavotti~\cite{gallavotti13}
and Campisi and Kobe~\cite{campisi10}.

 Consider a one-dimensional system with  Hamiltonian
 \begin{equation}
\mathcal{H}(q,p, V)=\frac{p^2}{2m} + \phi(q,V)
 \end{equation} 
 where $q$ and $p$ are the canonical coordinates, $m$ is the mass of the particle 
and $V$  is a control parameter, which can be varied. For instance, one may 
think to an oscillating pendulum, whose wire has a variable length $V$. These 
systems are called \textit{monocyclic}.
 Assume that, for each value of $V$, the potential energy $\phi(q,V)$ has a 
unique minimum; assume also that, for $|q| \to \infty$, $\phi(q,V)$ diverges as  
$|q|$.
In this system the motion is surely bounded once the value $E$ of the energy is 
fixed, i.e. it is possible to find suitable functions $q_{-}(E,V)$ and 
$q_{+}(E,V)$ of the external parameters such that
\begin{equation}
q_{-}(E,V) \le q(t) \le q_{+}(E,V)\,.
\end{equation} 
The motion is also periodic, with period $\tau(E,V)$. As a consequence, the 
system is ergodic in a trivial way:
during its dynamics, the system explores all the states at energy $E$, so that 
the time averages coincide with the averages computed with the microcanonical  
distribution.

Let us now define the temperature $T$ and the pressure $P$ in terms of time 
averages $ \langle \cdot \rangle_{\tau}$
computed on the period $\tau(E,V)$:
\begin{equation}
\begin{aligned}
T &\equiv {1 \over k_B} \left\langle {p^2 \over m} \right\rangle_{\tau} \,,\\
P &\equiv -{1 \over k_B} \left\langle {\partial \phi(q,V) \over \partial V} 
\right\rangle_{\tau} \,.
\end{aligned}
\end{equation}
 Let us notice that, according to the discussion in 
Section~\ref{sec:equipartition}, the above defined temperature is actually
 the Gibbs temperature $T_G$.
The following result, due to Helmholtz, holds: the function
\begin{equation}
S(E,V)\equiv k_B \ln 2 \int_{q_{-}{}}^{q_{+}} \sqrt{2m[E-\phi(q,V)]} dq
=k _B\ln \int_{\mathcal{H}(q,p,V)<E} dp dq \,,
\end{equation}
satisfies the following relations
\begin{equation}
\label{eq:ders}
\begin{aligned}
 \frac{\partial S}{\partial E} &={1 \over T}\\
\frac{\partial S}{\partial V} &={P \over T}\, .
\end{aligned}
\end{equation}
The above  results imply a rather interesting consequence, namely that the 
quantity
\begin{equation}
\label{eq:exdiff}
dS=\frac{dE + PdV}{T}\,,
\end{equation}
where $T$ and $P$ are expressed via time averages of mechanical observable, 
is an exact differential.  $S$ can be then interpreted as a mechanical analogue 
of the
thermodynamic entropy.

Boltzmann's idea was to generalise the above result, 
valid for monocyclics, to systems with many particles;
in other words he wanted to find a function
$S(E,V)$  such that relations~\eqref{eq:ders}  and~\eqref{eq:exdiff}  are still 
valid.
In a Hamiltonian system with $N$ particles, assuming ergodicity, it is indeed 
possible to prove
a {\it Generalised Helmholtz's theorem}~\cite{campisi10} for the function
\begin{equation}
S(E,V)=k_B \ln \int_{H({\bf q}, {\bf p},V)<E} \mathcal{D}{\bf Q} \mathcal{D}{\bf 
P} \, ,
\end{equation} 
i.e. for the Gibbs entropy.
As a consequence, the time averages 
can be replaced with the microcanonical averages, and from the Generalised   
Helmholtz's theorem one
can infer the Second Law of thermodynamics.


\subsection{The debate about negative temperature}
\label{sec:debate}

In the last years we have been facing a renewed dispute on the physical
significance of negative temperatures, with different parties forming in the
scientific community.

\subsubsection{Critics of Boltzmann entropy}
Dunkel and Hilbert~\cite{dunkel14}, relying upon the
assumption that thermostatistics can be derived from Gibbs entropy 
(along the lines
of the discussion in Section~\ref{sec:tgibbs}),
concluded that negative temperatures are not compatible with this assumption. 
In particular,
they contested the observation of negative temperature in motional degrees of 
freedom,
claimed by Braun \textit{et al.} in Ref.~\cite{braun13} (see also 
Section~\ref{sec:coldatoms}).
This point of view was not completely new, since a similar position 
had been already expressed
by Berdichevsky, Kunin and Hussain in Ref.~\cite{berdichevsky91}. In that paper, 
the authors
intended to criticize the use of a negative temperature formalism to refer to
the high-energy regime of systems of vortices, as those described in 
Section~\ref{sec:vortices}.
In a more general perspective Hilbert, H\"anggi and Dunkel~\cite{hilbert14} 
argued that
the zeroth, first and second principle of thermodynamics are fulfilled only by
Gibbs entropy.

\subsubsection{In defense of Boltzmann's formalism and negative temperature}
\label{sec:defense}
Frenkel and Warren~\cite{frenkel15} showed that Gibbs
entropy is conceptually inadequate, because it fails the basic principle (a direct consequence
of zeroth law of thermodynamics, see Section~\ref{sec:termometro}) that 
two
bodies at thermal equilibrium should be at the same temperature, at variance
with the definition of entropy due to Boltzmann. These conclusions were further confirmed by the general approach
due to Swendsen and Wang~\cite{swendsen16, swendsen18}.

More recently, Abraham and Penrose~\cite{abraham17}
argued that the main argument raised against the negative temperature
concept in the context of the Purcell and Pound experiment on nuclear spin
systems~\cite{purcell51} are not logically compelling.
Moreover, they showed that the
use of the so-called volume entropy leads to predictions inconsistent with the
experimental evidence. We can say that in this paper the authors aim at
providing clear answers to the main controversial points emerging from this
dispute. In this sense, we consider Ref.~\cite{abraham17} as an important 
reference for
sketching the state of the art in this field.

A first aspect that is clarified in Ref.~\cite{abraham17} is that the thermo-statistical
consistency condition
\begin{equation}
T dS = \delta Q = dE - dW
\end{equation}
which, for a magnetic system like the one of the Purcell-Pound experiment,
specializes to $dW = - Mdh$ ($M$ being the magnetization and $h$ the external
magnetic field) holds true also for the Boltzmann ``surface'' entropy and
not only for the ``volume'' Gibbs entropy, at variance with what previously
sustained by some authors (e.g., see~\cite{berdichevsky91,dunkel14}).
In a more general perspective, it has been shown that if the
averages appearing in Eq.~\eqref{eq:workmicro} are interpreted as canonical
ones (instead of microcanonical, as proposed in Ref~\cite{dunkel14}), the consistency
criterion is always satisfied by Boltzmann entropy~\cite{frenkel15, buonsante16}.

Another important point discussed by Abraham and Penrose concerns
an experimental test of the different entropy formulas. They elaborate on the
same conceptual direction raised by Frenkel and Warren~\cite{frenkel15} and 
Swendsen
and Wang~\cite{swendsen16} that Gibbs’ entropy contradicts the zeroth law 
of
thermodynamics, while making explicit reference to the experiment by Abragam and
Proctor~\cite{abragam57,abragam58}. In this old experiment two spin systems made 
of N nuclei
of magnetic spin $\mu$ were brought into thermal contact and the question is
if and how to eventually reach a thermal equilibrium state starting from a
out-of-equlibrium initial condition, where one set of N nuclei is prepared at
null energy and the other set of N nuclei is at maximal energy. Abraham and
Penrose show that making use of Boltzmann entropy one can predict that
the overall system will eventually evolve to an equilibrium state where the
total initial energy is equally shared between the two spin systems, which
is exactly what was observed in the experiments. But they also show that
the adoption of the Gibbs entropy does not allow to predict that the energy
equipartition state is eventually approached: the energy in the two subsystems,
after being put in thermal contact, is allowed to perform a sort of random
walk, characterized by arbitrary large fluctuation preventing the possibility
of equipartition.

Finally, Abraham and Penrose comment about the possibility of designing
superefficient Carnot engines, i.e. yielding $\eta >1 $ from negative 
temperature
sources, as asserted in~\cite{braun13,ramsey56}. Making also reference to 
previous careful
analyses~\cite{frenkel15,landsberg77}, they definitely clarify that, when 
dealing with a Carnot
engine exchanging heat with two sources at temperatures of different sign,
one cannot rely upon the standard hypothesis of adiabatic transformations.
Anyway, they also argue that one cannot overtake a Carnot efficiency $\eta=1$
and that such a situation may occur without contradicting any principle of
thermodynamics, see Section~\ref{sec:cycles} below.

\begin{figure}[h!]
 \centering
 \includegraphics[width=.48\linewidth]{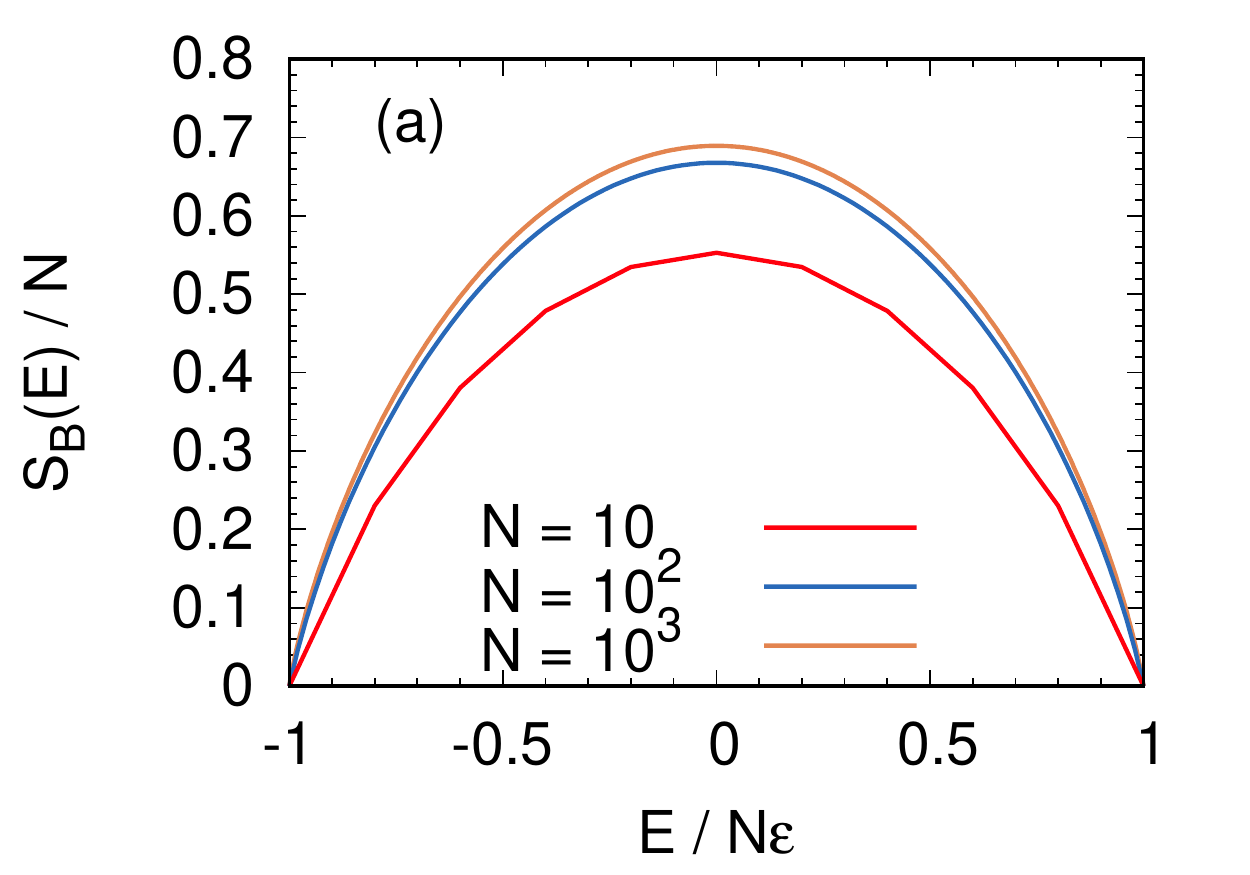}
 \includegraphics[width=.48\linewidth]{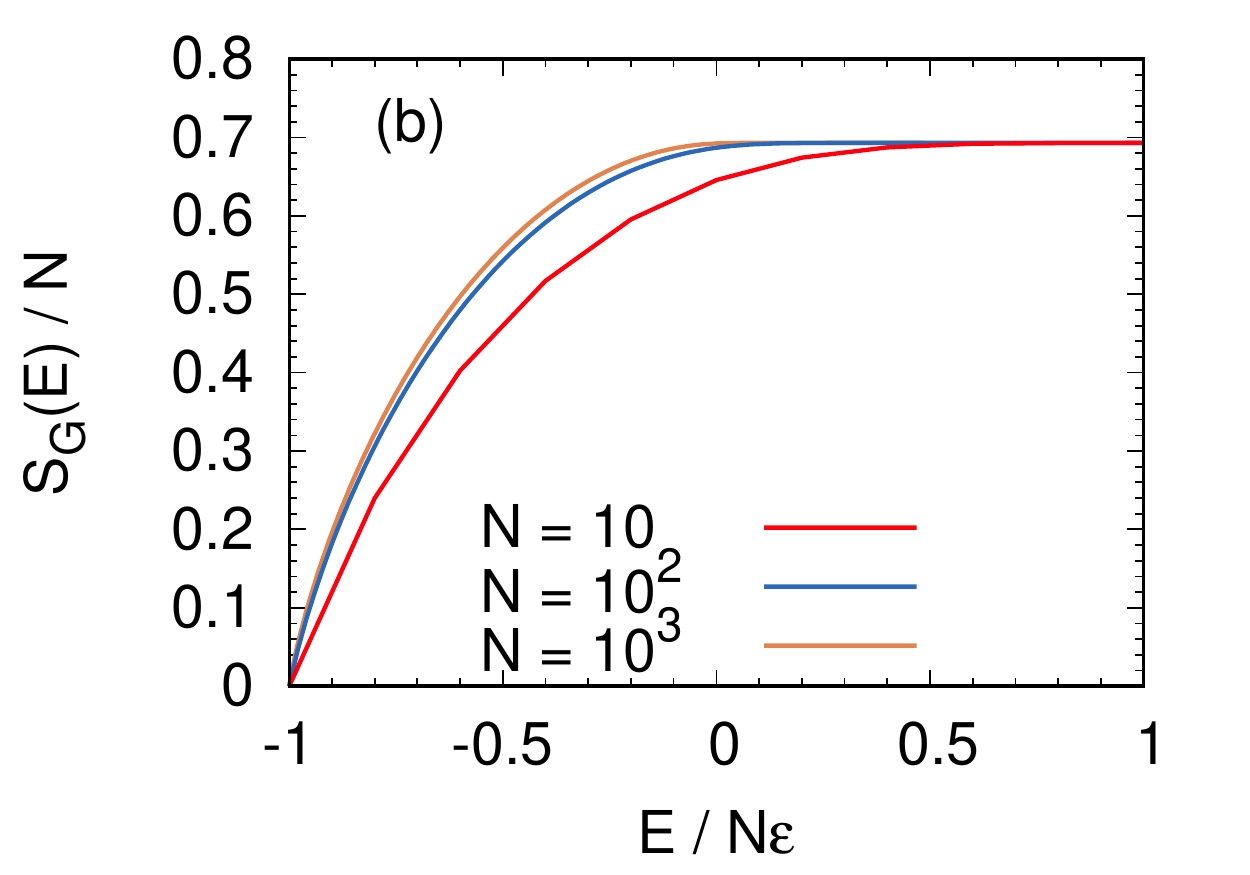}
 \caption{\label{fig:spinscaling} Entropy of the spin system~\ref{fig:spinscaling}, according (a) to the Boltzmann's and (b) to the Gibbs' definition, for different values of the number of particles $N$.}
\end{figure}

More arguments in favor of Boltzmann entropy can be found in other papers.
For instance, in Ref.~\cite{buonsante17} an example of phase transition at negative temperature is exhibited,
which could not be characterized in the Gibbs' formalism. In Ref.~\cite{cerino15} it is shown
that a correct description of thermodynamic fluctuations requires the use of Boltzmann temperature.
In Ref.s~\cite{vilar14, buonsante16, swendsen18}
the scalings of $S_G$ and $T_G$ in the thermodynamic limit are analysed,
and it is shown that they provide no useful information about the high-energy regime
of models living in bounded phase-space. To clarify this point, let us consider again the simple
spin model~\eqref{eq:hammag} introduced in Section~\ref{sec:mag_sys}.
For this system it is possible to compute explicit values for $S_B$ and $S_G$
for any fixed number of particles $N$, and the approach to the thermodynamic limit can be studied.
Indeed, denoting by $N_-=(N+E/\varepsilon)/2$ the number of negative spins in a
configuration with energy $E$, with $\varepsilon=\hbar \gamma|\mathbf{B}|/2$, the number of states
corresponding to that energy is given by
\begin{equation}
 W_{N_-}=\frac{N!}{N_-!(N-N_-)!}\,,
\end{equation} 
so that
\begin{equation}
 S_B=\ln W_{N_-}\,,\quad \quad S_G=\ln \sum_{n=0}^{N_-}W_{n}\,.
\end{equation} 
Both quantities are plotted in Fig.~\ref{fig:spinscaling}, for increasing values of $N$. As the thermodynamic limit is approached, Gibbs entropy
tends to a plateau for  $E>0$, i.e. for those values of the energy corresponding to negative temperatures in the Boltzmann description. As a consequence, Gibbs' formalism does not allow for a proper thermodynamic description of the properties of
such states, which are all identified by infinite temperature ($\beta=0$) in the limit of infinite $N$. See Section~\ref{ssec1} for a wider discussion on this model.

\subsubsection{The problem of thermodynamic cycles}
\label{sec:cycles}
The possibility to design a thermodynamic engine operating between two inverse temperatures $\beta_1$ and $\beta_2$ such that $\beta_2<\beta_1<0$, or even $\beta_2<0<\beta_1$, is another controversial point which has been being debated since the seminal paper by Ramsey~\cite{ramsey56}. In that work it is mentioned that Carnot cycles can be designed for spins systems subject to an external field $\mathbf{B}$ which is continuously varied, at a very slow rate, so that the system can be considered in a thermal equilibrium condition at every time. This external parameter plays the role assumed by the volume in the case of ideal gases. In principle, we can perform reversible transformations at fixed temperature, positive or negative, as well as adiabatic transformations in which the system does not exchange heat with the environment. It is worth noticing that this is true both at positive and at negative temperature, provided that the sign of temperature does not change during the transformation. 
The efficiency of such a Carnot cycle is defined as
\begin{equation}
\label{eq:efficiency}
 \eta=1-\frac{Q_2}{Q_1}=1-\frac{T_2}{T_1}
\end{equation}
where $Q_1$ is the heat absorbed at temperature $T_1$ and $Q_2$ is that released at temperature $T_2$ (we are assuming that $T_1$ is hotter than $T_2$, i.e. we are either in the $0<T_2<T_1$ or in the $T_2<T_1<0$ case). As argued by Ramsey, the fact that for $T_2<T_1<0$ the efficiency can be negative, and even less than $-1$, means that work must be done on the system in order to maintain the cycle: this behavior is opposite to the positive-temperature case, in which the absorbed heat is converted into work by the engine. The whole picture is consistent with the ``generalized'' Kelvin-Planck formulation of the second law of thermodynamics, enunciated by Ramsey himself, stating that at negative temperature it is not possible to completely transform work into heat (whereas, at positive temperature, this process is allowed and the reverse one is forbidden, see Section~\ref{sec:mag_sys}). In a later paper by Landsberg a precise correspondence is established between heat pumps at $T<0$ and heat engines at $T>0$~\cite{landsberg77}.

If $T_1$ and $T_2$ have different signs the scenario is a bit more involved. Ramsey noted that, for systems of nuclear spins, no way is known to operate adiabatic transformations between temperatures of opposite signs: indeed, as discussed in Section~\ref{sec:mag_sys}, the experimental protocol that is adopted to achieve negative values of $\beta$ in magnetic systems rely on a non-quasistatic inversion of the external field. However it is not clear, a priori, whether an adiabatic transformation between positive and negative temperature is forbidden in any kind of physical system, or in spin systems only~\cite{tykodi75}.
An answer to this question was provided by Sch\"{o}pf~\cite{schopf62}: after systematically analysing all non-paradoxical thermodynamic cycles, i.e. those which do not contradict the first and the second law of thermodynamics, he concluded that every adiabatic transformation linking states at negative and positive temperature would result in some inconsistencies. In a subsequent work~\cite{tykodi78} it has been shown that one of the proofs was incomplete, and that the third law of thermodynamics also needs to be assumed. A proof for the case with bounded energy is given in Ref.~\cite{tremblay76}. Under reasonable hypotheses, it is shown that if one considers a thin enough region around the $\beta=0$ surface, in the  state-variables space, entropy is constant, and maximal, at $\beta=0$. An immediate consequence is that such surface cannot be crossed by an isoentropic transformation.

While quasi-static Carnot cycles between temperatures with opposite signs are not allowed, their non-quasistatic versions can be still realised. A systematic discussion on this topic can be found in Ref.~\cite{landsberg80}. The efficiency of such engines, according to Eq.~\eqref{eq:efficiency}, is larger than one, an apparently paradoxical result. The contradiction is solved by noticing that Eq.~\eqref{eq:efficiency} is derived under the assumption that heat is absorbed by the system at temperature $T_1$ and released at temperature $T_2$, which is not possible if $T_2<0<T_1$ or $T_1<0<T_2$: in that case (as discussed in Ref.~\cite{campisi16}) heat is absorbed by both reservoirs and completely converted into work, so that the efficiency
\begin{equation}
 \eta=\frac{W}{Q}
\end{equation}
is actually always equal to 1. Obviously this does not mean that it is possible to completely convert an arbitrary large amount of heat into work: as correctly pointed out in Ref.~\cite{frenkel15, struchtrup18}, the work needed in order to realize a negative-temperature bath largely exceeds the one that can be obtained in this way. Such cycles have been experimentally realised in quantum systems~\cite{deassis19}.

\subsubsection{Critics of negative-temperature equilibrium}

In the light of all of these achievements one could have expected that
the dispute might have come to a happy end. This is not exactly the case,
because even more recently there was a revival, based on the criticism that
negative temperature states are not true equilibrium states~\cite{struchtrup18}, 
as assumed
by Ramsey in his influential paper of 1956~\cite{ramsey56}.
As recalled in Sec.~\ref{sec:nat}, if one wants to
discuss about negative temperature states, one has to deal with an isolated
thermodynamic system~\cite{abraham17}. 
Indeed, any coupling with the external environment, which is 
composed by ordinary
matter at positive temperature, would eventually lead to an equilibrium state at 
$T_B>0$, as
already understood by Ramsey~\cite{ramsey56}. See Section~\ref{sec:termometro} for further 
discussion on this point.
This is the reason why in all experiments on negative-temperature states it is 
crucial for the relaxation
time of the system to be much smaller than the typical time-scales of the 
thermalization with the environment~\cite{purcell51, pound51, braun13}.

Some authors claim that this is enough to consider states at NAT as out-of-equilibrium~\cite{romero13, struchtrup18}.
Of course this conclusion depends on which definition of ``thermodynamic equilibrium''
is chosen, but it does not affect the ability of equilibrium statistical mechanics to
properly characterize the considered states (as discussed in Section~\ref{sec:equilibrium} in some
detail).

\subsubsection{Remarks}
\label{sec:remarks}
In summary, we believe that the main merit of this dispute is that it has
re-raised the very long-standing problem about the equivalence of different
definitions of entropy and their relevance and consistency with thermodynamics
and related experimental facts. In fact, the kind of problems that
have been selected as relevant by the statistical mechanics community in the
last half a century could ignore the problem of its foundations, simply because
any definition of entropy, either Gibbs’ or Boltzmann’s, yielded the same
conclusions. Facing new questions about models with bounded energy spectra,
long-range interactions and condensation phenomena has determined a drastic
change of perspective, asking for crucial conceptual improvements in view
of a consistent and physically sound theory of thermodynamic phenomena.
By the way, it is important to mention that such a revision has to deal with
other basic concepts like ergodicity. Anyway, we want to state clearly that
the understanding of a physical phenomenon cannot depend on one's favorite
definition of entropy.

The above mentioned debate has suffered from taking a wrong perspective
from the very beginning, when it was claimed that negative temperatures
can be adopted as the right concept for referring to some peculiar physical
phenomena. Actually, the ambiguity in defining temperature by the volume
(i.e., Gibbs-like) or by the surface (i.e., Boltzmann-like) entropies is
essentially irrelevant when compared with experimental facts. For instance,
quite recent experimental achievements showed that the ``negative
temperature'' regime predicted by the Onsager’s model of point vortices does 
exist,
because of the special features of the interactions among such 
vortices~\cite{gauthier19,johnstone19}.

Also in the light of these recent achievements, it seems to us that
a general method to be adopted for identifying conditions corresponding
to negative temperature phenomena has to based on the properties of the
density of states $\omega(E)$ of a given isolated system described by 
a Hamiltonian, rather than by the ambiguous choice of the preferred 
entropy. Note that
the density of states is not prone to any ambiguous interpretation, it 
is relevant for classical as well as for quantum systems and it allows one to 
establish the basic conditions to observe unusual thermodynamic behavior, 
irrespectively of the system at hand. For instance, in Sec. 1.3 we have
already discussed how, looking at $\omega(E)$, one can easily conclude
that Hamiltonian models with extensive standard kinetic energy cannot 
exhibit negative temperatures, and also why Hamiltonian models with bounded 
phase space have to. The models discussed in Sec. 2 belong to the latter class 
of systems, with the remarkable exception of the DNLS one (see Sec. 2.5). 
We want to point out here that for this model the presence of negative
temperatures can rely upon Machlup's criterion [29]: peculiar physical
phenomena (i.e., negative temperatures) can be observed in systems 
where at infinite temperature the typical energy is finite, in formulae
\be
\int_{0}^{+\infty} e^{-\beta E} E \omega(E) dE |_{\beta=0} = 
\int_{0}^{+\infty} E \omega(E) dE = \langle E \rangle_0 < \infty
\label{Machlup_crit}
\ee 
In particular, if the system can access values of $E$ larger than 
$\langle E \rangle_0$ we have to face a new kind of thermodynamic phenomena.
It can be easily argued that condition (\ref{Machlup_crit}) applies to the
DNLS problem. In fact, as discussed in Sec. 2.5, for any fixed value of the
conserved total norm $A$ one has $\langle E \rangle_0 = 2a A$. 
This result stems from two basic physical
properties of the DNLS model: the hopping term is  not a standard kinetic 
one (actually, on the line $\beta=0$ it is subextensive, because the
phases of the complex state variables $z_j$ have to be random) and, 
beyond total energy $E$, it has an additional conserved extensive quantity 
$A$, because of the gauge symmetry $z_j \to e^{i \pi j} z_j$. 
In retrospect it is quite a peculiar fact that nowhere in the papers cited 
in Sec. 2.5 it was realized that it is would be enough to reconsider Machlup's 
criterion for concluding that negative temperatures pertain the DNLS model, 
as well as systems with population inversion, for which it was originally 
formulated (see [29]). Note also that Machlup's criterion (\ref{Machlup_crit}) 
is a stronger bound than assuming that $\omega(E)$ is just a decreasing 
function of $E$, which, as already mentioned, is a sufficient criterion for 
negative temperatures, once Boltzmann entropy is taken as a suitable 
definition.

 However, since $|\gamma|$ 
is usually proportional to the number of degrees of freedom of the system,
we remark that that condition might
be only realized in systems with very few particles.

To conclude this Section, let us stress once again that the existence of physical systems which show the 
 high-energy behaviour identified by Machlup's criterion (as those discussed in Section~\ref{sec:examples}) is an experimental fact;
 most authors (with some exceptions: see e.g. Ref.~\cite{struchtrup18})
 agree that such systems, when isolated from the external environment,
 reach thermal equilibrium even if the energy $E$ of the system corresponds to a
 decreasing branch of the density of states $\omega(E)$.
 The choice between Boltzmann or Gibbs temperature is, of course, a matter 
of taste; however  the former, for this kind of systems, has the practical advantage to assume 
opposite signs  in presence of the two, qualitatively different, thermodynamic regimes.
In our opinion, as already pointed out by Montgomery~\cite{montgomery91} in a 
comment
to Ref.~\cite{berdichevsky91}, the last point provides a very good reason to 
choose Boltzmann's
formalism when dealing with systems satisfying Machlup's criterion.

\newpage
\section{Equilibrium at $\beta < 0$ }
\label{sec:equilibrium}
In this Section we review some fundamental topics about the equilibrium statistical mechanics
of systems which admit negative absolute temperatures.
We aim at clarifying in which sense temperature is a physical observable, also for systems
where no equipartition theorem is expected to hold, and how it is related to the statistical
properties of the system. The problem of ensemble equivalence at negative temperature
is discussed in some detail: even if, in principle, the possibility to achieve negative temperature
does not imply a failure of ensemble equivalence, several relevant systems show both features.
We investigate this link by considering some particular examples.

\subsection{The problem of measuring temperature }
\label{sec:measuring}

In Section~\ref{sec:entropy} we have recalled and justified the notion of 
temperature as it is usually meant in statistical mechanics. Since two 
alternative definitions of entropy are commonly used in literature, two distinct 
quantities with the meaning of temperature, $T_B$ and $T_G$, can be defined as well: as discussed 
in Section~\ref{gibbstemperature}, in the thermodynamic limit such observables assume nearly identical 
values for most physical systems, but they may show 
very dissimilar behaviours if the system lives in bounded phase space.
From a physical point of view, a very important question to address is whether (and how)
these quantities can be measured -- at least in principle -- in actual experiments
or in numerical simulations. In other words, one would like to have  a
protocol to determine $T_B$ and $T_G$ by only sampling a system at equilibrium.
In this Section we address this problem, devoting particular attention to models
which admit NAT.

The temperature of a macroscopic object whose internal degrees of freedom are 
in thermal equilibrium is determined, in everyday practice as well as in experiments,
by putting a measuring device (a thermometer) at contact with a small portion of the system. 
The thermometer has only access to information about the state of this small 
subsystem, which we can identify with a vector ${\bf X} \in \mathbb{R}^{2dN_1}$, where 
$N_1$ is much smaller than $2dN$, the dimension of the full phase space $({\bf 
Q}, {\bf P})$. Let us also indicate with $\widetilde{{\bf X}} \in 
\mathbb{R}^{2d(N-N_1)}$ 
the remaining variables. The Hamiltonian can be decomposed into
\begin{equation}
 \mathcal{H}({\bf Q}, {\bf P})=\mathcal{H}_1({\bf 
X})+\mathcal{H}_2(\widetilde{{\bf X}})+\mathcal{H}_I({\bf X}, \widetilde{{\bf 
X}})
\end{equation} 
where $\mathcal{H}_1$ and $\mathcal{H}_2$ include all terms only depending on 
${\bf X}$ and $\widetilde{{\bf X}}$ respectively, while $\mathcal{H}_I$ 
accounts 
for the interaction terms between the two subsystems. In 
the microcanonical ensemble, where the total energy is fixed to $E$, the 
probability density function (pdf) for the full phase space  is
 \begin{equation}
 p({\bf Q}, {\bf P})=\frac{1} {\omega(E,N)} \delta (\mathcal{H}({\bf Q}, {\bf 
P})-E) \,.
\end{equation}
Let us now specialize to the physically relevant case $1 \ll N_1 \ll N$.
To determine the pdf $p_1({\bf X})$, which describes the stationary state of 
the 
subsystem ``seen'' by the thermometer, one has to marginalize $p({\bf Q}, {\bf 
P})$ by integrating over the unaccessible variables $\widetilde{{\bf X}}$. The  
derivation is quite simple, and it can be found in many textbooks~\cite{ma85, 
huang88}: assuming that $\mathcal{H}_I({\bf X}, \widetilde{{\bf X}})$  is 
negligible with respect to $\mathcal{H}_1$ and $\mathcal{H}_2$, an hypothesis 
which can be rigorously proved if the system has short-range 
interactions~\cite{campa09}, we have
 \begin{equation}
 p_1({\bf X})\simeq \frac{\omega_2(E- H_1({\bf X}), N-N_1)} {\omega(E,N)}\,,
\label{EA}
\end{equation}
where $\omega_2$ is the density of states of the subsystem 
$\widetilde{\mathbf{X}}$. By  writing $\omega$ in terms of $S_B$, the Boltzmann 
entropy defined by Eq.~\eqref{eq:boltentropy}, we have
\begin{equation}
\omega(E,N)= e^{S_B(E,N)/k_B} 
\end{equation}
and
\begin{equation}
\omega_2(E-\mathcal{H}_1({\bf X}),N-N_1)= e^{S_B(E-\mathcal{H}_1({\bf 
X}),N-N_1)/k_B} \,.
\label{EC}
\end{equation}
  Let us remark that in the previous equation the functional form of $\omega_2$ 
is assumed to be the same as that of $\omega$, which is a reasonable 
approximation in the limit $N_1 \ll N$ even if the system is not homogeneous. 
Since  $\mathcal{H}_1({\bf X}) \ll E$, we can exploit the expansion 
\begin{equation}
  S_B(E-\mathcal{H}_1({\bf X}),N-N_1)\simeq 
S_B(E)- \frac{\partial S_B(E,N)}{\partial E} \mathcal{H}_1({\bf X}) + const\,.
\label{ED} 
\end{equation}
Using equations (\ref{EA},  \ref{EC}) and (\ref{ED}), together with the 
definition of Boltzmann (inverse) temperature~\eqref{eq:beta}, one obtains the 
pdf of the small subsystem
\begin{equation}
 p_1({\bf X})=\frac{1}{Z_{\beta}} \, e^{-\beta H_1({\bf X})}\,,
\end{equation} 
$Z_{\beta}$ being a normalization factor.
In other words, the thermometer has access to a portion of the system, whose 
state is described by
a canonical distribution corresponding to the Boltzmann inverse temperature
\begin{equation}
 \beta=\frac{\partial S_B}{\partial E}\,.
\end{equation} 
Since this temperature can be negative, there is no obstacle, in principle, to 
the design of a thermometer able to measure negative temperature, as we will 
show in the next section.
 
The above discussion can be also applied to the case in which $\mathbf{X}$ 
represents one degree of freedom only, for instance a ``generalized'' momentum 
of the Hamiltonian
\begin{equation}
\label{eq:minimal}
\mathcal{H}=\sum_{n=1}^N \mathcal{K}(P_n) + \mathcal{U}(\mathbf{Q})
\end{equation}
where the variables $\{ P_n \}$, as well as the functions $\mathcal{K}(P)$ and 
$\mathcal{U}(\mathbf{Q})$, are bounded.
Using the same argument presented above, one obtains the generalised 
Maxwell-Boltzmann distribution:
\begin{equation} 
\label{eq:maxbol}
\rho(P) \propto  e^{-\beta \mathcal{K}(P)},
\end{equation}
which is valid for both positive and negative $\beta$.  The analysis of the 
momentum pdf can then be used as a  practical  recipe to measure both positive 
and negative Boltzmann temperatures in real experiments~\cite{braun13}, as 
well as in numerical simulations.
\begin{figure}[h!]
\centering
\includegraphics[width=.6\linewidth]{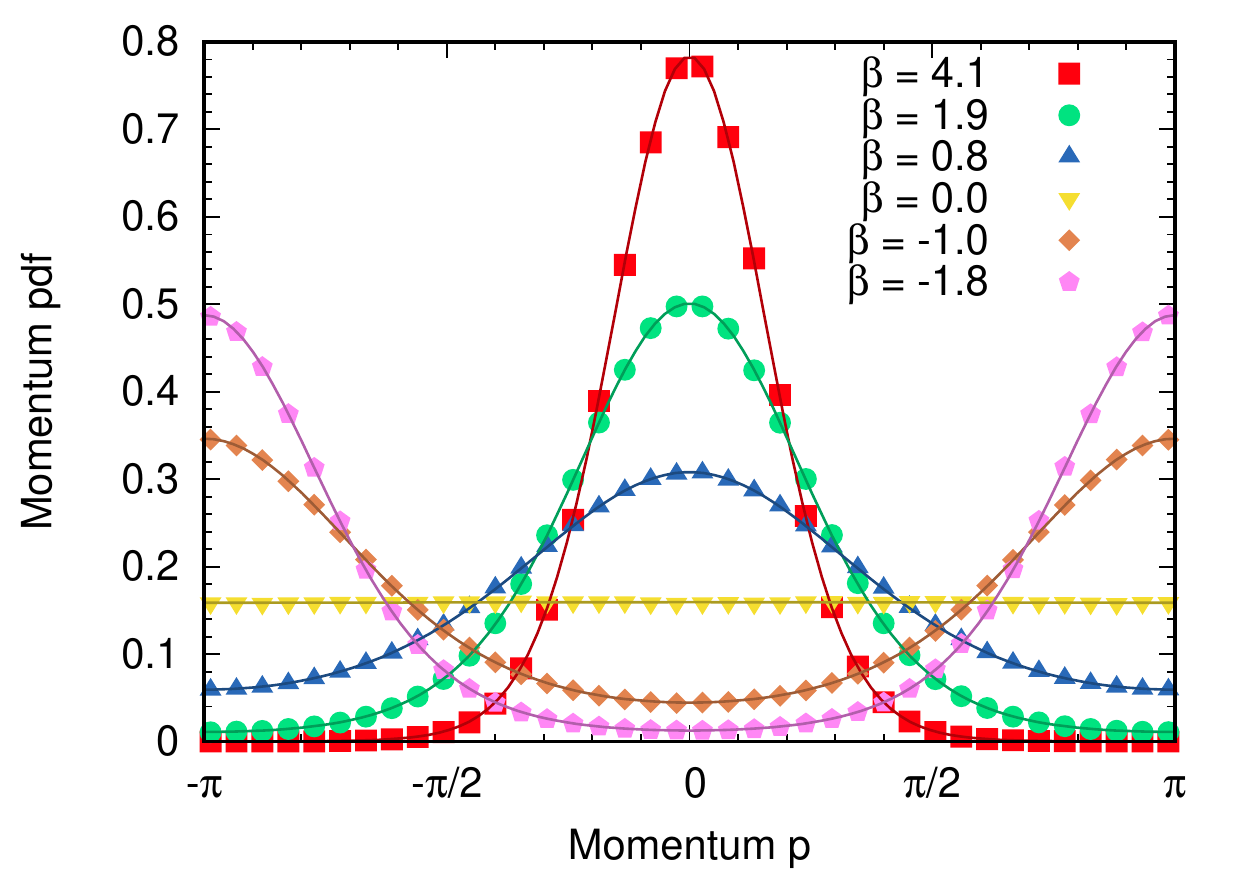}
\caption{Single-particle momentum distribution for a model with bounded 
Hamiltonian. Points: histograms of $P$ obtained from molecular dynamics 
simulations of model~\eqref{eq:minimal} with kinetic energy~\eqref{eq:minimalk} 
and potential~\eqref{eq:minimalu}, for different values of the total energy. 
Solid lines: best fit of the pdf with the functional form~\eqref{eq:maxbol}; the 
resulting values of $\beta$ are reported in the key. All simulations have been 
executed through a second order Velocity Verlet integration scheme with time 
step $\Delta t=10^{-3}$, over a total integration time (for each plot) of 
$\mathcal{T}=10^5$. Here $N=128$.}
\label{fig:profili}
\end{figure}
A minimal example is presented in 
Fig.~\ref{fig:profili}, where the pdf of the single particle momentum $P$ 
is shown, for several values of 
$\beta$, in a Hamiltonian system of the form~\eqref{eq:minimal} with
\begin{equation}
\label{eq:minimalk}
\mathcal{K}(P)=1-\cos(P)                                                         
\end{equation}
and
\begin{equation}
\label{eq:minimalu}
 \mathcal{U}(\mathbf{Q})=\frac{1}{2}\sum_{n=1}^N [1-\cos(Q_{n}-Q_{n-1})]\quad 
\quad Q_0 \equiv Q_{N} \equiv 0\,.
\end{equation}
This model has been introduced in Ref.~\cite{cerino15} as an example of 
Hamiltonian with bounded phase space; the peculiar form of the kinetic energy 
resembles the semi-classical limit of the bosonic system studied in 
Ref.~\cite{braun13} and it has been investigated also in Ref.~\cite{hanggi15}. 
In the low-energy limit the functional form of the kinetic terms tends to the 
usual one, $\mathcal{K}(P)\simeq P^2/2$, while the potential interaction reduces 
to that of a chain of harmonic oscillators; in this limit, the chain behaves as 
a classical Hamiltonian system at positive temperature, and $\rho(P)$ resembles 
the usual Gaussian distribution. The scenario is different at high energies: in 
this case, the dynamics typically selects configurations in which the angular 
distance between two neighbour oscillators is maximized~\cite{cerino15} and the 
distribution of the momentum is peaked around $P\simeq \pi$ (see 
Fig.~\ref{fig:profili}), a state corresponding to a negative value of $\beta$.
The knowledge of the functional form of the pdf~\eqref{eq:maxbol} allows to infer 
the value of the (Boltzmann) temperature through a fit over the distribution, 
even in cases with negative temperature.


An alternative way to measure the microcanonical Boltzmann temperature has 
been proposed by Rugh~\cite{rugh97, rugh98}. It can be shown that $\beta$ is 
correctly estimated by the following observable:
\begin{equation}
\beta=\frac{1}{\omega(E)}  \frac{\partial \omega(E)}{\partial E}=
\left\langle \Phi \right\rangle_E
\label{R1}
\end{equation}
where
\begin{equation}
\label{eq:defphi}
\Phi=\nabla \cdot \left( \frac{\nabla \mathcal{H}}{ || \nabla \mathcal{H} ||^2} 
\right) \,.
\end{equation}
and the average is  computed according to the microcanonical distribution, 
over the energy constant hypersurface $ \Gamma_E=\{{\bf X}: \mathcal{H}({\bf 
X})=E \}$.
A proof of the above result can be found in Ref.~\cite{puglisi17}.
The  conceptual and practical relevance of formula~\eqref{R1} relies on the 
fact that it links temperature to a mechanical observable~\cite{giardina98}; 
since the considered average is microcanonical, one can 
compute temperature through a time average along a trajectory (assuming ergodicity):
\begin{equation}
\beta=\lim_{\cal T \to \infty} { 1 \over {\cal T}}
 \int_0^{\cal T} \Phi({\bf Q}(t),{\bf P}(t)) \, dt\,.
\end{equation} 
If the system is very large ($N\gg1$), Eq.~\eqref{R1} can be simplified in the 
following way~\cite{puglisi17}:
\begin{equation}
\beta \simeq
\left\langle  {\nabla^2 \mathcal{H} \over || \nabla \mathcal{H} ||^2}  \right>_E 
\Big( 1 + O\Big({ 1 \over N}\Big)  \Big) \simeq
{ \left<  \nabla^2 \mathcal{H}  \right\rangle_E  \over  \left< || \nabla 
\mathcal{H}||^2 \right>_E }\Big( 1 + 
O\Big({ 1 \over N}\Big) \Big)\,.
\end{equation} 
If the Hamiltonian is quadratic, in the limit of large $N$ we recover an 
expression which is consistent with the equipartition theorem.
Eq.~\eqref{R1} has been further generalised~\cite{jepps00,rickayzen01}: it is 
possible to show that
\begin{equation}
\beta=
\left<   \nabla \cdot \Big( { {\bf B} \over  {\bf B} \cdot \nabla \mathcal{H}} 
\Big) \right>_E
\label{G1}
\end{equation}
where ${\bf B}$ is a vector of arbitrary continuous and differentiable functions 
of the phase-space.
Such a generalization has been successfully used in numerical simulations,
for instance to check the algorithmic correctness of 
Monte Carlo computer programs, and  to design  
new thermostats~\cite{morriss99, jepps00}. Rugh's approach may lead to 
particularly interesting equivalences for systems whose Hamiltonian structure is 
different from~\eqref{eq:minimal}, as for instance the $2D$-model for
 inviscid hydrodynamics discussed in Section~\ref{sec:vortices}. 
 The result has been also extended to the case of one or more conserved quantities~\cite{Franzosi,franzosi12}.
 Let us also notice that from the microcanonical definition~\eqref{eq:defphi} it follows that a change of sign of $H$ implies a change of sign of $\beta$, and vice versa. This general property has profound implications also for experiments, as it corresponds to the main  protocol employed to produce NAT states in~\cite{purcell51,schreiber15}.

  
So far, we have shown that it is generally possible to measure the Boltzmann 
temperature, even if it assumes negative values, by studying the distribution of 
some degrees of freedom of the system or by looking at proper observables.  
Critics of Gibbs' definition of entropy stress that $T_G$, on the contrary, is 
unphysical in the following sense: it is a function of a phase-space volume 
region which includes states that the system is not allowed to visit, because of 
energy conservation~\cite{schneider14,buonsante17}. By looking at an actual trajectory of
the system in the phase space one cannot sample such volume, the dynamics
being constrained on the constant-energy hypersurface.
The problem of measuring $T_G$ can be 
actually approached in an alternative way, often discussed in textbooks. As 
discussed in Section~\ref{sec:equipartition}, the Equipartition Theorem sets a 
link between a microcanonical average and the Gibbs temperature
\begin{equation} \label{eqpart}
\left\langle x_i \frac{\partial \mathcal{H}}{\partial x_j} \right\rangle_E = 
\delta_{ij} T_G\,;
\end{equation}
again, if one assumes ergodicity, Eq.~\eqref{eqpart} allows to compute 
temperature by mean of time averages. However, it is important to notice that 
the usual derivation of Eq.~\eqref{eqpart} implies the possibility to neglect 
boundary terms in an integration by parts~\cite{cerino15,buonsante16}. This possibility is challenged when 
the phase space is bounded, and in particular it can be shown that Equipartition 
Theorem does not hold under the simultaneous realization of the following 
conditions~\cite{cerino15}:
\begin{enumerate}
 \item bounded space of the canonical variables;
 \item bounded derivatives of the Hamiltonian $ \frac{\partial \mathcal{H}}{\partial x_j} 
$;
 \item bounded energy from above and below: $E_m\le E \le E_M$;
 \item vanishing density of states at the boundaries, i.e. $\omega(E_M)=0$. 
\end{enumerate}
Given such conditions, one has that, on the one side,
$T_G(E)={\Sigma(E)}/{\omega(E)}$
diverges when $E \to E_M$; on the other side,
$\langle x_i \frac{\partial \mathcal{H}}{\partial x_j} \rangle$ is limited,
resulting in a contradiction.
In the light of the above, Eq.~\eqref{eqpart} can
fail even in systems where there are no negative temperatures and
 $T_G \simeq T_B >0$ for all $E$. Consider, for instance, the following
Hamiltonian
\begin{equation}
\mathcal{H}=\sum_{n=1}^N {p_n^2 \over 2} +\epsilon \sum_{n=1}^N [ 1- \cos 
(q_n-q_{n-1})]
\end{equation}
where $q_n \in [-\pi , \pi)$.  For large $E$, i.e.
$E \gg \epsilon N$, the contribution to $\Sigma(E)$ of the variables
$\{ q_n \}$ is almost independent of the value of $E$, so that
$\Sigma_{\epsilon}(E)\simeq \Sigma_{0}(E)\propto \, E^{N/2}$,
and  $T_G\simeq 2E/N$. In the limit $N \gg 1$, we have that $T_B \simeq 
T_G+O(1/N)$; on the other hand it is easy to see that
\begin{equation}
\left|  q_n {\partial \mathcal{H} \over \partial q_n} \right| \le 2 \pi \epsilon 
\,\, ,
\end{equation}
and, therefore, formula~\eqref{eqpart} cannot be valid 
for large values of $E$ and $N$.

\subsection{Zeroth law and thermometry}
\label{sec:termometro}

From a thermodynamic point of view the notion of temperature is based on the 
Zeroth Law: if two systems $\cal A$  and $\cal B$ are each at thermal 
equilibrium with a third, then they are at thermal equilibrium with each other. 
This implies that being at thermal equilibrium is an equivalence relation, and 
temperature is the thermodynamic quantity that determines to which ``class of 
equivalence'' a system belongs~\cite{kardar07, puglisi17}. This principle 
justifies the usage of thermometers to measure temperature: it is sufficient to 
verify that $\cal A$ is at thermal equilibrium with a thermometer at a certain 
temperature to conclude that it belongs to the class of equilibrium identified 
by that temperature.

As mentioned in Section~\ref{sec:measuring}, in principle it is possible to use 
a thermometer to measure the temperature of systems with bounded phase space,
even when $\beta < 0$. However we know from the discussion in 
Section~\ref{sec:nat} that coupling a system $\mathcal{A}$ at negative 
temperature with a system $\mathcal{B}$ composed by ordinary matter will lead 
after a certain time to an equilibrium state at positive temperature.  Therefore 
if we try to  use the system $\cal B$ as a thermometer  to read the absolute    
temperature of $\cal A$,  we shall observe the rather disappointing result that the 
thermometer  heavily perturbs the measured system, changing the sign of its 
original  temperature. The above result is in obvious disagreement with the 
natural idea that a  thermometer should be able to measure a system's 
temperature within an acceptable time and without perturbing it significantly. 
The above difficulty comes from the fact that once $\cal A$ and the thermometer 
are coupled together, the resulting system lives in a unbounded phase space, due 
to the degrees of freedom of the thermometer~\cite{ramsey56}. The problem is solved as soon as 
the measuring device also admits NAT states: as shown in Ref.~\cite{buonsante16},
if two systems admitting negative temperature are coupled together, they will eventually
reach the same inverse Boltzmann temperature, whose value will be intermediate
between the two $\beta$'s of the subsystems. In that work the Hamiltonian dynamics of
systems inspired by a (quantum) bosonic dynamics is numerically simulated; in the following
we shall discuss a similar numerical experiment on Hamiltonian systems with the generalized kinetic energy~\eqref{eq:minimalk}
introduced in the previous Section. Further details can be found in Ref.~\cite{baldovin17}.

Let us consider
 \textit{(i)} a sample with Hamiltonian $\mathcal{H}_S({\bf X})$ where ${\bf  X}\in \mathbb{R}^{N_S}$ 
denotes the system's mechanical variables, living in a $N_S$-dimensional
phase space with $N_S  \gg 1$; \textit{(ii)} a ``thermometer'',  with Hamiltonian $\mathcal{H}_T({\bf Y})$ 
where ${\bf Y}\in \mathbb{R}^{N_T}$ denotes the vector of the Hamiltonian coordinates,
with $N_T\ll N_S$ (i.e. the  size of the thermometer  $N_T$ is small with respect to that of the sample).
The total Hamiltonian reads
\begin{equation}
\mathcal{H}({\bf X}, {\bf Y})=\mathcal{H}_S({\bf X})+\mathcal{H}_T({\bf Y})+\varepsilon \mathcal{H}_I({\bf X}, {\bf Y})\,,
\end{equation} 
where $\varepsilon \ll 1$ and
\begin{equation}
\label{eq:termint}
\mathcal{H}_I({\bf X}, {\bf Y})=\sum_{i=1}^{N_T}\mathcal{V}(q_i -Q_i),
\end{equation} 
accounts for the (weak) interactions between the two subsystems; here the
$\{ q_i \}$ ($i=1,2,.. , N_T$) denote the positions of the
particles of the thermometer and $\{ Q_j \}$ ($j=1,2, .., N_S$) those of the system to be measured. In this simple
model, each particle of the thermometer interacts with only one particle of the sample through a generic
smooth potential $\mathcal{V}$.
Before $t=0$ the parameter $\varepsilon$ is set equal to zero, so that there is no interaction between the thermometer and the sample;
the two subsystems are at equilibrium with different temperatures. At $t=0$
the coupling is switched on and we measure the common value of $\beta$ eventually
reached by the whole system.

We run numerical simulations where the sample is described by
\begin{equation}
\label{eq:tn}
\mathcal{H}_S=\sum_{i=1}^{N_S}\left( 1- \cos P_i\right)+ 
 K \sum_{i=1}^{N_S}\left( 1- \cos (Q_{i}-Q_{i-1})\right)+\frac{J}{2 N_S} \sum_{i,j=1}^{N_S}\left(1-\cos(Q_i-Q_j)\right)\,.
\end{equation}
The kinetic energy and the first interaction term resemble those introduced by Eqs.~\eqref{eq:minimalk} and~\eqref{eq:minimalu},
describing a nonlinear chain of generalized oscillators in bounded phase space; the last potential
interaction is a ``long-range'' contribution inspired by the Hamiltonian Mean Field model~\cite{antoni95, campa06, campa09},
which has only been introduced to increase the velocity of the thermalization between the degrees of freedom of the system.
The qualitative scenario discussed in the following is not expected to depend crucially on the details of the sample.

\begin{figure}[h!]
\centering
\includegraphics[width=.6\linewidth]{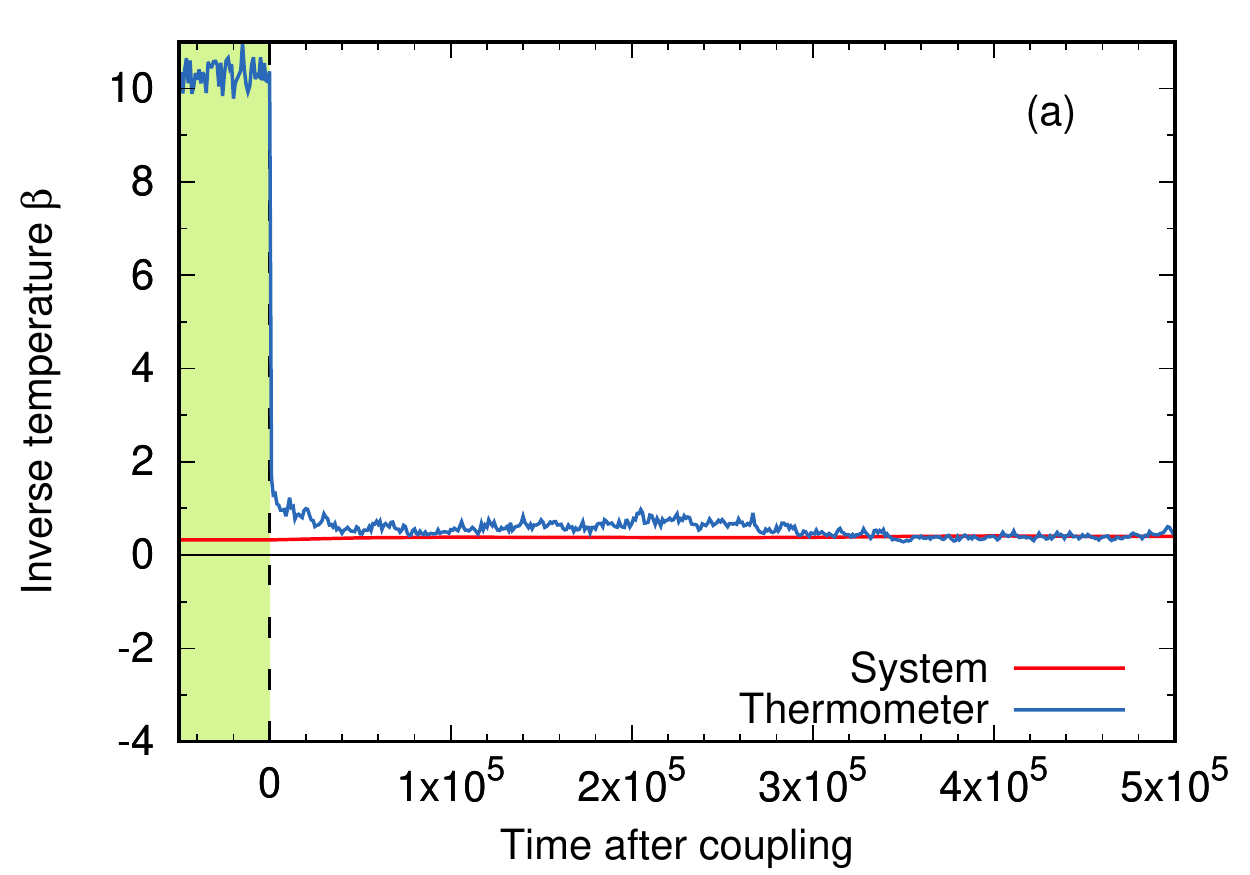}
\quad\\ \quad\\
\includegraphics[width=.6\linewidth]{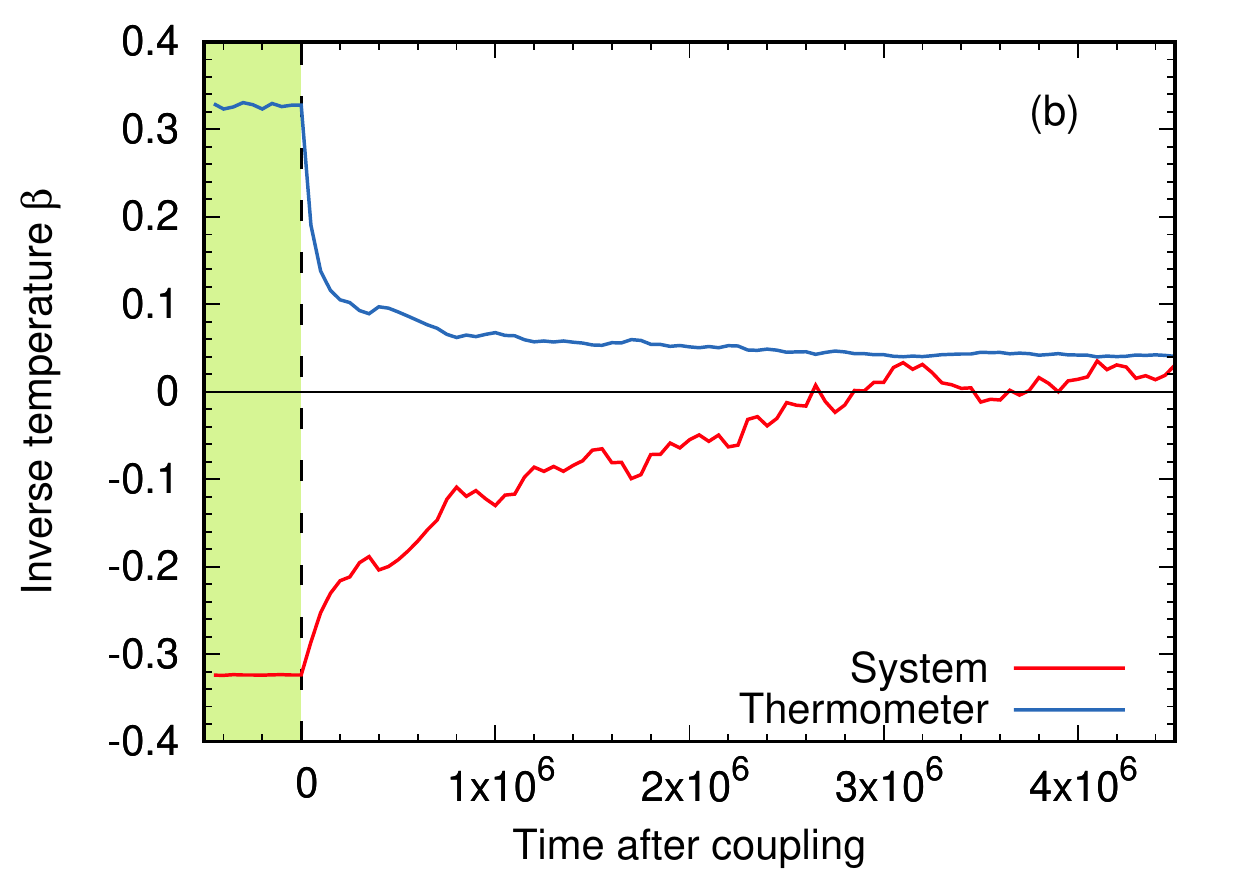}
\caption{Inverse temperature $\beta$ as a function of time, for the sample~\eqref{eq:tn}
and the positive-temperature thermometer~\eqref{eq:term_ang}.
In panel (a) a case in which both subsystems start at positive temperature is shown, while in panel (b)
the sample is prepared in a NAT equilibrium state.
The inverse temperature $\beta(t)$ of the sample is computed from a fit on the single particle momentum 
pdf: we consider the histogram of the measured momenta from time $t$ to time 
$t+\Delta t$, with $\Delta t=5\cdot10^4$ in this case, and we obtain the value of 
$\beta$ from the slope of $\log[\rho(p)]$, as explained in Ref. \cite{cerino15}. 
Thermometer's inverse temperature has been determined, as usual, by 
$\beta=\langle p^2\rangle ^{-1}$. The shape of the interaction potential 
between the sample and the thermometer appearing in Eq.~\eqref{eq:termint} is 
$\mathcal{V}(x)=1-\cos(x)$. Parameters: $N_S=100$, $N_T=30$, $K=\gamma=0.5$, 
$J=0.05$ $\varepsilon=0.1$. Figure adapted from Ref.~\cite{baldovin17}.}
\label{fig:tp_time}
\end{figure}

First, let us consider a case in which the thermometer does not admit negative temperature:
\begin{equation}
 \label{eq:term_ang}
 \mathcal{H}_T=\sum_{i=0}^{N_T}\frac{p_i^2}{2 
}+\sum_{i=1}^{N_T}m\gamma^2\left[1-\cos(\phi_i-\phi_{i-1})\right]\,;
\end{equation} 
the kinetic energy of the thermometer is the usual quadratic one, so that NAT cannot be reached.
Figure~\ref{fig:tp_time} shows that, if the sample is initially prepared in a positive-temperature state,
the thermometer is able to reach its temperature and measure it without relevant perturbations.
On the other hand, if the sample is in a NAT state for $t<0$, when the coupling is switched on
its inverse temperature starts increasing, due to an energy flux from the sample to the thermometer.
This process only ends when both systems reach a common, positive temperature.
The amount of the exchanged energy is huge even if the size of the
thermometer and its coupling  with the system are small (i.e. $N_T/N_S \ll 1$,
$\epsilon \ll 1$). The final state of the sample is significantly different from the initial one,
so that we must conclude, as it could be expected from the beginning, that this kind of thermometer is not suitable 
at all to measure the temperature of systems at negative temperature.

\begin{figure}[h!]
\centering
\includegraphics[width=.6\linewidth]{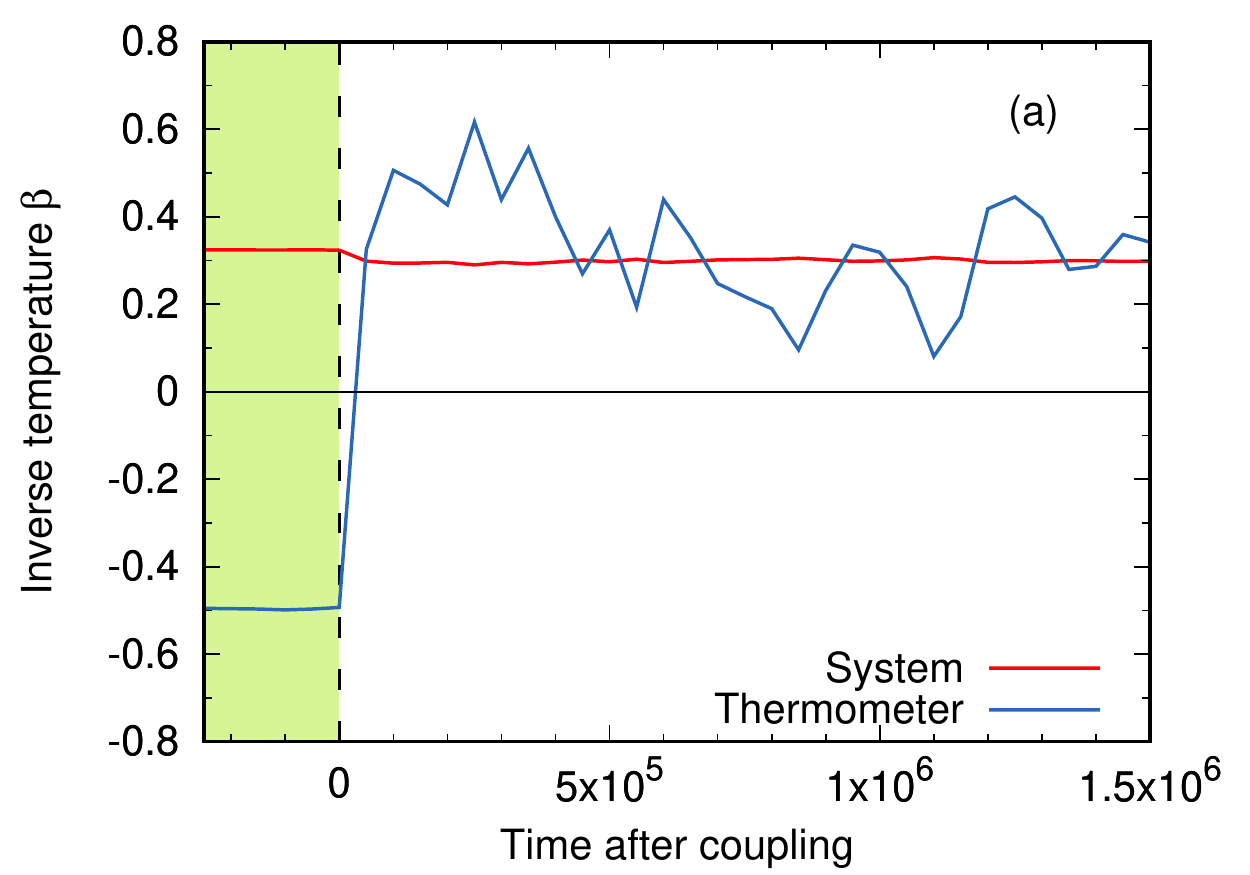}
\quad\\ \quad\\
\includegraphics[width=.6\linewidth]{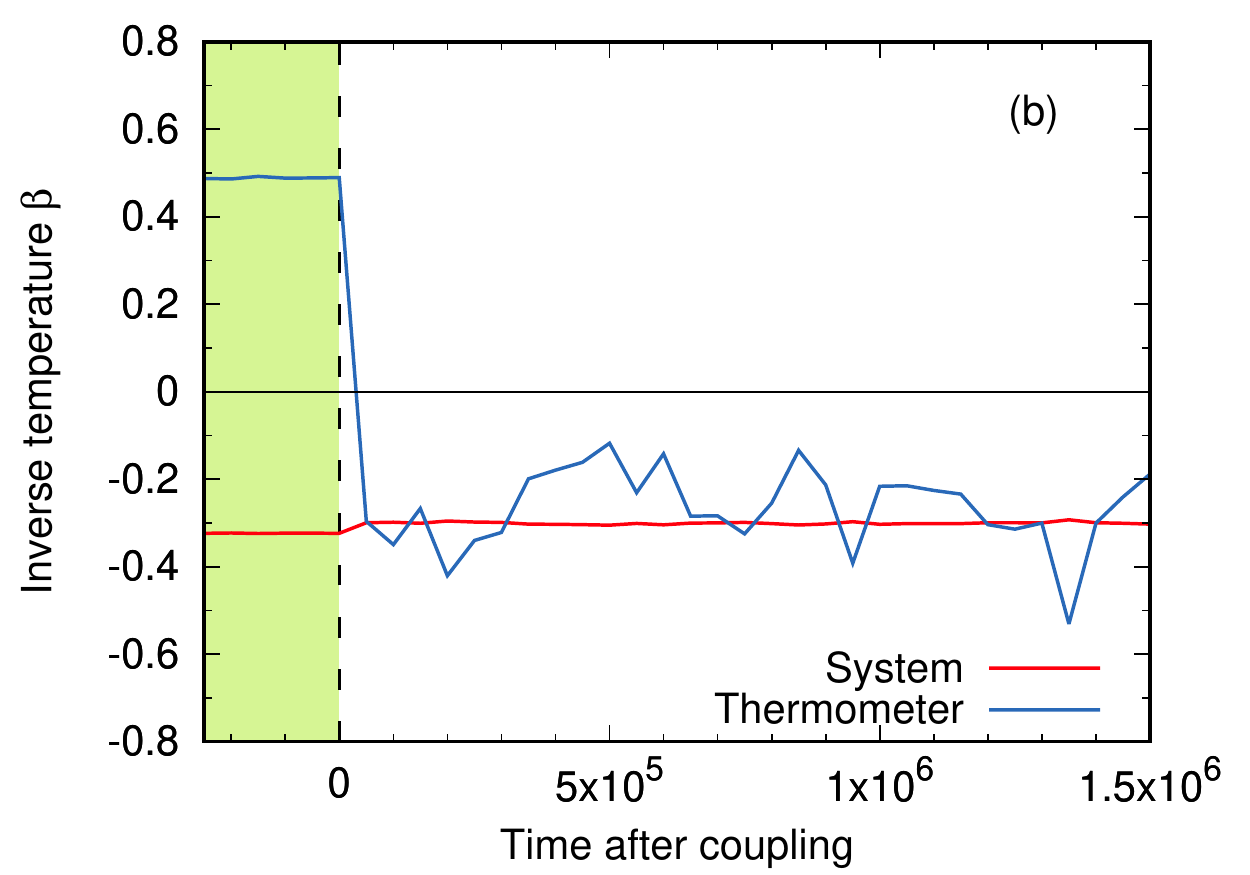}
\caption{Inverse temperature $\beta$ as a function of time, for the
  system \eqref{eq:tn} and the thermometer \eqref{eq:tn_term}, both 
  measured by the distribution fitting procedure explained in 
Fig~\ref{fig:tp_time}. In panel (a) the sample is initially prepared in a positive-temperature state,
in panel (b) its initial $\beta$ is negative. The shape of the interaction potential 
between the sample and the thermometer appearing in Eq.~\eqref{eq:termint} is 
$\mathcal{V}(x)=1-\cos(x)$. Parameters:
  $N_S=100$, $N_T=30$, $K=\gamma=0.5$, $J=0.05$ $\varepsilon=0.1$.
   Figure adapted from Ref.~\cite{baldovin17}.}
\label{fig:tn_time}
\end{figure}

Let us now study what happens if the thermometer is properly chosen, i.e. if 
its Hamiltonian coordinates, as well as those of the sample to be measured, live in a bounded
phase space:
\begin{equation}
\label{eq:tn_term}
\mathcal{H}_T=\sum_{i=0}^{N_T}(1-\cos p_i)+
\sum_{i=1}^{N_T}\gamma\left[1-\cos(\phi_i-\phi_{i-1})\right]  \, .
\end{equation}
Now the temperature cannot be estimated by the mean kinetic energy,
so we have to adopt an appropriate methodology, e.g. use Rugh's definition or
look at the shape of the probability distribution
of the single particle momentum, the generalized Maxwell-Boltzmann 
$\rho(p) \propto e^{-\beta [1-\cos(p)]}$, 
see Eq.~\eqref{eq:maxbol}.
With this kind of thermometer a correct measurement  of the temperature 
can be performed, as shown in Fig.~\ref{fig:tn_time}.

We conclude by noticing that the details of $\mathcal{H}_T$ are not particularly important:
the relevant  aspects are  the bounded nature of the phase space and the fact that $N_T$ is 
small with respect to $N_S$. These conditions should be sufficient, in general, to
insure that the thermometer equilibrates with the sample within an acceptable time and without perturbing it
significantly,  allowing therefore to measure the system's temperature
by looking at accessible observables.

\subsection{Ensemble equivalence (and its violations) }
\label{sec4.3}
The relation between ensemble equivalence and negative temperatures is far from 
straightforward
and deserves a careful consideration. 
As a general remark, let us briefly consider the role of $T_B$ in the context of
the problem of the equivalence of the ensembles.
If we limit our attention to systems with short-range interactions,
we can assume, following the usual arguments that can be found in textbooks,
that
\begin{equation}
S_B(E,N)=N s(e)\,,
\end{equation} 
i.e. the entropy is an extensive quantity (here $e=E/N$ and $s(e)$ is
the entropy per particle); moreover, $s(e)$ cane be assumed to be
concave, so that, performing a steepest descent analysis in the
large-$N$ limit, one obtains the canonical description from the (Boltzmann) 
microcanonical one, e.g.:
\begin{equation}
T_B(e) s(e) = e-f(T_B(e)),
\end{equation}
where $f(T)$ is the free energy per particle in the canonical ensemble.
In such a derivation,
the only relevant point is  the convexity of $s(e)$, and nothing about
its first derivative is asked: therefore, the equivalence of ensembles
naturally holds even when $T_B$ assumes negative values. Since
$T_B$ and $T_G$ can be different even for large $N$, it is evident
that $T_G$ is not relevant for the determination of equivalence
between statistical ensembles.

In this Section we aim at summarizing the 
overall scenario, while illustrating in some detail 
the following points:
\begin{itemize}
\item there are examples of simple statistical models defined on a discrete 
lattice exhibiting negative temperature equilibrium states without 
any violation of the equivalence of statistical ensembles, provided the 
definition of Boltzmann's microcanonical temperature $\beta^{-1}_B$
(see Sec.~\ref{gibbstemperature})  is adopted;
\item for these simple models the Gibbs' microcanonical temperature 
$\beta^{-1}_G$ is found to be inconsistent with
ensemble equivalence;
\item there is evidence that in more elaborated lattice models the equivalence 
between equilibrium statistical ensembles may fail in the negative temperature 
phase, because of 
specific physical reasons.
\end{itemize}

\subsubsection{Simple models}
\label{ssec1}
Let us first discuss the classical model of the Discrete Linear Schr\"{o}dinger 
Equation, i.e. the noninteracting version
of the model discussed in Sec.~\ref{sec:esdnls}, defined on a 1d lattice of $N$ sites with 
periodic boundary conditions, whose Hamiltonian reads
\begin{equation}
\label{Hdlse}
{\mathcal H} = - \sum_{j=1}^N ( z_j \, z^*_{j+1} + c.c.)  \, ,
\end{equation}
where $z_j$ is the complex amplitude of the wave function at site $j$.
The choice of the minus sign in front of the r.h.s. of  this equation is  
essentially irrelevant, because of the
symmetry associated to the gauge transformation $z_j \to e^{i \pi j} \, z_j$, 
with $i$ representing the imaginary constant. 
This Hamiltonian model can be interpreted as a system of noninteracting 
particles, hopping on a lattice: it conserves not only the energy ${\mathcal 
H}$, 
but also the quantity $A=\sum_{j=1}^N |z_j|^2$. If one interprets $ 
|z_i|^2$ as the number of particles at site $j$, the total norm $A$ is 
the total number of
particles in the lattice. For practical reasons it is worth introducing the 
energy and norm densities
\begin{equation}
\label{densi}
h = \frac{ {\mathcal H}}{N} \, , \quad \quad a = \frac{ {\mathcal 
N}}{N} \, .
\end{equation}

The grand-canonical partition function of this simple model is defined as 
follows
\begin{equation}
\label{gcpf}
{\mathcal Z} = \int \prod_{j=1}^{N} dz_j e^{- \beta N (h - \mu a)} \, ,
\end{equation}
where $\mu$ is the chemical potential. The explicit calculation of ${\mathcal 
Z}$ can be easily performed 
(e.g., see \cite{buonsante16,buonsante17}) by introducing the
Fourier-transformed variables
\begin{equation}
\label{ftz}
\tilde z_k = (1/N) \sum_{j=1}^N e^{i \, kj} z_j \, , \quad\quad  k= 0, 
\dots, N-1\, ,
\end{equation}
and one can write
\begin{equation}
\label{consk}
{\mathcal H} =  \sum_{k=0}^{N-1} \epsilon_k |z_k|^2 \,, \quad\quad 
A = \sum_{k=0}^{N-1} |z_k|^2 \, ,
\end{equation}
where
\begin{equation}
\label{epermode}
\epsilon_k = -2\cos(2\pi k /N)  
\end{equation}
is the energy associated to the $k$-th Fourier mode. The grand-canonical 
partition function turns out to have the explicit expression
\begin{equation}
\label{gcpf2}
{\mathcal Z} = \prod_{k=0}^{N-1} \frac{\pi}{\beta(\epsilon_k -\mu)} \, ,
\end{equation}
The average norm density and the average energy density are given by the 
expressions \cite{buonsante16}
\begin{equation}
\label{avq}
a = \frac{N}{N} =  \frac{1}{N}\sum_{k=0}^{N-1}\langle |\tilde z_k|^2 \rangle= 
\frac{1}{N} \sum_{k=0}^{N-1} \frac{1}{\beta (\epsilon_k -\mu)} \, , 
\quad \quad h = \frac{1}{\beta N} 
\sum_{k=0}^{N-1}  \frac{\epsilon_k}{\epsilon_k -\mu} \, .
\end{equation}
In the limit of large values of $N$ one can obtain the explicit form of the 
inverse temperature and of the chemical potential in terms
of $a$ and $h$:
\begin{eqnarray}
\mu &=& \frac{h^2 + 4 a^2}{2 a h}  \\
\label{muacca}
\beta&=& - \frac{2 h}{4a^2 - h^2}\,.
\label{muacca1}
\end{eqnarray}
Notice that $\beta$ is positive for $-2a < h < 0$, but it is negative for $0 < h 
< 2a$. On the other hand there is no thermodynamic difference
between the two phases at positive and negative temperatures, because they are 
related by the gauge symmetry intrinsic
to this model: in fact, changing the sign in front of (\ref{Hdlse}) amounts to 
exchange these two
phases, i.e. $\beta \to -\beta$. 

For this simple model one can compute explicitly also the Boltzmann 
microcanonical entropy density
\begin{equation}
\label{mced}
s_B(h,a) = \frac{1}{N} k_B \ln(\omega(h,a) \, C) \, ,
\end{equation}
where $\omega(h,a)$ is the density of states for given values of $h$ and 
$a$, in formulae
\begin{equation}
\label{mcdB}
\omega(h,a) = \int \prod_{k=0}^{N-1} {\mathrm d}z_k \, \delta(Nh - {\mathcal H}) 
\, \delta(Na -A)
\end{equation}
and $C$ is a suitable constant with the physical dimensions of $h$.
The complete expression of $\omega(h,a)$ is reported in \cite{buonsante16}. For our 
purposes it is enough to observe that
for large values of $N$ this expression boils down to 
\begin{equation}
\label{mcdB1}
\omega(h,a) \propto \Big(\frac{4 a^2 - h^2}{a^2}\Big)^N
\end{equation}
By recalling the definition of the Boltzmann inverse temperature $\beta_B = 
\frac{1}{k_B} \frac{\partial s(h,a)}{\partial h}$ one
can immediately check that $\beta_B$ coincides with the inverse temperature 
$\beta$ obtained for the grand-canonical ensemble in Eq.(\ref{muacca1}).

These easy calculations certify that, despite the thermodynamics of this simple 
model admits equilibrium states at negative temperature (which are
just symmetric to states with positive temperature), one has ensemble 
equivalence between the grand-canonical ensemble and the microcanonical one, 
computed by
the Boltzmann entropy density (\ref{mced}). 

Conversely, if one adopts the the definition of Gibbs microcanonical entropy 
density
\begin{equation}
\label{mcedG}
s_B(h,a) = \frac{1}{N} k_B \ln(\Omega(h,a) \, C') \, ,
\end{equation}
where 
\begin{equation}
\label{mcdG}
\Omega(h,a) =  \int_{-2a}^h \, {\mathrm d} h' \,\omega(h',a) \, ,
\end{equation}
one obtains that the Gibbs inverse temperature verifies $\beta_G (h,a) = \beta_B (h,a)$ 
only for 
$h \in [-2a,0)$ (i.e. in the region where $\beta_B (h,a)$ is positive), while it 
vanishes identically for 
$ h \in (0, 2a]$, where $\beta_B$ and $\beta$ coincide and are both negative 
(see \cite{buonsante17}). This testifies that
Gibbs' entropy violates ensemble equivalence. Let us remark that the lower bound 
$-2a$ of the integral 
in Eq.(\ref{mcdG}) comes from observing that, for a given value $a$ of the 
norm density, the energy density $h$ is bounded
in the interval $[-2a , 2a]$.

Another simple model exhibiting analogous features is described by the 
Hamiltonian
\begin{equation}
\label{Hspin}
{\mathcal H} = - \varepsilon \, \sum_{j=1}^N \sigma_j  \, ,
\end{equation}
already introduced in Sections~\ref{sec:mag_sys} and~\ref{sec:defense} (see also \cite{dunkel14,campisi15,campisi16}). This model
represents a lattice of $N$ uncoupled spin variables $\sigma_j=\pm 1$;
$\varepsilon$ is a constant depending on the gyromagnetic ratio of the spins and
the applied field. It is the 
simplest example of the class of models discussed 
in the celebrated paper by Ramsey about negative temperature equilibrium states 
in nuclear spin systems \cite{ramsey56}.
As reported in \cite{buonsante16}, the calculation of the canonical partition function 
is straightforward, yielding the expression
\begin{equation}
\label{partf}
Z = \Big[ 2 \cosh (\beta \, \varepsilon)\Big]^N  = e^{- \beta \, F}\, ,
\end{equation}
where, again, $\beta$ is the inverse temperature in units of the Boltzmann 
constant $k_B$ and $F$ is the Helmholtz free energy. Standard
thermodynamic relations allow to obtain the explicit expressions of the internal 
energy $E$:
\begin{equation}
\label{intE}
E = - \frac{\partial}{\partial \beta} \log Z = - N \varepsilon \tanh(\beta \varepsilon)
\end{equation}
and of the entropy $S$:
\begin{equation}
\label{centr}
S = k_B \, \beta^2 \, \frac{\partial F}{\partial \beta}  = k_B \, N\, \Big[ \log 
2 + \log(\cosh(\beta \varepsilon)) - \beta \varepsilon \tanh(\beta \varepsilon)\Big]\,.
\end{equation}
It is worth introducing the dimensionless quantity $\lambda = E/(N \varepsilon)$ ($-1 < 
\lambda <1$) to write the expression of the inverse temperature (which
is obtained by inverting Eq.(\ref{intE}) ) 
\begin{equation}
\label{betac}
\beta = - {\mathrm {arctanh}} (\lambda)
\end{equation}
and of the entropy
\begin{equation}
\label{entroc}
S = \frac{k_B\, N}{2} \Big[2\log 2 - (1+\lambda)\log (1+\lambda)  - 
(1-\lambda)\log (1-\lambda)  \Big]
\end{equation}
The first equation indicates that negative temperature equilibrium states are 
obtained for $0< \lambda <1$, while the second equation
tells us that $S$ is a concave function of $\lambda$  and exhibits a maximum at 
$\lambda = 0$. It is worth pointing out that $\beta$ and $S$
obtained for the canonical ensemble satisfy the same relation, that is expected 
to hold for the microcanonical ensemble
$\beta = 1/k_B \, \frac{\partial s}{\partial \lambda}$, with $N \, s = S$. 

For this model one can compute explicitly also the Boltzmann entropy $S_B$, 
which reads (see \cite{buonsante16})
\begin{equation}
\label{entroB}
S_B = k_B \log (\omega (\lambda)) \, ,
\end{equation}
where
\begin{equation}
\label{omegaeps}
\omega (\lambda)= \frac{N !}{(\frac{1+\lambda}{2} N)! \, (\frac{1-\lambda}{2} 
N)! } 
\end{equation}
By applying the Stirling approximation of the factorial one can easily check 
that $S_B$ coincides with $S$ for large values of $N$.
Conversely, the Gibbs entropy $S_G$ reads (see \cite{buonsante16})
\begin{equation}
\label{entroG}
S_G = k_B \log (\Omega (\lambda)) \, ,
\end{equation}
where
\begin{equation}
\label{omegaGeps}
\Omega (\lambda)= \frac{N}{2} \int_{-1}^{\lambda} {\mathrm d}\lambda'  \, 
\omega(\lambda')\,.
\end{equation}
It can be easily verified that in the region of  positive temperatures,  $-1< 
\lambda <0$, one has $\beta_G \equiv 1/(k_B\,N) \, \frac{\partial S_G}{\partial 
\lambda} =\beta_B\approx\beta$, while in the region of negative temperatures,
$0 < \lambda <1$, $\beta_G=0$. Accordingly we are facing the same scenario 
previously described for the free-hopping-particle model (\ref{Hdlse}).

\subsubsection{A wider scenario}
\label{ssec2}
The pedagogical examples discussed in the previous section point out that the 
equivalence of equilibrium statistical ensembles, which is expected to hold for 
standard short-range 
interacting systems at positive temperature, may be extended also to 
negative-temperature equilibrium states, provided the Boltzmann definition of 
microcanonical entropy is
adopted (e.g., see (\ref{entroB})).  Accordingly, these examples strongly 
challenge the very  conceptual consistency of the Gibbs microcanonical entropy 
(e.g., see (\ref{entroG})), 
which merely reduces to an approximate formula for estimating the Boltzmann 
microcanonical entropy for large values of $N$ and for positive temperatures, 
only. On the other hand,
if one would like to attribute to Gibbs' entropy the role of the cornerstone of 
equilibrium thermodynamics 
one might argue that negative temperature equilibrium states are just an 
artifact, suggested by pathologically simple models, while they have no room in 
real physical phenomena. 
In such a perspective, the experimental evidence of genuine negative-temperature 
equilibrium states observed by Purcell and Pound~\cite{purcell51} in nuclear-spin 
systems and,
much more recently, by various researchers in superfluid Bose-Einstein 
condensates \cite{gauthier19,johnstone19} could be reinterpreted as a manifestation of 
long-living
metastable states. As discussed in Sec.~\ref{sec:debate}, these different points of view have 
raised a long-standing dispute about the physical relevance of negative 
temperature equilibrium states, 
which has eventually come to the conclusion that their existence does not 
contradict any basic principle of thermodynamics.

In order to enforce the physical interest for these states, in what follows we 
want to shortly illustrate more elaborated models, where they appear also when 
the equivalence
between equilibrium statistical ensembles is violated.

\begin{figure}[h!]
 \centering
\includegraphics[width=.6\linewidth]{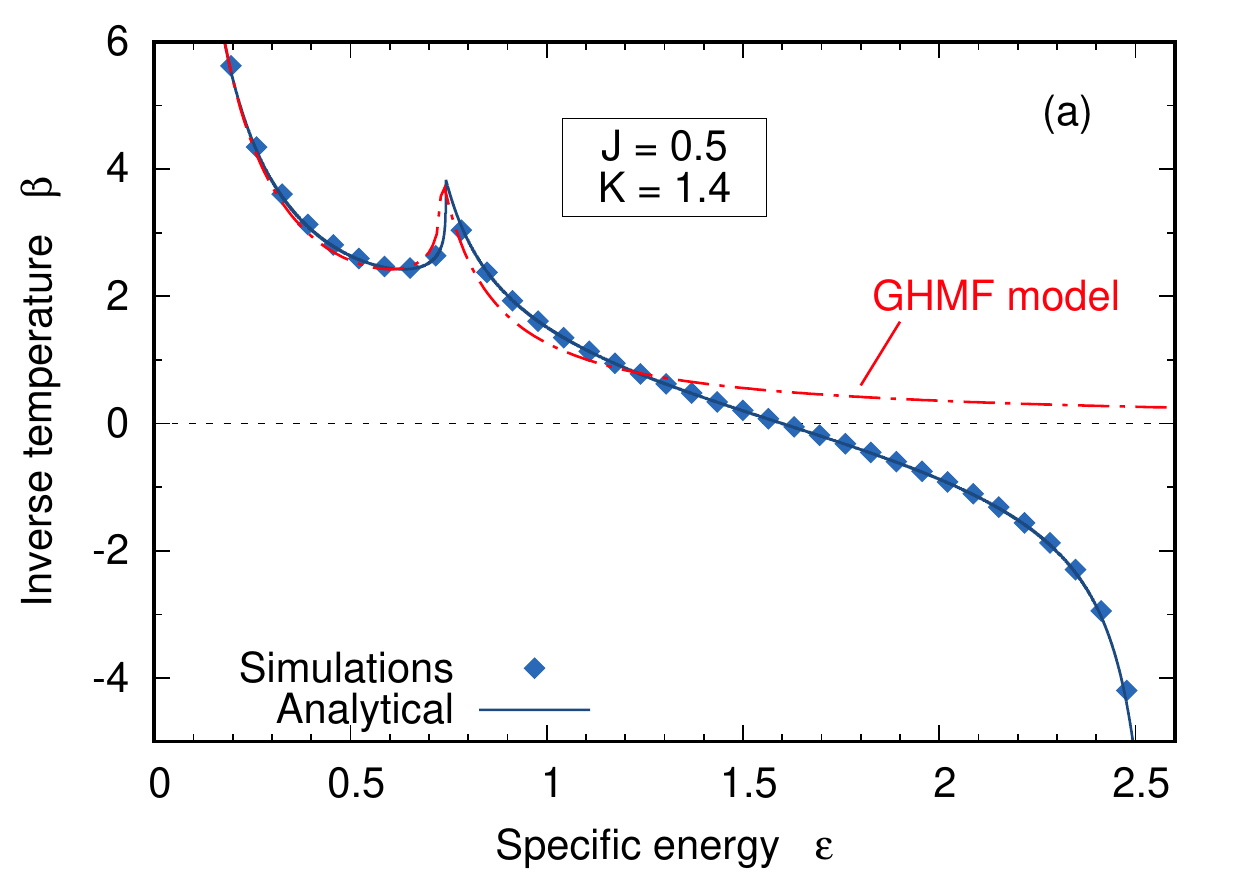}
\quad\\ \quad\\
\includegraphics[width=.6\linewidth]{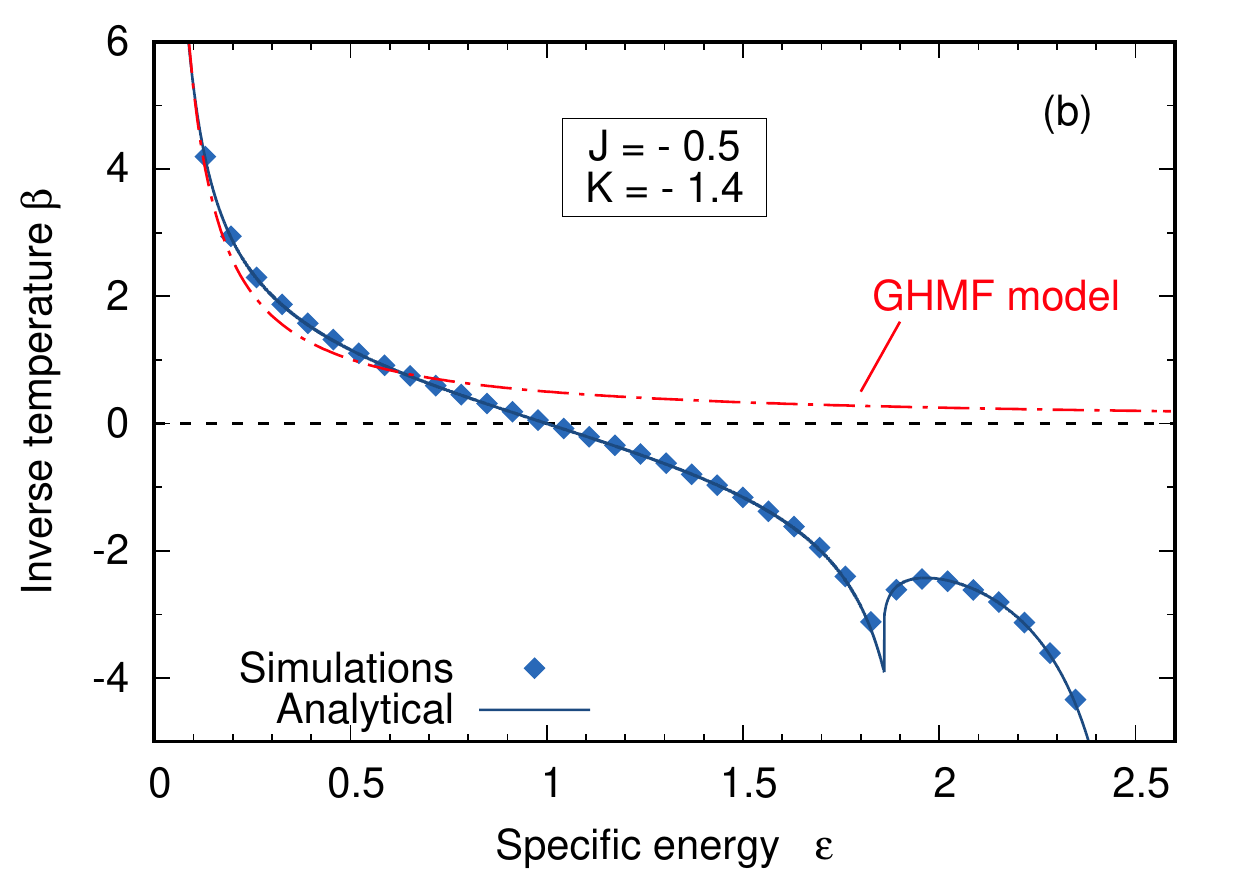}
\caption{\label{fig:miceli12} Dependence of $\beta$ on the specific energy in the Hamiltonian model~\eqref{GHMF_B}, living in bounded phase space, and in the corresponding GHMF model introduced in Ref.~\cite{debuyl05}. Two choices of the parameters are considered: $J=0.5$, $K=1,4$ (panel (a)) and $J=-0.5$, $K=-1,4$ (panel (b)). For both cases, numerical simulations in the microcanonical ensemble (light blue diamonds) are compared with the analytical equilibrium prediction (blue solid line), obtained with large-deviations techniques~\cite{miceli19}. The values of $\beta$ are obtained from the analysis of averages of suitable observables, depending on the single-particle momentum distribution. Red dash-dotted lines represent the equilibrium behaviour of the GHMF model with the same parameters: the two models show similar features at low energy, but they differ in the high-energy limit. Size of the system $N = 10^3$.
Figure adapted from Ref.~\cite{miceli19}.}
\end{figure}

\begin{figure}[h!]
 \centering
\includegraphics[width=.6\linewidth]{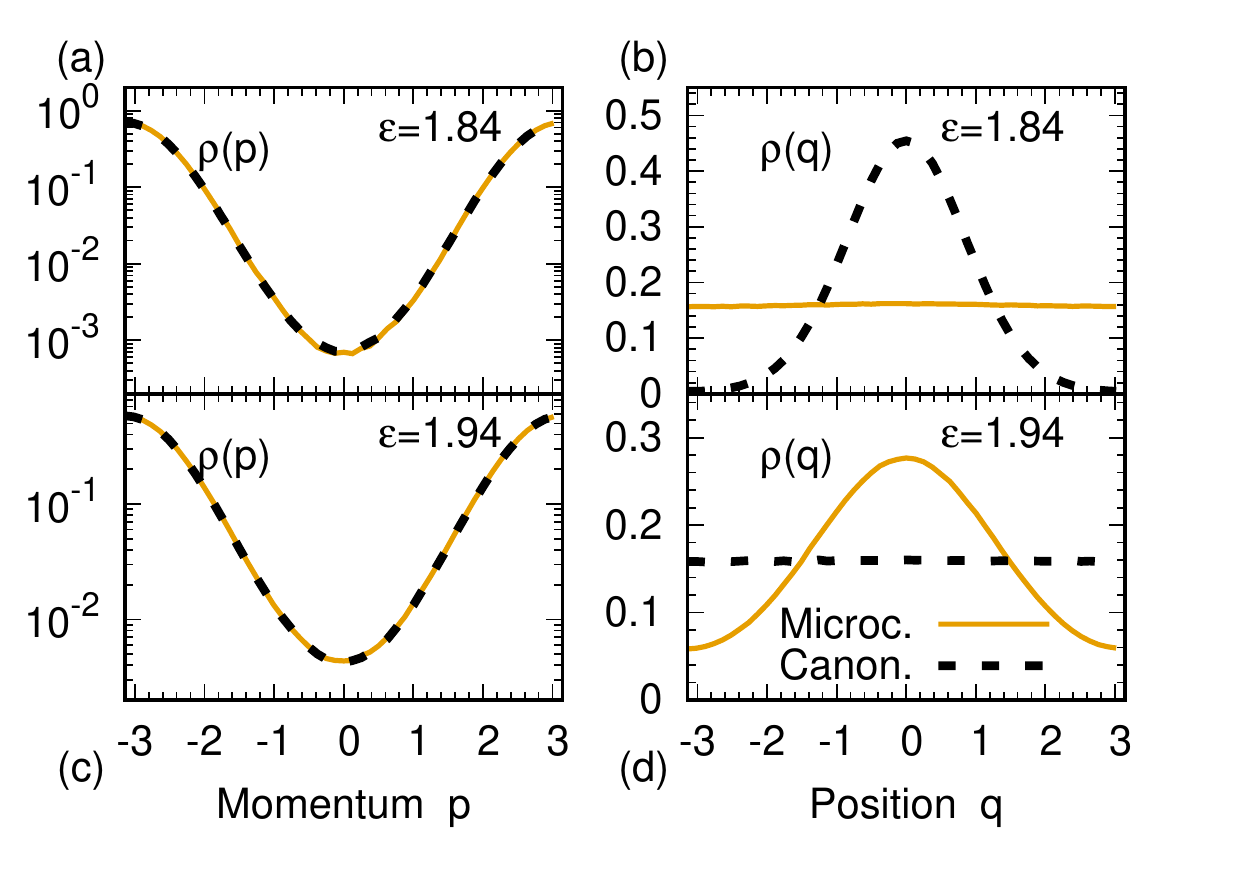}
\caption{\label{fig:miceli4} Inequivalence between canonical and microcanonical ensemble for model~\eqref{GHMF_B}
with $J=-0.5$, $K=-1.4$. The microcanonical equilibrium state described by a given specific energy $\epsilon=\mathcal{H}/N$ is compared to the corresponding canonical state at the same inverse temperature. Two cases are shown in which, even if the pdf of the single-particle momentum is identical in the two ensembles (panel (a): $\epsilon=1.84$; panel (c):$\epsilon=1.94$), the positions of the particles behave in a completely different way (panel (b): $\epsilon=1.84$; panel (d): $\epsilon=1.94$). The canonical ensemble has been
reproduced by mean of a Langevin-like stochastic bath, see Ref.~\cite{miceli19} for details, and the general discussion in Section~\ref{sec:langevin} of the present review. Figure adapted from Ref.~\cite{miceli19}.}
\end{figure}

The first example is a model of a 1d lattice of rotators, whose Hamiltonian 
reads \cite{miceli19}
\begin{equation}
\label{GHMF_B}
{\mathcal H} = - \sum_{j=1}^N ( 1 - \cos (p_j)) - N v(m)  \, ,
\end{equation}
where
\begin{equation}
\label{GHMF_B_pot}
v(m) =  \frac{J}{2} m^2 +  \frac{K}{4} m^4 + C \, ,
\end{equation}
with
\begin{equation}
\label{GHMF_B_m}
m =   |{\bf m}| \,, \quad {\bf m} = \frac{1}{N} \Big(\sum_{j=1}^N 
\cos q_j , \sum_{j=1}^N \cos q_j \Big) \,.
\end{equation}
The  canonically conjugated angular variables $q_j$ and $p_j$ are 
constrained to the interval $(-\pi, \pi]$; the long-range interaction among 
rotators is due to the nonlinear mean-field
potential $v(m)$, which depends on the modulus of the magnetization vector 
${\bf m}$. This is a suitably modified version of the Generalized Hamiltonian 
Mean-Field (GHMF)
model \cite{debuyl05}, which exhibits a standard quadratic kinetic term of the form 
$ \sum_{j=1}^N p^2_j$ (here each $p_j$ is a true unbounded action variable) 
replacing the sum on the r.h.s. of Eq.(\ref{GHMF_B}). The GHMF model has been 
widely investigated as a paradigmatic example of a long-range interacting 
system which exhibits inequivalence between statistical
ensembles in correspondence of a magnetic transition~\cite{debuyl05}.  As discussed 
in~\cite{miceli19}, the modified kinetic term in (\ref{GHMF_B}) allows one to obtain 
negative-temperature equilibrium 
states, because of the boundedness of the phase-space yielding an energy range 
with decreasing entropy. It is worth recalling again here that, as discussed in 
other sections of this review
(see~\ref{sec:nat},~\ref{sec:debate}), this peculiar feature of the phase-space is a basic ingredient for 
observing negative-temperature equilibrium states in similar models. The 
behaviour of the caloric curve of model (\ref{GHMF_B}), i.e. $\beta \, {\mathrm 
vs}. \, \epsilon = {\mathcal H}/N$,  has been obtained by large-deviation 
techniques and further checked by numerical simulations of the
Hamiltonian dynamics (see \cite{miceli19}: as discussed  in this reference the 
non-quadratic form of the kinetic term demands the use of a 
suitable relation for obtaining the  numerical estimate of $\beta$).  For 
positive values of the parameters $J$ and $K$ the magnetic transition is found 
to occur at positive temperature
(see Fig.~\ref{fig:miceli12}(a)), while the 
equivalence between equilibrium statistical ensembles if found to extend out of 
the transition region, including the
negative temperature phase which is observed for large enough values of 
$\epsilon$. For negative values of $J$ and $K$ the magnetic transition moves to 
the region of negative 
temperatures and also in this case inequivalence between equilibrium statistical 
ensembles is observed in the transition region (see Fig.~\ref{fig:miceli12}(b)).  In particular, this feature has 
been tested
numerically by comparing the probability distributions of the angular variable 
$q_j$ obtained by the dynamics of an isolated chain (microcanonical setup) 
and by the stochastic 
Langevin-like dynamics, simulating the interaction with a thermal reservoir: in 
the transition region where the microcanonical specific heat is predicted to be
negative these two probability distributions do not match each other (see Fig.~\ref{fig:miceli4}). In summary, this model 
provides us with additional
relevant information about the relation between negative temperature equilibrium 
states and equivalence between statistical ensembles:
\begin{itemize}
\item apart the region of the magnetic transition
we are facing the same scenario provided by the  simple models discussed in the 
previous section, which rules out a consistent description of the thermodynamics 
of this model
making use of the Gibbs' entropy;
\item inequivalence between statistical ensembles may occur also at negative 
temperature around a phase transition, but this is due to the same mechanism
occurring at positive temperature, i.e. the  metastability  of microcanonical 
states, yielding negative specific heat.  
\end{itemize}

The last example we want to discuss here is the Hamiltonian lattice model 
yielding the Discrete Nonlinear Schr\"odinger Equation:
\begin{equation}
\label{DNLS}
{\mathcal H} = \sum_{j=1}^N ( z_j \, z^*_{j+1} + c.c.) + \sum_{j=1}^N |z_j|^4
  \, .
\end{equation}
Without prejudice of generality, here the Hamiltonian~\eqref{eq:Hdnsle} has been turned to the simplified form where $\Lambda =2$.
It can be viewed as the nonlinear version of model (\ref{Hdlse}) with a quartic 
nonlinearity. This model has been extensively studied, because of its 
phenomenological interest for many physical applications, e.g. Bose-Einstein 
condensates in optical lattices \cite{TS,franzosi11} and
light propagating in arrays of optical waveguides \cite{ESMBA}. As well as 
(\ref{Hdlse}), this nonlinear model conserves the total energy and the total 
number of
particles in the lattice, $A=\sum_{i=1}^N |z_i|^2$. The problem of 
the equivalence between equilibrium statistical ensembles and negative 
temperature states
in this model is discussed in detail in Section~\ref{sec:dnls}. Here we just summarize the 
specific situation emerging in this problem, which further enriches the overall 
scenario.
The thermodynamics of model (\ref{DNLS}) is summarized by the phase diagram in 
Fig.~\ref{fig:phase_diagDNLS}. No physical states are accessible
below the $(\beta=+\infty)$-line, $h=a^2 - 2a$,
$h$ and $a$ being respectively the densities $H/N$ and $A/N$. Between this line and the 
$(\beta=0)$-line, $h=2a^2$, one has positive temperature states and equivalence 
between
statistical ensembles. At variance with the previous model, the phase-space is 
unbounded. This notwithstanding, for energies above the $(\beta=0)$-line 
the microcanonical
Boltzmann entropy is found to decrease with $h$. In particular, it has been 
shown \cite{GILM1} that  large-deviation techniques allow one to obtain an 
analytic estimate of $S_B$, 
which indicates that for large values of $N$, $\beta_B$ is actually negative. On 
the other hand, the same estimate rules out the possibility of the equivalence 
of the
microcanonical ensemble with the grand-canonical one in the phase where 
$\beta_B <0$, because of the very analytic structure of $S_B$, which forbids to 
express the 
grand-canonical partition function ${\mathcal Z}$ as a Fenchel-Legendre 
transform of $S_B$ (for details see \cite{GILM1, GILM2}). What makes the 
difference between this model
and all the previously described ones is the physical mechanism ruling the phase 
at negative temperature, i.e. the region above the line $h=2a^2$. Actually, 
numerical simulations
of the Hamiltonian dynamics (e.g. see \cite{franzosi11,ICOPP}) indicate the 
formation of localized excitations in the form of breathers, which may live over 
extremely long time spans (the larger their
amplitude the less they interact with the radiation background) and then 
may be born again by large fluctuations at some lattice site. Thermodynamics 
predicts that, apart a small region just above the line $h=2a^2$, which vanishes 
in the thermodynamic
limit,  all breathers should coalesce into a single giant breather collecting a 
macroscopic portion of the total energy \cite{GILM1, GILM2}.  In fact, at 
variance with
solitons, breathers interact between themselves, like in a fish-species 
dynamics, where the larger typically eats the smaller. On the other hand, this 
coalescence mechanism 
seems to never come to the expected end in the dynamics of an isolated DNLS 
chain, even for moderate system size $N$: 
the transient to equilibrium is strongly inhibited by localization. Conversely, 
stochastic numerical simulations conserving both ${\mathcal H}$ and $A$ yield
the expected equilibrium configuration through a coarsening process ruled by 
predictable scaling properties~\cite{Iubini14, Iubini17}.
As a concluding remark about microcanonical thermodynamics, it is 
worth
mentioning that in the thermodynamic limit negative temperature states boil down 
to infinite-temperature ones~\cite{GILM1}.

We want to conclude this section by pointing out  that one can perform also
canonical simulation in the $\beta<0$ region of the DNLS model by exploiting 
inverse Monte Carlo protocols or Langevin-like heat reservoirs acting on a DNLS 
chain. As one can easily guess,
the localization mechanism of energy into breather states allows one to pump 
from thermodynamic fluctuations induced by the reservoirs an arbitrary amount on 
energy and norm into the system, 
which eventually ``explodes'', thus providing evidence 
that, as already reported in the seminal book by D. Ruelle~\cite{Ruelle}, 
physical condensation mechanisms may play a major role in the nonequivalence of 
statistical equilibrium
ensembles. This point will be further discussed in Section~\ref{sec:fourierdnls}.

\newpage
\section{Fluctuation-dissipation and response theory }
\label{sec:fdr}

In the last Section we discussed in which sense the notion of equilibrium at 
NAT 
is physically meaningful, and to what extent it is consistent with statistical 
mechanics. In what follows we focus instead on fluctuation-dissipation 
relation (FDR) and response theory: this can be seen as a first step toward a 
generalization of out-of-equilibrium statistical mechanics to cases with NAT, a 
topic which will be also faced in Section~\ref{sec:fourier}. First we 
discuss a stochastic differential equation (a ``generalized'' 
Langevin equation) to describe the motion of particles interacting with NAT 
heat baths. Such thermal reservoirs may be composed, for instance, by many 
``lighter'' particles with smaller inertia, or by a lattice of Ising spins with 
their own stochastic dynamics; the overall result does not depend significantly 
on these details. The validity of FDR,  linking spontaneous correlations to the 
linear response induced by an external perturbation, is then investigated when
NAT is present.

\subsection{Langevin equation with $\beta < 0$}
\label{sec:langevin}

One of the pillars of out-of-equilibrium statistical mechanics is the
possibility to write a Langevin equation, i.e. an effective description for
the slow motion  of  a ``heavy'' particle (mass $M$) interacting
 with many ``light'', fast degrees of freedom (mass $m\ll M$).
 Such a  problem was originally faced  with  a phenomenological approach  
by Paul Langevin 
 in his seminal paper on   the Brownian motion of a colloidal particle in a 
fluid~\cite{langevin}.
 Langevin's analysis grounds on the assumption that
the heavy particle's behaviour is determined by two competing elements,
whose combined effect allows for the thermal equilibrium at temperature $T$:
\textit{(i)} the viscous force  given by Stokes law; \textit{(ii)} the presence of random thermal 
fluctuations.
The resulting differential stochastic equation reads
 \begin{equation}
 \label{eq:langevin}
\frac{dP}{dt}= -\frac{\gamma}{M} P(t) + \sqrt{2\gamma k_B T} \xi(t) \,,
\end{equation}
where $P(t)$ is the momentum of the heavy particle,
 $T$ is the temperature of the system  and
 $\xi$ is a Gaussian white noise: $\langle \xi(t) \rangle=0$,  $\langle 
\xi(t)\xi(t')\rangle=\delta(t-t')$.

Actually, only in few cases  it is possible to obtain the Langevin 
equation~\eqref{eq:langevin} analytically, determining the friction parameter 
$\gamma$ from first principles. Important examples are the motion of a heavy 
intruder in a chain of harmonic oscillators~\cite{rubin60, turner60, 
takeno-hori62, ford-kac-mazur65, zwanzig73} and that of a tracer in a gas of elastic hard 
disks (or beads, in 3-dimensional space)  with very low 
density~\cite{smoluchowski06, dorfman70, dorfman09}. Usually 
Eq.~\eqref{eq:langevin} is instead assumed to be a valid model for the 
considered phenomenon, and its effectiveness is checked \textit{a posteriori}.

Langevin-like equations hold also in cases with NAT. For instance, it has been 
shown that it is possible to determine drift and a diffusion coefficients 
accounting for the effective motion of a single vortex in two-dimensional 
hydrodynamics~\cite{chavanis98, chavanis07, yatsuyanagi15explicit}; as 
discussed 
in Section~\ref{sec:vortices}, NAT can be observed, as a consequence of the 
bounded nature of the Hamiltonian coordinates space, as soon as the system is 
constrained in a finite domain. The drift term of this Langevin equation 
explicitly depends on $\beta$, in such a way that states with opposite 
temperature correspond to qualitatively different dynamics: if $\beta>0$ the 
vortex is repelled from the center of the domain, whereas if $\beta<0$ it is 
attracted~\cite{chavanis98}. 

In what follows we discuss a generalized Langevin equation, valid for 
mechanical 
models of the form~\eqref{eq:minimal} already encountered in the previous 
Section. In this kind of systems, at variance with the case of vortices, the 
Hamiltonian includes  generalized kinetic terms $\mathcal{K}(p)$ explicitly 
depending on the single particle momentum: in this sense the resulting 
stochastic differential equation is a direct generalization of 
Eq.~\eqref{eq:langevin}, which is recovered as a particular case.
We will consider systems composed by a large number $N$ of fast, microscopic 
degrees of freedom
(the ``bath'') and a slow, mesoscopic one (the ``tracer''). The physical meaning 
of this scale separation
and its relation to a generalized notion of inertia for Hamiltonians of the 
form~\eqref{eq:minimal}
will be outlined in the following subsection. By now we shall just assume, as a 
work hypothesis,
that the typical times of the tracer dynamics result much longer than those of 
the bath.

The Hamiltonian of the whole systems reads
 \begin{equation}
 \label{eq:completeham}
  \mathcal{H}(P,\{p_n\},Q,\{q_n\})=\mathcal{K}(P)+\sum_{n=1}^N 
{\mathcal{K}'}(p_n)+ \mathcal{U}(Q) + \mathcal{V}(Q,\{q_n\})
 \end{equation}
 where $(P,Q)$  are the canonical coordinates of the mesoscopic tracer, 
$(p_n,q_n)$ those of the $n$-th particle of the bath.
 Of course in the equations of motion for the tracer there appear terms 
depending on the coordinates of the fast particles
 constituting the bath:
 \begin{equation}
 \label{eq:motion}
   {d \over dt} Q=\partial_PK(P) \,\,\, , \,\,\,
   {d \over dt} P=-\partial_QU(Q)-\partial_QV(Q,\{q_n\}) \, .
 \end{equation}
 Within the mentioned time-scale separation hypothesis it is reasonable to write 
down an effective stochastic equation
 for the tracer: 
 \begin{equation}
 \begin{aligned}
    {d \over dt} Q&=\partial_P\mathcal{K}(P)\\
   {d \over dt} P  &=-\partial_Q\mathcal{U}(Q)+\Gamma(P)+\sqrt{2D}\xi(t)\,,
 \end{aligned}
 \end{equation}
 where the interactions with the microscopic degrees of freedom of the bath are 
mimicked  by a ``friction'' term $\Gamma(P)$
 and a noisy force depending on a diffusivity $D$. If this effective description 
actually holds,
 we are only left with the problem of determining the shape of  $\Gamma(P)$ and 
the value of $D$.

In this respect, a simple phenomenological argument can be outlined as follows.
Let us  define the steady probability density $f(Q,P)$ for the coordinates of 
the tracer, and the currents
\begin{equation}
 \label{eq:currentq}
 J_Q(Q,P)=f(Q,P)\partial_P\mathcal{K}(P)
 \end{equation}
 \begin{equation}
 \label{eq:currentp}
 J_P(Q,P)=-f(Q,P)\partial_Q\mathcal{U}(Q,P)+ \Gamma(P)f(Q,P)-D\partial_P 
f(Q,P)\,.
\end{equation}
The associated Fokker-Planck equation reads:
\begin{equation}
 \partial_QJ_Q(Q,P)+\partial_PJ_P(Q,P)=0\,.
\end{equation}
Assuming detailed balance (i.e. thermal equilibrium), the irreversible part of 
current~\eqref{eq:currentp} must vanish:
\begin{equation}
\Gamma(P)f(Q,P) -D\partial_Pf(Q,P)=0\,;
\end{equation} 
as a consequence, by exploiting the factorization
\begin{equation}
  f(Q,P) = f_Q(Q) f_P(P) \propto e^{-\beta[\mathcal{K}(P)+\mathcal{U}(Q)]}
   \end{equation}
    one finds
 \begin{equation}
  \Gamma(P)=-D \beta \partial_P \mathcal{K}(P) \,.
\end{equation}

The general version of the Langevin equation for the tracer is then
\begin{equation}
\label{eq:genlangevin}
 \frac{d}{dt}P=-D \beta \partial_P \mathcal{K}(P) + \sqrt{2 D}\xi(t) \,,
\end{equation}
where $D$ is a diffusivity term, which fixes the time scale of the dynamics. 
No assumption has been made about the sign of $\beta$, which can assume,
in principle, negative values. This has no influence on the diffusivity, but 
changes the sign of the drift term.
In the usual case $ \mathcal{K}(P)=P^2/2M$ we recover the linear form for the 
drift term,
with a viscous coefficient satisfying the familiar Einstein relation
\begin{equation}
\label{eq:geneinstein}
 \gamma=D \beta\,.
\end{equation} 

The validity of Eq.~\eqref{eq:genlangevin} can be numerically tested in the 
following way. First, one
runs molecular dynamics simulations of a Hamiltonian system of the 
form~\eqref{eq:completeham},
including both tracer and thermal reservoir, according to a deterministic 
evolution rule.
The particles composing the thermal bath should have very fast decorrelation 
times,
so to avoid memory effects; in this way the evolution of $P$ may be expected to 
be Markovian.
The tracer's coordinates are measured with high sampling frequency, as a 
function of time,
disregarding the dynamics of the fast particles; the resulting trajectory 
$(P(t), Q(t))$ is then
analyzed, \textit{a posteriori}, with usual Langevin equation reconstruction
techniques~\cite{peinke11, peinke19}, in order to find the stochastic 
differential
equation which reproduces the dynamics to the best extent. In other words, one
has to find, from numerical data, the drift $F(P)$ and diffusivity $D$ such that
\begin{equation}
 \frac{dP}{dt}= F(P) + \sqrt{2D} \xi(t)
\end{equation}
is a good model for the coarse-grained dynamics of the slow particle.
This task can be achieved, in principle, by exploiting the very definition of 
$F$ and $D$ as 
limits of the conditioned moments~\cite{gardiner85}
 \begin{equation}
 \label{eq:ded}
 \begin{aligned}
    F(P)=&\lim_{\Delta t \to 0}\, \frac{\langle \Delta P | P(t_0)=P 
\rangle}{\Delta t}\\
  D(P)=&\lim_{\Delta t \to 0}\, \frac{ \langle \Delta P^2 | P(t_0)=P \rangle}{2 
\Delta t}\,.
 \end{aligned}
 \end{equation}
When dealing with numerical simulations the notion of $\Delta \to 0$ limit has 
to be handled
with some care, since the time scale of $\Delta$ at which such limit is 
evaluated has to be small with
respect to typical time-scales of the dynamics, but still much larger than the 
time-step of the
numerical integration.

 \begin{figure}[h!]
 \centering
  \quad \quad \includegraphics[width=.6\linewidth]{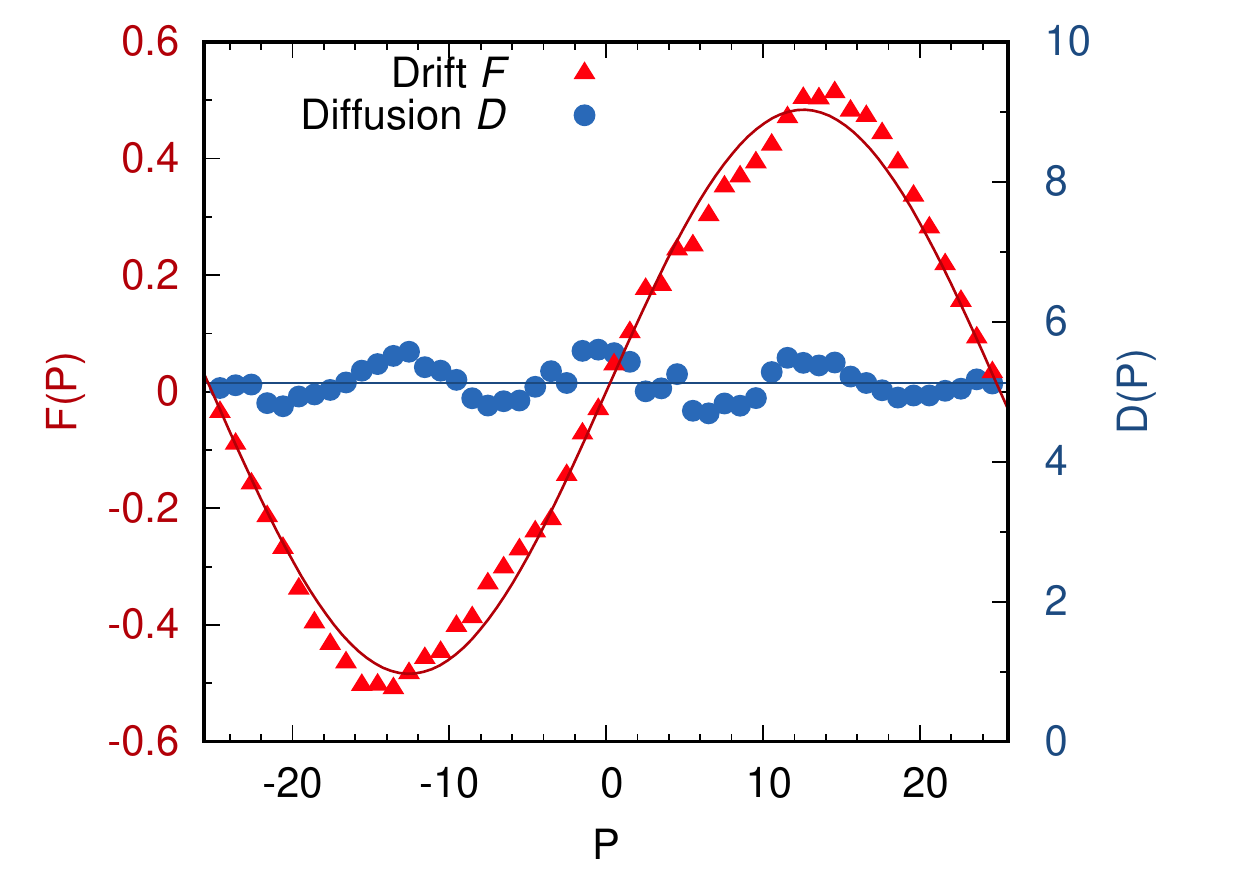}
   \caption{\label{fig:drift1} Drift and diffusivity reconstruction for the Langevin equation of $P$ for a model with
   kinetic energy~\eqref{eq:kintracer}, in a case with negative temperature. Based on numerical simulations of the whole system, the observables~\eqref{eq:ded}
   are plotted (points) and compared with the functional forms (solid lines) which can be guessed from the argument in the main text, see Eq.~\eqref{eq:genlangevin}. Here $\beta=-0.1$; the considered particle, with generalized ``mass'' $M=8$ (see also Section~\ref{sec:baths}), interacts with a bath composed by $N=600$ particles with mass $m=1$. Figure adapted from Ref.~\cite{baldovin18}.}
 \end{figure} 
  
  \begin{figure}[h!]
 \centering
  \includegraphics[width=.6\linewidth]{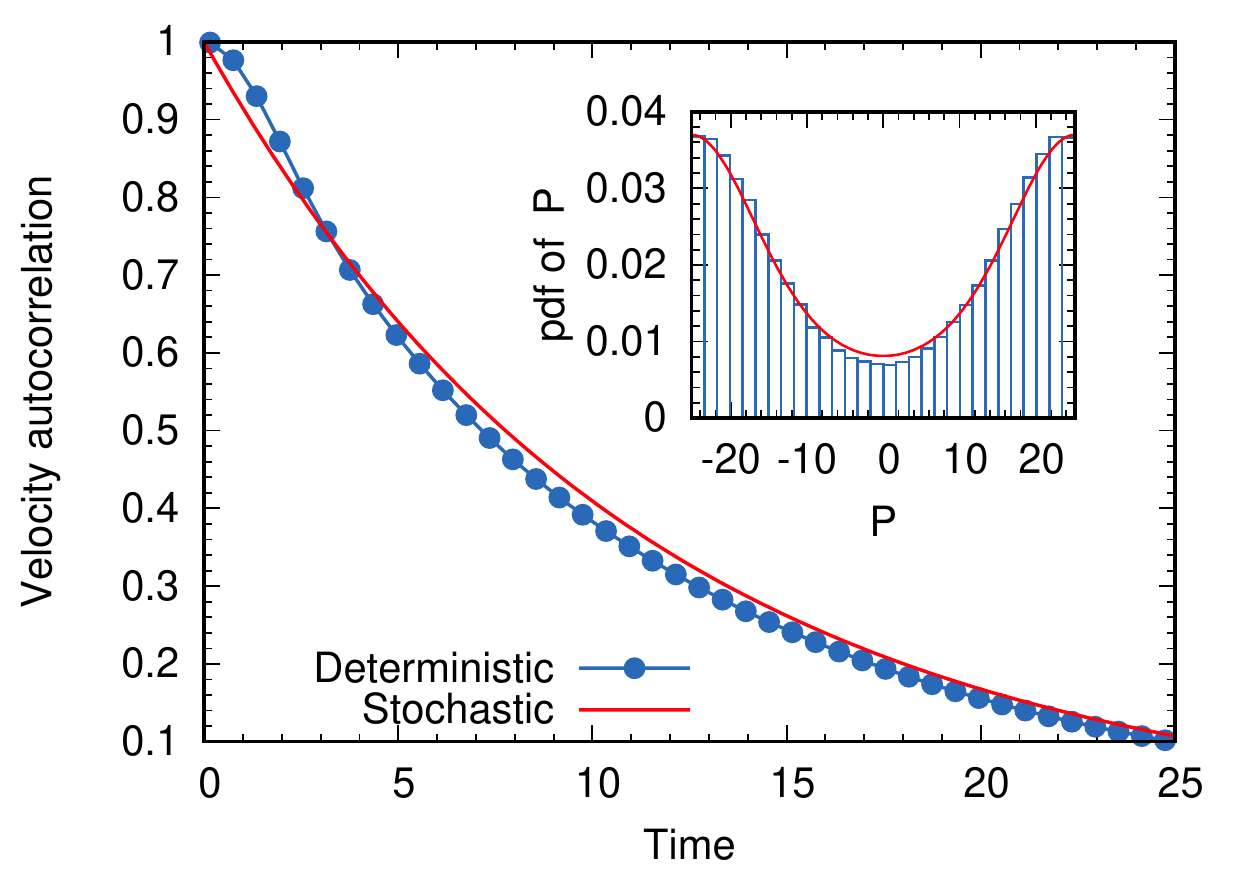}
  \caption{\label{fig:drift2} In the same conditions of Fig.~\ref{fig:drift1}, velocity autocorrelation function (main plot) and momentum distribution (inset) for the slow particle. Points and the histogram refer to the values measured in the original, complete dynamics; red solid lines show the predictions of the coarse-grained dynamics obtained as described in the main text. Figure adapted from Ref.~\cite{baldovin18}.}
 \end{figure}

 The proposed check has been done in Ref.~\cite{baldovin18} for the particular 
case
 \begin{equation}
 \label{eq:kintracer}
  \mathcal{K}(P)=1-\cos(P)\,.
 \end{equation} 
 Leaving apart, for a moment, the problem of modeling the thermal bath (we shall 
discuss it in the following subsection),
 we can already anticipate the results of this analysis. In 
Fig.~\ref{fig:drift1} the reconstructed functional forms of drift and 
diffusivity are shown to be
 consistent with Eq.~\eqref{eq:genlangevin}, in a case at negative temperature. 
Let us notice that the sign of $\beta$ influences the qualitative behaviour of 
the drift term, which in this case is positive when $P>0$ and negative when 
$P<0$ (in a positive-temperature scenario the behaviour would have been 
opposite). Figure~\ref{fig:drift2} shows instead the good agreement between 
the original, complete dynamics and the Langevin 
equation~\eqref{eq:genlangevin}, both for static and dynamic observables 
(namely, equilibrium p.d.f. and velocity autocorrelation function).

  \subsection{Thermal baths at negative temperature }
  \label{sec:baths}
 
The mesoscopic motion of a particle with generalized kinetic energy can be 
described by a Langevin equation of the form~\eqref{eq:genlangevin} with 
$\beta<0$ when the tracer is coupled to a thermal bath at NAT.  Such a  
reservoir should be composed by a large set of microscopic degrees of freedom 
living in bounded phase spaces, isolated from ordinary-matter environment, but 
coupled with the mesoscopic particle under study. To avoid memory effects in the 
noise term, the dynamics of the microscopic particles belonging to the bath 
should be much faster than that of the tracer. It is natural to wonder whether 
this thermal bath is physically meaningful, i.e. if it is possible to think of a
high-dimensional dynamics whose effect is actually described by
Eq.~\eqref{eq:genlangevin} also when $\beta<0$.

A minimal model can be implemented, for Hamiltonian systems of the 
form~\eqref{eq:completeham}, by a direct generalization of the notion of 
inertial mass to systems with generic kinetic terms. We ask a very natural 
physical condition to be satisfied, i.e. that the momentum and the kinetic 
energy of particles moving with the same velocity are proportional to the 
masses. This extensivity criterion is quite reasonable: indeed, if $n$ 
particles with mass $m$ and momentum $p$ are stucked together (and, therefore, 
they move with the same velocity), we expect their behaviour to be 
indistinguishable from that of a single particle with mass $nm$, momentum $np$ 
and total kinetic energy equal to $n$ times the one of the single particle. In 
formulae
\begin{equation}
\mathcal{K}(np,nm)=n\mathcal{K}(p,m)\,.
\end{equation} 
It is easy to check that the above condition is verified by kinetic terms of the 
form
\begin{equation}
\label{eq:kineticform}
 \mathcal{K}(p,m)= mc^2 k(p/cm)\,,
\end{equation} 
where $k(x)$ is an even function and $c$ is a constant with the physical 
dimensions of a velocity. The function $k$ needs to be even in order to 
preserve spatial isotropy. For instance, let us remark that choosing 
$k(x)=x^2/2$ leads to the usual kinetic energy of classical mechanics (and no 
dependence on $c$ appears), while $k(x)=\sqrt{1+x^2}$ gives the relativistic one 
($c$ being the light velocity in that context).

With this picture in mind, we can simulate a thermal bath for the tracer with 
kinetic energy~\eqref{eq:kintracer} by considering a large set of microscopic 
particles of the same nature, but with smaller masses:
\begin{equation}
  \mathcal{H}=M[1-\cos(P/M)]+\sum_{n=1}^N m[1-\cos(p_n/m)]+ \mathcal{U}(Q) + 
\mathcal{V}(Q,\{q_n\})\,,
\end{equation} 
 where $m \ll M$ and we have set $c=1$. To obtain the results shown in 
Fig.s~\eqref{fig:drift1} and ~\eqref{fig:drift2} one can choose $\mathcal{V}(Q,\{q_n\})$ in such a way 
that the particles of the bath exchange heat between themselves and some of 
them are also coupled to the tracer: this should prevent the presence of memory 
effects in the noise term, and insure as well thermal equilibrium among the 
particles of the bath~\cite{baldovin18}.
  
Of course the above discussed Hamiltonian approach is not the only possible choice
 to model a bath at NAT. In many cases a thermal bath can be
 constituted by a lattice of Ising spins interacting with the tracer~\cite{bonilla10,bonilla15,ruiz-garcia17}; since
 Ising spins can be also found, in principle, in states with $\beta<0$ (see Section~\ref{sec:mag_sys}), they are
 a promising candidate.
 
 In Ref.~\cite{baldovin19} the compound physical system
 \begin{equation}
 \label{eq:hamprados}
  \mathcal{H}(P,Q, \boldsymbol{\sigma})=\mathcal{K}(P)+\mathcal{U}(Q)-\mu \lambda(Q) \sum_{n=1}^N \sigma_n
 \end{equation} 
 has been investigated, where the tracer $(P,Q)$ interacts with $N$ Ising spins $\{\sigma_n=\pm1\}$. Here $\lambda(Q)$ is a function of the tracer position and $\mu$ is a constant, scaling as $N^{-1/2}$. 
 The spin lattice follows a stochastic Glauber dynamics~\cite{glauber63}, so that each spin ``flips'' (i.e., changes its sign) with a rate $\omega$ depending both on $\beta$ and on the external field:
 \begin{equation}
  \omega(\boldsymbol{\sigma}\to\boldsymbol{\sigma}'_{n}|Q)=\frac{\alpha}{2}\left(1+\sigma_n \tanh[\beta \lambda(Q)] \right)\,,
 \end{equation}
 where $\boldsymbol{\sigma}'_{n}$ is the spins configuration which is obtained from $\boldsymbol{\sigma}$ by switching only the $n$-th spin, and $\alpha$ is a typical frequency of the bath. With the above stochastic evolution, the stationary pdf of the $n$-th spin, at fixed $Q$, is given by the equilibrium distribution
 \begin{equation}
  \mathcal{P}^{eq}(\sigma_n|Q)= \frac{e^{\sigma_n \beta \mu \lambda(Q)}}{2 \cosh[\beta \mu \lambda(Q)]}\,,
 \end{equation} 
and detailed balance is satisfied.

\begin{figure}[h!]
 \centering
\includegraphics[width=.6\linewidth]{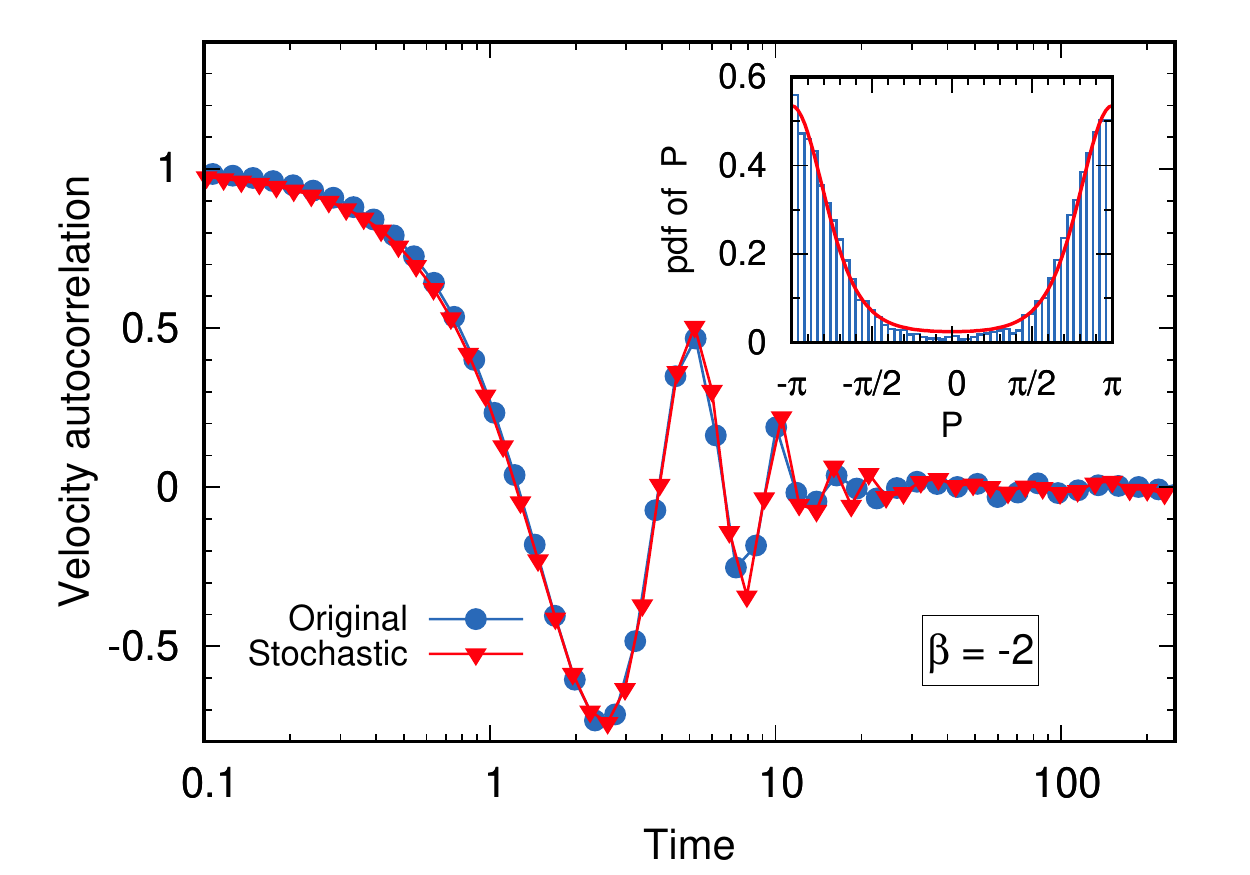}
\caption{\label{fig:prados} For model~\eqref{eq:hamprados}, velocity autocorrelation function (main plot) and  momentum distribution (inset) of the oscillator. Blue circles and the histogram refer to the original dynamics, red squares and solid line to the coarse-grained stochastic dynamics deduced by mean of the Chapman-Enskog expansion. Here $\beta=-2$. Parameters: $N=10^4$, $\alpha=10$, $\mu=10^{-2}$. Figure adapted from Ref.~\cite{baldovin19}.}
\end{figure}

The complete dynamics of the system is given by a Liouville-Master equation which evolves both the continuous pdf of the tracer and the discrete distribution of the spins. If the motion of the latter is fast enough (i.e., $\alpha \gg 1$), a Chapman-Enskog expansion can be performed, with parameter $\varepsilon=\alpha^{-1}$, in order to find an approximate equation for the time-dependent marginalized pdf of the tracer, $f(Q,P,t)$: this amounts to expanding both the total pdf
\begin{equation}
 \mathcal{P}(P,Q,\boldsymbol{\sigma},t)=f(P,Q,t)\mathcal{P}^{eq}(\boldsymbol{\sigma}|Q)+ \sum_{l=1}^{\infty}\varepsilon^l \mathcal{P}^{(l)}(P,Q,\boldsymbol{\sigma},t)
\end{equation} 
and the time derivative of $f(Q,P,t)$,
\begin{equation}
 \partial_t f(Q,P,t)=\sum_{l=0}^{\infty}\varepsilon^l F^{(l)}(P,Q,t)\,.
\end{equation} 
The resulting equations can be solved order by order, leading to an approximate form very similar to those obtained with the ``mechanical'' bath. In particular, Eq.~\eqref{eq:genlangevin} is recovered, with the only differences that a ``renormalization'' of the potential occurs in this case, due to the interaction with the spins, and that the diffusivity $D$ depends on $Q$ in this case~\cite{baldovin19}.
Fig.~\ref{fig:prados} shows that the coarse-grained dynamics obtained with the Chapman-Enskog expansion is characterized by a behaviour quite similar to that of the original evolution: the stationary pdfs are very similar, both at positive and negative temperature, and also dynamical observables as correlation functions show a good agreement.

It is important to stress that the validity of a coarse-grained stochastic description for the dynamics of particles at negative temperature has an immediate consequence, very useful from a computational point of view, i.e. the possibility to simulate
the effect of a thermal bath at negative temperature by mean of standard methods of numerical stochastic integration. An application is discussed, e.g., in Section~\ref{sec:fouriersimple}. However this is not the only way to implement numerically
a thermal bath at negative temperature.

\begin{figure}[h!]
 \centering
 \includegraphics[width=.6\linewidth]{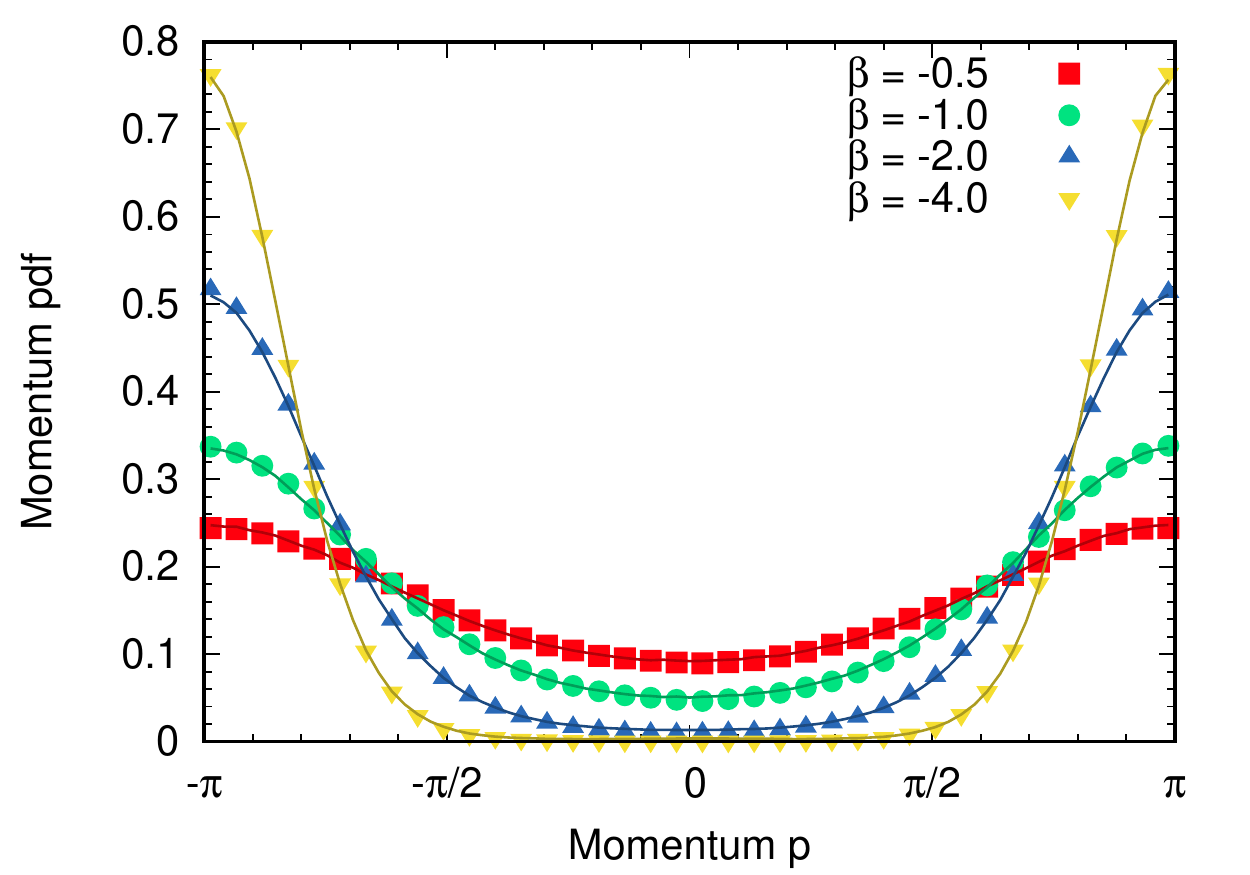}
 \caption{\label{fig:bath} Comparison between Langevin stochastic bath and Monte Carlo algorithm. Model~\eqref{eq:minimal} with kinetic energy~\eqref{eq:minimalk} 
and potential~\eqref{eq:minimalu} is considered. Points represent histograms of the single-particle momentum distribution for Monte Carlo simulations at the (negative) inverse temperatures reported in the key. Solid lines show the results of corresponding simulations in which the extremal points of the chain are coupled to Langevin stochastic baths, as discussed in the main text. Parameters: $\delta=0.1$ for Monte Carlo, $D=0.5$ and time-step $\Delta t=0.3$ for simulations with Langevin baths. Here $N=100$.}
\end{figure}

In Ref.~\cite{iubini12,iubini17entropy} a Monte Carlo algorithm is introduced to study the equilibrium dynamics of the Discrete Nonlinear Schr\"odinger equation; such numerical scheme  aims at reproducing the statistical properties of an equilibrium state at a given inverse temperature $\beta$; in principle, $\beta$ can be either positive or negative. As in the usual Monte Carlo, at each step a small random perturbation of the considered system (a ``move'') is proposed; the move is extracted  according to some fixed rule, whose optimization is usually a non-trivial task, clearly depending on the details of the considered system.
The energy variation $\Delta E$  corresponding to the proposed move is then computed; the criterion to accept or reject the move is basically a generalization of the Metropolis algorithm:
\begin{itemize}
 \item if the desired temperature is positive, the move is always accepted when $\Delta E<0$, and accepted with a probability $e^{-\beta \Delta E}$ when $\Delta E >0$;
 \item if, instead, the desired temperature is negative, the move is always accepted when $\Delta E >0$, and accepted with a probability $e^{-\beta \Delta E}$ when $\Delta E <0$.
\end{itemize}
The above procedure insures that the detailed balance condition between the transition rate of a move and that of its reverse is always satisfied; indeed, it is given by construction by $e^{-\beta \Delta E}$, i.e. by the ratio between the stationary probabilities of the corresponding initial states in the canonical ensemble. The possibility of negative temperature does not affect this general principle: at a practical level, it only implies that one has to pay attention to the sign of $\beta$ and apply the above rule accordingly.

In Fig.~\ref{fig:bath} the single particle momentum pdfs which can be obtained with the above discussed Monte Carlo method are shown for the already mentioned Hamiltonian chain~\eqref{eq:minimal} with kinetic energy~\eqref{eq:minimalk} and potential~\eqref{eq:minimalu}. At each step one particle is randomly chosen, and the proposed move is a shift of its position and its momentum by a quantity randomly extracted from a Gaussian distribution with zero mean and variance $\delta^2$. A nice agreement is found with the outcomes of quasi-symplectic simulations, in which the extremal particles of the chain are coupled to Langevin-like baths reproducing Eq.~\eqref{eq:genlangevin}.

\subsection{Response theory }

As mentioned before, FDR allows to predict the response of a system to (small) 
time-dependent external fields, by just analyzing time correlations in absence 
of perturbations.

The best known formulation of FDR is due to Kubo~\cite{kubo66}, and it holds for systems at equilibrium subjected to the action of a small, time-dependent field $h(t)$. Let us denote by $\mathcal{H}_0 ({\bf x})$ the ``unperturbed'' Hamiltonian, ${\bf x}$ being the vector of canonical coordinates. If at time $t=0$ we turn on the external field, a small perturbation
\begin{equation}
\mathcal{H}_0 ({\bf x}) \to \mathcal{H}_0 ({\bf x}) -h(t) \mathcal{M}({\bf x})\,,
\end{equation}
is induced on the system. The linear dependence on the perturbation is a mere consequence of the assumption that $h(t)$ is small, so that higher-order terms can be neglected. Response theory insures then that the quantity
$$
 \delta \mathcal{A}(t) = \overline{ \mathcal{A}({\bf x}(t)) - \mathcal{A}({\bf x}(0)) }\,,
$$
where $\mathcal{A}({\bf x})$ is a mechanical observable and the average $\overline{\,\cdot\,}$ is computed by
repeating the experiment many times, can be expressed as~\cite{kubo66}
\begin{equation}
\label{eq:responsekubo}
\delta \mathcal{A}(t) = \beta \int_0^t d t'\left\langle \mathcal{A}({\bf x}(t-t')) \dot{ \mathcal{M}}({\bf x}(0))   \right\rangle  h(t')\,,
\end{equation} 
where average $\langle \cdot \rangle$ is computed instead on the unperturbed system.
In the above formula the inverse temperature $\beta$ appears, and no assumption on its sign is made in the
derivation of Eq.~\eqref{eq:responsekubo}.

\begin{figure}[h!]
 \centering
\includegraphics[width=.6\linewidth]{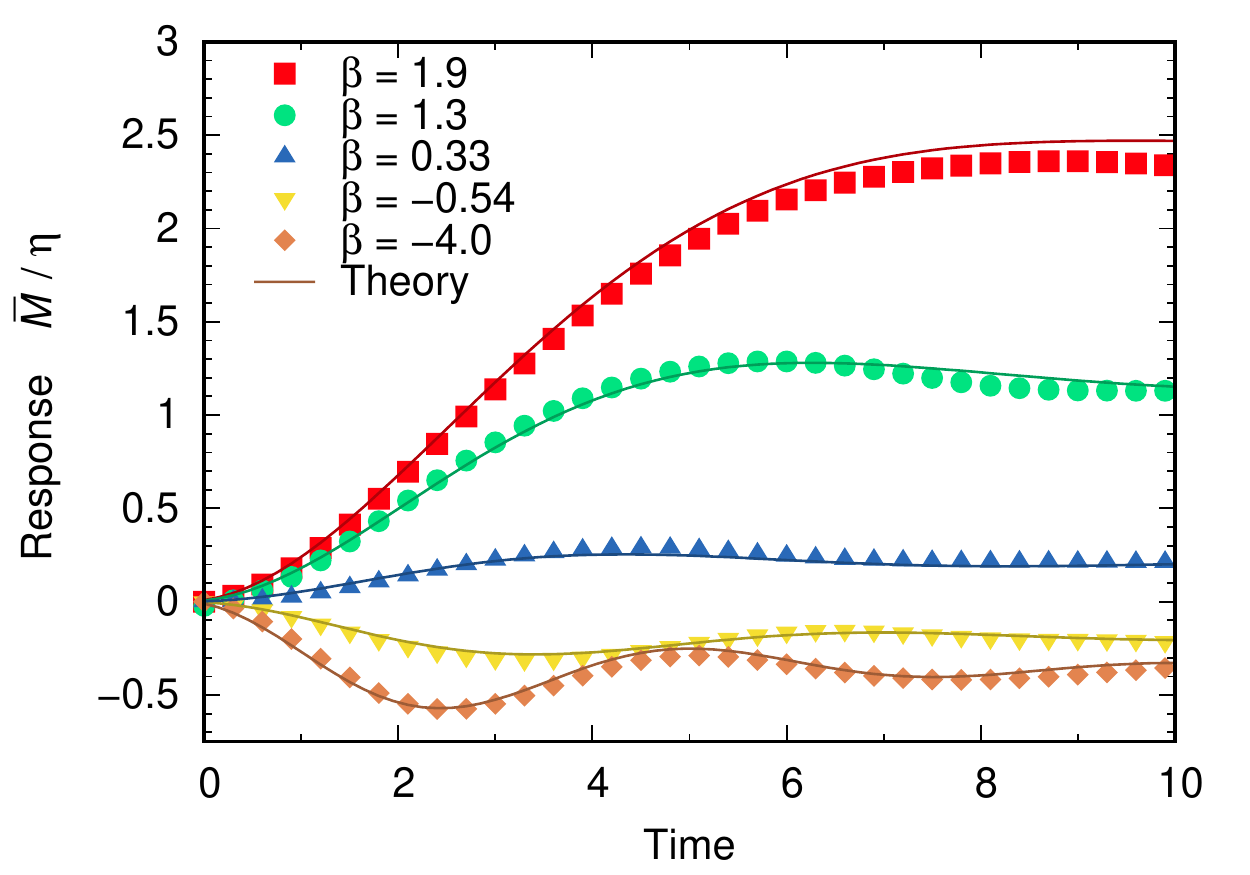}
\caption{\label{fig:resp1} Check of Kubo formula in a system with NAT. For various choices of inverse temperature, both positive and negative, the measured value of the time-dependent magnetization $M$ of model~\eqref{eq:hamresp} after the sudden switch of an external field of intensity $\eta$ is shown. Here $N=128$, $\eta=0.02$, average over $10^4$ realizations.}
\end{figure}

To check the validity of Kubo formula also in systems with negative temperature, 
we can rely again on the simple Hamiltonian model in bounded phase space 
discussed in Section~\ref{sec:measuring}, introducing, this time, an explicit 
dependence on an external field. The Hamiltonian reads
\begin{equation}
\label{eq:hamresp}
 \mathcal{H}(\mathbf{q},\mathbf{p},t)= \mathcal{H}_0(\mathbf{q},\mathbf{p})-h(t) \mathcal{M}(\mathbf{q})\,,
\end{equation} 
where
\begin{equation}
\label{eq:hamunp}
 \mathcal{H}_0(\mathbf{q},\mathbf{p})= \sum_{i=1}^N[1-\cos(p_i)]+ \frac{1}{2} \sum_{i=1}^{N+1}[1-\cos(q_i-q_{i-1})]\quad \quad q_{0}\equiv q_{N+1}\equiv 0\,
\end{equation} 
is the unperturbed part and 
\begin{equation}
 \mathcal{M}(\mathbf{q})=\sum_{i=1}^N \sin(q_i)\,
\end{equation} 
can be seen as the ``magnetization'' associated to the external field $h(t)$.

In Fig.~\eqref{fig:resp1} we show the results of numerical simulations on system~\eqref{eq:hamresp}, where the external field is switched on to a value of $\eta$ at time 0, i.e.
\begin{equation}
 h(t)=\eta\, \Theta(t)\,,
\end{equation}
$\Theta(t)$ being the Heavyside step-function; magnetization $\mathcal{M}$ is then measured as a function of time.
The experiment is repeated for several values of the inverse temperature $\beta$, both positive and negative, and a nice agreement is found  with the Kubo formula~\eqref{eq:responsekubo}, which in this case simply reads
\begin{equation}
 \overline{\mathcal{M}(t)}=\beta \eta \int_0^t dt'\, \langle \dot{\mathcal{M}}(0)\mathcal{M}(t')\rangle\,.
\end{equation}

Since FDR theory was initially developed within the context of
equilibrium statistical mechanics of Hamiltonian systems, there has been some confusion
over the years about its range of applicability. As a matter of fact,
a generalized version of FDR can be shown to hold also in out-of-equilibrium
conditions (i.e. in absence of detailed balance) and even for non deterministic
dynamics, provided that \textit{(i)} the dynamics is mixing and \textit{(ii)} the system reaches an absolutely continuous and
differentiable invariant probability distribution~\cite{marconi08}.

The key idea is to consider the behaviour of the state ${\bf x}$ of the system when,
at time  $t=0$, a non-random perturbation is applied, such that 
\begin{equation}
{\bf x}(0) \to {\bf x}(0) + \delta {\bf  x}_{0}\,.
\end{equation} 
It is easy to understand that 
this instantaneous ``kick'' modifies the initial (stationary) density $\rho({\bf x})$ of the system into
a new $\rho'({\bf x})$, related to the invariant distribution by 
\begin{equation}
\label{eq:pertrho}
\rho'({\bf x}) = \rho ({\bf x} - \delta {\bf x}_0)\,.
\end{equation}  

For  an infinitesimal perturbation $\delta {\bf x}_0 =
(\delta x_1(0) \cdots \delta x_N(0))$, if $\rho({\bf x})$ is
non-vanishing and differentiable, it can be shown that perturbation~\eqref{eq:pertrho} leads to (see Ref.~\cite{marconi08} for the details of the derivation):
\begin{equation}
\label{eq:fdrgen}
\overline{\delta x_i} \, (t) =
- \sum_j \Biggl
\langle x_i(t) \left. \frac{\partial \ln \rho({\bf x})}{\partial x_j} 
\right|_{t=0}  \delta x_j(0)  \Biggr \rangle  \, ,
\end{equation}
where the meaning of the two averages is the same as before.
If the protocol of the experiment is fixed and $\delta {\bf x}_0$ is the same for each trial, Eq.~\eqref{eq:fdrgen} can be simplified into
\begin{equation}
\overline{\delta x_i} \, (t) = 
\sum_j R_{i,j}(t) \delta x_j(0),
\end{equation}
where we have introduced the linear response matrix  
\begin{equation}
R_{i,j}(t) = - \Biggl \langle x_i(t) \left.
 \frac{\partial \ln \rho({\bf x})} {\partial x_j} \right|_{t=0}
\Biggr  \rangle \,.
\end{equation} 
The computation can be easily repeated 
for a generic observable $\mathcal{A}({\bf x})$, yielding 
\begin{equation}
\label{eq:responsevulp}
\overline{\delta \mathcal{A}(t)}=- \sum_j \left\langle \mathcal{A}({\bf x}(t)) \left.\frac{\partial \ln 
\rho({\bf x})} {\partial x_j} \right|_{t=0}  \delta x_j(0)  \right\rangle \,.
\end{equation} 

Let us also notice that considering an impulsive  perturbation is not a
limitation. Indeed, in the linear regime, from the (differential) linear
response one can understand the effect of a generic perturbation
\begin{equation}
 \mathbf{F}(\mathbf{x}) \to \mathbf{F}(\mathbf{x}) + \delta\mathbf{F}(t)
\end{equation} 
of the dynamic evolution
$$
\mathbf{F}(\mathbf{x})=\frac{d\mathbf{x}}{dt}\,;
$$
it is enough to  notice that
\begin{equation}
\label{eq:nonistresp}
\overline{\delta x_i(t)}=
\sum_j \int_0^t R_{ij}(t-t') \delta F_j(t') \, dt' \,.
\end{equation} 
From Eq.~\eqref{eq:nonistresp} it is easy to recover, for Hamiltonian systems, the 
already mentioned Kubo formula~\eqref{eq:responsekubo}.

In this general framework, it is quite clear that the presence of $\beta$ in the FRD
for systems at equilibrium is related to the fact that the stationary distribution $\rho(\mathbf{x})$
explicitly depends on the inverse temperature. This is also valid, of course, for systems whose
equilibrium distribution is correctly described by a negative value of $\beta$.

\begin{figure}[h!]
 \centering
\includegraphics[width=.6\linewidth]{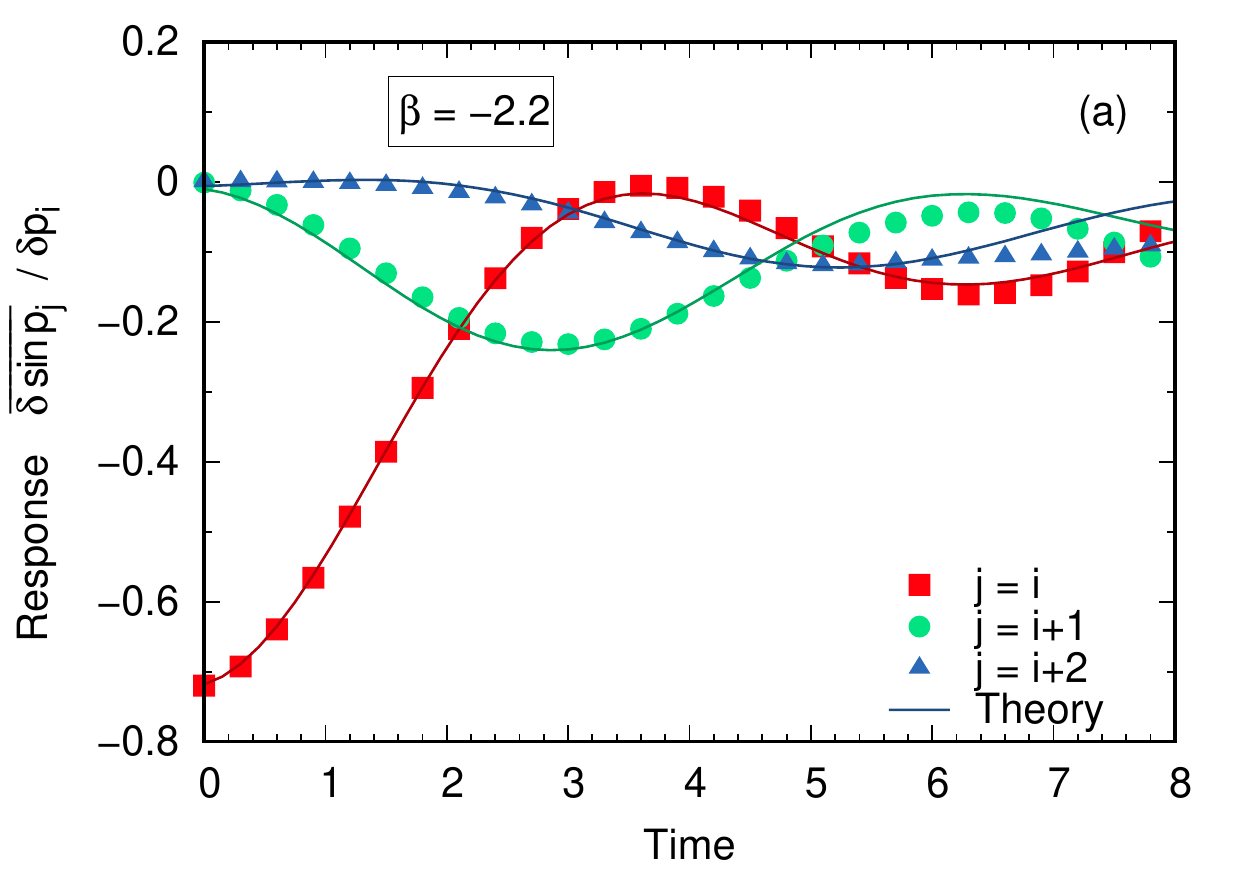}
\includegraphics[width=.6\linewidth]{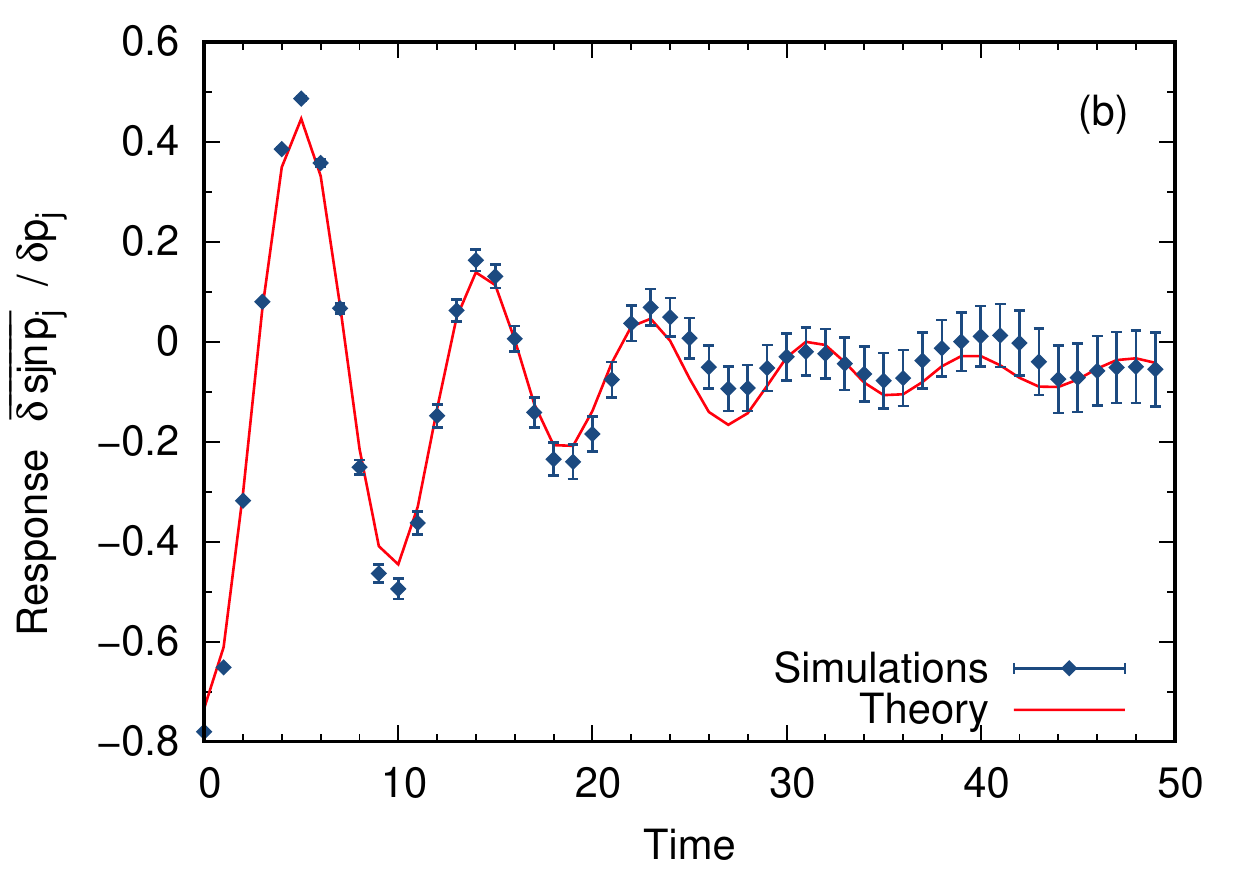}
\caption{\label{fig:resp2} Check of the FDR by a direct computation of the mean response
  and comparison with the theory in model~\eqref{eq:hamunp} (panel (a)) and model~\eqref{GHMF_B} (panel (b)).
 In panel (a), the inverse temperature of the considered system is $\beta=-2.2$, $N=128$ and $\delta p_i(0)=0.01$, average over $10^4$ realizations. As for the plot in panel (b), the parameters are $J=-0.5,\, K=-1.4, \, N=250, \,
 E/N=1.9$ (corresponding to $\beta \approx -2.6$) and $\delta p_j(0)=0.01$. Average over $10^5$ realizations. This panel is adapted from Ref.~\cite{miceli19}. }
\end{figure}

As a first example of the application of Eq.~\eqref{eq:responsevulp}, in Fig.~\ref{fig:resp2}(a) we show
the effect of a local perturbation in system~\eqref{eq:hamunp}, i.e. a chain of Hamiltonian particles
in bounded phase space. At time $t=0$ the momentum of a particle is instantaneously modified,
\begin{equation}
 p_i \to p_i + \delta p_i
\end{equation} 
and the effect on the velocities of the neighbour particles is considered. The FDR reads in this case
\begin{equation}
\label{eq:respcos}
 \overline{\delta \dot{q}_j}=\beta \langle \sin(p_j(t)) \sin(p_i(0))\rangle \delta p_i\,,
\end{equation} 
and it is nicely verified by numerical simulations.

In Fig.~\ref{fig:resp2}(b) a case with  long-range interactions is considered, 
discussed in Ref.~\cite{miceli19}. The system is the bounded space-phase 
Hamiltonian with mean field interactions already introduced in 
Section~\ref{ssec1}. Also in this case an infinitesimal perturbation on the 
$i$-th momentum induces a response on the velocity given by 
Eq.~\eqref{eq:respcos}, where the equilibrium average is now computed according 
to the mean field Hamiltonian~\eqref{GHMF_B}. As a side remark, let us notice 
that this result holds even though, at the considered value of specific energy, 
there is no equivalence between microcanonical and canonical ensemble in the 
considered system. Indeed, even if the pdf of the total system is not 
proportional to $\exp(-\beta \mathcal{H})$ in this case, it can be shown (see 
Ref.~\cite{miceli19}) that the marginal distribution for $p_i$ still verifies
\begin{equation}
 \rho(p_i) \propto e^{-\beta[1-\cos(p_i)]}\,.
\end{equation} 
This is actually the only requirement to obtain Eq.~\eqref{eq:respcos} from the 
generalized FDR~\eqref{eq:responsevulp}, which holds also if the state of the 
system cannot be described by the canonical ensemble.

\newpage
\section{Non-equilibrium and localization in the DNLS chain }
\label{sec:dnls}

\subsection{Ensemble inequivalence}
\label{sec_6.1}
In Sec.~\ref{sec:esdnls} we have discussed the phenomenology of the DNLS problem and its 
relation with negative temperatures.
In particular, we have pointed out that the existence of negative temperature 
equilibrium states in the high-energy region of the
phase diagram shown in Fig.~\ref{fig:phase_diagDNLS} has been longly debated by several authors. 
Already in the first paper where this problem was tackled \cite{rasmussen00} the
authors observed that a thermodynamic approach to the DNLS problem, based on the 
canonical 
equilibrium ensemble, yields a negative temperature in the high-energy phase, 
which
contradicts the very existence of a Gibbsian equilibrium measure. Moreover, they
guessed that a consistent definition of negative temperatures compatible
with a grand-canonical representation could be obtained  only by transforming 
the original short-range
Hamiltonian model into a long-range one.  Later a thermodynamic interpretation 
was
proposed by Rumpf in the grand-canonical ensemble and he reached the
conclusion  that negative-temperature states are
not compatible with thermodynamic equilibrium conditions~\cite{R3, R4}.
More recently, the statistical mechanics of the disordered DNLS
Hamiltonian has been analyzed making use of the grand-canonical
formalism~\cite{BM18}: the authors conclude that for weak disorder the
phase diagram looks like the one of the non-disordered model, while
correctly pointing out that their results apply to the microcanonical
case, whenever the equivalence between ensembles could be established.
In a more recent paper the thermodynamics of the DNLS model and
of its quantum counterpart, the Bose-Hubbard model, has been analyzed
by the canonical ensemble~\cite{CEF}. The main claim of these authors is that 
the Gibbs canonical ensemble is, conceptually, the most
convenient one to study this problem and conclude that the
high-energy phase is characterized by the presence of non-Gibbs
states, that cannot be converted into standard Gibbs states by
introducing negative temperatures. 

It seems quite hard drawing any clear and consistent conclusion from this
contradictory and  confused scenario, where it seems that the choice of the
adopted statistical ensemble can be made by assuming  {\sl a priori}  the 
preferred one. Conversely, the key question to
be answered concerns the equivalence between equilibrium statistical ensembles
in the high-energy phase of the DNLS model.

In order to  tackle this problem let us first recall some robust and carefully 
tested numerical results.
Extended numerical simulations pointed out that
in the high-energy phase the Hamiltonian evolution of an isolated DNLS chain
is characterized by the spontaneous formation of long-living multi-breather 
states, that last over
astronomical times~\cite{franzosi11,ICOPP}.  The microcanonical temperature $T = 
\Big(\frac{\partial S} {\partial E}\Big)^{-1} $
(for its rigorous definition see~\cite{Franzosi}) was found numerically to be 
negative
in this high-energy phase~\cite{Iubini13}. 
On the other hand,  if one assumes the thermodynamic arguments raised by Rumpf 
and Newell~\cite{RN} and later
by Rumpf~\cite{R1,R2,R3,R4}, these ``negative-temperature'' dynamical states 
should
eventually relax to an ``equilibrium state'', characterized by an
extensive background at infinite temperature with a superimposed
localized breather, containing  all the excess of energy initially stored
in the system. In fact,  an equilibrium state is usually expected to occur when 
entropy is at its maximum value,
compatible with   the conservation constraints of the model. It is 
straightforward to realize that this maximum
value lies  on the line  $T=+\infty$ $(\beta=0)$, where the value of the entropy 
for energies $E> E(T=+\infty)$  is further 
increased  by the ``degeneracy'' of any giant localized breather collecting a 
macroscopic fraction
of the total energy at any lattice point.  As we are going to discuss in the 
following section, this is
exactly what has been observed by simulating the evolution of the DNLS chain by 
a stochastic dynamics, which conserves energy and total norm
\cite{Iubini14,Iubini17}.  We want to point out that this numerical 
result (obtained for finite chains, as in the studies of deterministic  
Hamiltonian evolution)  cannot be considered as an
argument supporting  the equivalence of statistical ensembles, for the very 
reason that 
in this stochastic dynamics both energy and norm do not fluctuate, as it should 
happen
in a system in contact with standard heat and particle reservoirs. More 
precisely, the observed equilibrium state could be
compatible with a microcanonical measure on the hyper-surface of constant total 
energy and norm, through a random sampling of the states
laying on it. This manifestly has nothing to deal with a grand-canonical 
measure.

Beyond this heuristic considerations based on numerics, there is a deeper reason 
to doubt about equivalence
of statistical ensembles in the high energy phase. In fact, as pointed out by D. 
Ruelle in his seminal book
~\cite{Ruelle}, in the presence of condensation phenomena, like the localization 
of a macroscopic fraction of 
energy into the breathers of the DNLS system in the high-energy phase, 
equivalence between statistical ensembles does not apply.  

Only recently a consistent thermodynamic interpretation of the phase diagram in 
Fig. 1 has been proposed \cite{GILM1,GILM2}.
In the low-temperature phase, i.e. in between the line at  $T=0$ and the line at 
$T=+\infty$, equivalence between
statistical ensembles is granted, meaning that one can always find a 
Fenchel-Legendre transform connecting
the microcanonical entropy $S(E,A;N)$ to the Gibbs potential ${\mathcal G} 
(T,\mu;N)$ ( $\mu$ is the chemical potential and $N$ is the number of lattice
sites, i.e. the spatial extension of the DNLS chain). On the other hand, 
equivalence
between statistical ensembles does not hold  
in the thermodynamic phase extending above the line at $T=+\infty$, where only a 
microcanonical description is allowed.
For the mathematical details we address the reader to the references 
\cite{GILM1,GILM2}. Here we just summarize the
main conceptual steps used to reach this conclusion. By elaborating large 
deviation techniques (e.g., see \cite{largedev}) one can  
compute explicitly the microcanonical entropy $S(E,A;N)$. The passage to the 
grand-canonical description can be performed by
Laplace- transforming $S(E,A;N)$ with respect to $E$ and $A$ by introducing the 
dual quantities $T$ and $\mu$, respectively.
This task is accomplished every time one can find a real solution for both $T$ 
and $\mu$ in a one-to-one correspondence with
$E$ and $A$, respectively. It has been shown that this is always possible for 
any finite $N$ in the low-energy phase, but in the
high-energy phase $E > E(T=+\infty)$ a real solution does not exist for $T$,  
despite a real solution for $\mu$ still exists for any value of $A$.
Moreover, it has been shown that for finite values of $N$ the microcanonical 
equilibrium states for $E > E(+\infty)$ have a true negative temperature.
Only in the thermodynamic limit $N\to +\infty$, while keeping constant the 
energy and norm densities $E/N$ and $A/N$, the
microcanonical equilibrium states boils down to an infinite temperature state 
with a superimposed giant breather collecting the
residual macroscopic fraction of energy $\Delta E = E - E(+\infty)$ 
\cite{GILM2}.

\subsection{Slow relaxation to equilibrium}
\label{sec_6.2}
Besides the properties of the equilibrium state, the nature of the condensation 
transition in the DNLS equation
raises important questions about how the system {\it relaxes} to the localized 
state above the infinite-temperature line.
As a matter of fact, the relaxation of a macroscopic DNLS system in the 
presence of localized states turns out to be
extremely slow. 
In the following, we shall discuss the main mechanisms that are known to be 
responsible of this slow relaxation.
For the sake of simplicity, we  will not consider here the effect of intrinsic 
disorder, thus making reference to the 
standard DNLS Eq.~\eqref{eq:Intro_DNLS}, where the nature of the localization process is 
exclusively nonlinear. 
Recent studies on the
interplay between nonlinearity and disorder can be found 
in~\cite{flach16,mallick20,kati20}.

Assuming to initialize the system on a generic  initial condition far from the 
final localized state, one can identify two different kinds of sources of 
slowness, namely entropic and dynamical. 
\paragraph{Entropic effects} Entropic effects originate  from the presence of 
entropic barriers in the phase space that slow
down the relaxation to equilibrium.  Their role is essential for the physics of 
glassy and disordered systems~\cite{biroli11}
while their influence for the condensation phenomenon of the DNLS equation was studied 
in~\cite{Iubini13,Iubini14,Iubini17} in a purely
stochastic version of the model.
In these studies
it was considered a simplified DNLS model obtained from Eq.~\eqref{eq:Hdnsle} by 
taking the limit of vanishing hopping term. Accordingly,
the two conserved quantities of the model read
\begin{eqnarray}
H &=&\sum_{j=1}^N a_j^2\nonumber \\
A &=&\sum_{j=1}^N a_j \,,
\end{eqnarray}
where $a_j=|z_j|^2$ are the local norms.
One can realize that the equilibrium phase diagram of this simplified model 
preserves all the relevant features of the DNLS one. In particular the 
infinite-temperature  line separates a low-energy
delocalized region from a high-energy localized one~\cite{Iubini17,SEM14}.
A stochastic short-range conservative dynamics was defined by introducing local 
microcanonical
Monte-Carlo moves. In detail, given a random triplet of consecutive sites 
$(j-1,j,j+1)$ and assuming periodic 
boundary conditions, it is performed a random update
of the norms $(a_{j-1},a_j,a_{j+1})$ such that the total energy and norm of the 
triplet are conserved~\cite{Iubini14}.

Above the infinite-temperature line, the microcanonical relaxation dynamics 
manifest itself in the form of a {\it coarsening} of breathers. At early times, 
several breathers with small amplitude are created, while over longer time 
scales they  are found to 
progressively disappear. During this process, the  average energy of the 
survived breathers increases until a single breather localizing all the excess 
energy remains.  Overall, the average breather distance $\lambda$ is found to 
increase subdiffusively  as $\lambda=t^\alpha$, with $\alpha=1/3$. The physical 
interpretation of such a subdiffusive coarsening
is that breathers exchange energy among themselves diffusively through the 
background with an effective diffusion constant that
scales as $\lambda^{-1}$. As a result, the fewer the breathers are, the slower 
the relaxation process is.

To summarize, the study of conservative stochastic versions of the DNLS equation unveils 
the presence of a first source of
slowness in the form of a coarsening of localized states. Such an effect is 
purely entropic and originates from the
properties of the two conserved quantities of the model and of the constraint on 
the positivity of local norms $a_j$.

\paragraph{Dynamical effects} 
Despite its simplicity and analytic tractability~\cite{GILM1,Iubini17}, the 
uncoupled limit of the DNLS equation is rather peculiar 
because {\it (i)} the original coupling mechanism is replaced by
an effective stochastic interaction and {\it (ii)} the phase dynamics of the 
local variables $z_j$ does not play any role.   
When the full (deterministic) DNLS dynamics is considered, an essential 
physical effect must be taken into account: high-energy breathers are 
characterized by a large local frequency $\omega_j$ which scales as the breather's
local norm $a_j$. 
As a result of this fast rotation, higher breathers are effectively {\it 
dynamically} decoupled from the rest of the system. 

The intrinsic dynamical stability of breather states makes the localization 
process of the full DNLS model even slower. Indeed, microcanonical
simulations slightly above the $\beta=0$ line have not shown any 
evidence of coarsening. Instead, the system was observed to relax on a 
long-lasting metastable state characterized by a finite and stationary breather 
density and a negative (Boltzmann) temperature. In practice, a dynamical 
mechanism prevents breathers from becoming too large.  These findings have been 
confirmed
by more recent numerical studies, which identify a subregion of the localized 
phase where the DNLS dynamics is ergodic~\cite{mithun18}. Except
for this subregion, the nucleation of breather solutions in the localized 
phase was observed to be compatible with
a (weakly) nonergodic dynamics. In this respect, we mention that  a similar 
behaviour was found in
related studies of rotor chains and Klein-Gordon lattices, where the concept of 
{\it dynamical glass} was introduced to 
identify the onset of extremely long ergodization time 
scales~\cite{mithun19,danieli19}.

The dynamical freezing of relaxation due to breather states in the DNLS equation was 
studied in detail in~\cite{ICOPP,iubini20} in a simplified setup.
In particular, it was performed a statistical study of the relaxation of a 
single breather  on a background thermalized at positive
temperature. 
 While thermodynamics predicts that the breather needs to disappear at 
equilibrium~\cite{rasmussen00}, its average lifetime (i.e. the time needed to be 
completely absorbed by the background) was found to be exponentially large in 
the breather initial norm~\cite{ICOPP}. 
In short, the breather relaxation dynamics was found to be characterized by 
extremely long ``laminar'' periods 
interrupted by sporadic and sudden energy jumps. During the laminar phase, the 
breather amplitude displays small fluctuations with
no clear evidence of specific drifts.  A direct measure of a suitable diffusion 
coefficient reveals that this quantity decreases 
exponentially with the initial breather norm, in agreement with the evidence of 
a dynamical freezing mechanism. 
Energy jumps 
were found to be correlated to the transient creation of localized dimer states 
or the occurrence of resonances between the breather and the surrounding 
background. Both these processes are essentially rare events, whose probability 
of
occurrence was estimated to be exponentially or even super-exponentially small 
with the breather norm.

The above results bring to the fore several open problems and future 
perspectives. On the one hand,  an exponentially
small coupling between breathers and background seems compatible with the 
absence of a direct numerical observation of
the coarsening of breathers. More precisely, one should expect that 
the dynamical freezing mechanism
induces an extremely slow logarithmic coarsening~\cite{Iubini17}, although no 
direct evidence has been found. 
On the other hand, the dynamical origin of the freezing mechanism is  not clear 
at all. In this respect, in~\cite{ICOPP} it 
was  argued that the dynamics is slow because an adiabatic invariant localized 
on the breather site blocks energy diffusion. An approximate expression of the 
adiabatic invariant was derived perturbatively at the lowest orders in the 
inverse breather norm. However, the properties of this adiabatic invariant at 
higher perturbative orders have still to be explored.  

It seems natural to conclude this Section by remarking that the onset of 
negative-temperature states in the DNLS equation 
involves a very rich scenario where fundamental aspects of statistical 
mechanics, dynamics and thermodynamics 
mix together in a quite uncommon way. 
Despite for this model negative-temperature states do not exist in equilibrium 
and 
in the thermodynamic limit, they turn out to be physically meaningful and 
observable either as robust equilibrium
finite-size effects or as  extremely long metastable states. 
These peculiar features  have deep implications also in stationary 
out-of-equilibrium regimes, as it will be discussed 
in  Section~\ref{sec:fourier}.

\newpage
\section{Fourier transport }
\label{sec:fourier}

Among the out-of equilibrium situations which are most commonly found in 
statistical mechanics, the problem of heat transport along one-dimensional 
systems has a particular relevance.  It is well known that the flux of energy in 
presence of a temperature gradient is well described, in most systems, by the 
phenomenological Fourier law
\begin{equation}
\label{eq:fourier}
 \mathbf{J}=-\kappa(T) \nabla T\,,
\end{equation} 
where $\kappa(T)$ is the conductivity tensor, which is assumed to depend only on 
temperature; despite its apparent simplicity, a rigorous derivation of Fourier 
law from first principles is still lacking. The topic was widely studied in 
recent years by considering many simplified models: an exhaustive account of 
known results and open questions can be found in Ref.~\cite{lepri03, dhar08}.

A typical scenario is represented by a one-dimensional body whose boundaries are 
kept at fixed temperatures by thermal baths. If the system is close to 
equilibrium (i.e., there is only a slight difference between the temperatures of 
the two baths), $\kappa(T)$ can be approximated by a constant and the resulting 
temperature profile is linear. In this respect, many questions can be raised when
systems which admit also negative temperatures are taken into account. What happens
when the body is coupled to a bath with $T>0$ at one end and to a bath with $T<0$ at the other?
Does it reach a stationary state with well-defined local temperatures? Is Fourier
law still valid in its form~\eqref{eq:fourier} even in this case?

In the following we shall first try to answer such questions by reviewing recent results
on this topic. In particular we shall focus our attention on a minimalistic spin chain,
ruled by a stochastic dynamics which insures local conservation of energy. This simple
toy model provides a sketch of the main features of heat transport in presence of
negative temperature; other systems are found to be characterized by qualitatively
similar properties~\cite{baldiub}.

We will finally consider a more complex situation, i.e. the DNLSE already mentioned in
Section~\ref{sec:esdnls} and widely discussed in Section~\ref{sec:dnls}. For reasons
which will be clarified in the following, in this case it is not possible to couple the system
to a negative-temperature bath; however, due to the additional conservation law which
characterize this model, it is possible to observe negative temperature \textit{locally} even
if the system is coupled to thermal baths with $\beta \ge 0$. This nontrivial behaviour
is discussed in some detail.

\subsection{Simple one-dimensional models}

\subsubsection{Spin chain: equilibrium properties}
\label{sec:fourierspin}

The first model we shall discuss is somehow inspired by the pioneering experiments
on nuclear spins performed by Purcell, Pound and Ramsey in the early '50s and recalled
in Sections~\ref{sec:mag_sys},~\ref{sec:defense} and~\ref{sec:debate}. As already discussed, in that case
the spins are subjected to
the action of a large external magnetic field, so that the energy of the system is given by
Eq.~\eqref{eq:hammag}. Such relation can be conveniently adimensionalized, yielding
\begin{equation}
\label{eq:hamspin}
 \mathcal{H}=-\varepsilon \sum_{n=1}^N \sigma_n\,,
\end{equation} 
where $\varepsilon$ is proportional to the intensity of the external field and $\sigma_n = \pm 1$ are Ising spins.

Let us recall the basic equilibrium properties of this model. Denoting by $N_+$ and 
$N_-=N-N_+$ the number of positive and negative spins, respectively, and making use
of Stirling's approximation, total entropy can be written as 
\begin{equation}
\begin{aligned}
 & S(N_+)=\ln\left( \frac{N!}{N_+! N_-!}\right)\\
 & \approx -N_+ \ln \left(\frac{N_+}{N}\right)+N_- \ln \left(\frac{N_-}{N}\right) + O(\ln N)\,.
\end{aligned}
\end{equation} 
Defining the density of positive spins
\begin{equation}
 p=N_+/N\,,
\end{equation} 
the specific entropy reads then
\begin{equation}
\label{eq:spinentropp}
 s(p)=-p\ln(p) -(1-p) \ln \left( 1-p \right)\,.
\end{equation}
In order to infer an explicit expression for temperature, we have to substitute
into Eq.~\eqref{eq:spinentropp} the relation between $p$ and the specific energy $e$,
\begin{equation}
\label{eq:evsn}
 e=\varepsilon(1-2p)\,,
\end{equation} 
as it can be deduced from Eq.~\eqref{eq:hamspin}.
The resulting equation
\begin{equation}
 s(e,\varepsilon)= -\frac{\varepsilon- e}{2\varepsilon} \ln \left( \frac{\varepsilon- e}{2\varepsilon}\right) - \frac{\varepsilon+ e}{2h} \ln \left(\frac{\varepsilon+ e}{2\varepsilon}\right) \,
\end{equation} 
can be derived with respect to $e$, at fixed $\varepsilon$, so to obtain the inverse temperature:
\begin{equation}
\label{eq:betaspin}
 \beta(e,\varepsilon)=\frac{\partial s}{\partial e} = \frac{1}{2\varepsilon}\ln\left(\frac{\varepsilon-e}{\varepsilon+e}\right)\,.
\end{equation}

The value of $\beta$ tends to $-\infty$ when $e=\varepsilon$ (all spins are positive) and it tends to
$+\infty$ when $e=-\varepsilon$ (all spins are negative). The following relation holds between $p$ and $\beta$
\begin{equation}
\label{eq:posdensity}
 p=\frac{e^{\beta \varepsilon}}{2 \cosh(\beta \varepsilon)}\,,
\end{equation} 
which is consistent with the statistical interpretation of $p$ as the 
probability that, at inverse temperature $\beta$, a given spin is positive.

\subsubsection{Spin chain: heat transport}
The above recalled properties hold for any spin system at equilibrium, provided 
that its energy can be written as in Eq.~\eqref{eq:hamspin}. Now we shall focus 
our attention to the specific model discussed in Ref.~\cite{baldiub}, 
characterized by a linear geometry. We assume that the spins are placed along a 
one-dimensional lattice, and energy is locally conserved; in practice this means 
that when one spin is ``flipped'', i.e. it changes its sign, also one of its 
neighbours, initially with opposite sign, has to flip as well.
A simple stochastic dynamics which provides such behaviour is the following.
At each time step one of the $2N-2$ oriented pairs of neighbour spins is
randomly extracted; if the first spin is positive and the second one is negative,
they are both flipped; otherwise, nothing happens. It can be useful to notice that
this dynamics can be mapped into an exclusion process such that each positive
spin is replaced by a particle carrying an amount of energy equal to $2\varepsilon$, while
each negative spin is replaced by a hole.
With this analogy it is clear that the local energy flux in the stationary state reads
\begin{equation}
\label{eq:jspin}
j_n=-2\varepsilon \left(p_n w_{n\to n+1} - p_{n+1} w_{n+1\to n}\right)\,,
 \end{equation} 
 where $p_n$ is the stationary probability to find a particle in site $n$, while
 $w_{n \to k}$ is the transition rate from site $n$ to its neighbour $k$. Assuming
 that $p_n$ and $p_k$ are independent, such rate can be written as 
 \begin{equation}
\label{eq:wspin}
  w_{n\to k}=\frac{1}{\tau}\, \text{Prob}(\,k \text{ empty } |\, n \text{ occupied } ) \approx \frac{1-p_k}{\tau}\,,
 \end{equation} 
 where $\tau$ is a typical time of the dynamics.
 
To study heat transport in this chain, we need to couple its ends to
two thermal baths able to keep the local temperatures of the boundaries at 
fixed values. The simplest way is to extract periodically the value of the rightmost and
of the leftmost spin, according to the distribution~\eqref{eq:posdensity}. A natural choice
for the rate of such extraction is $1/\tau$.

If a stationary state is reached, due to local conservation of energy, the average
energy flux will be constant along the chain; from Eqs.~\eqref{eq:jspin} and~\eqref{eq:wspin}
we obtain in particular
 \begin{equation}
 p_n (1-p_{n+1})- p_{n+1} (1-p_{n})=p_n-p_{n+1}=const.\,,
\end{equation} 
i.e., the probability of finding a positive spin linearly depends on the
site position. Assuming that local equilibrium holds, so that we can define a
local inverse temperature
\begin{equation}
\label{eq:betaonp}
 \beta_n = -\frac{1}{2\varepsilon}\ln \left( \frac{1-p_n}{p_n}\right)\,,
\end{equation} 
consistently with Eq.~\eqref{eq:posdensity}, we have an explicit expression
for the inverse temperature profile along the chain.

\begin{figure}[h!]
 \centering
 \includegraphics[width=.6\linewidth]{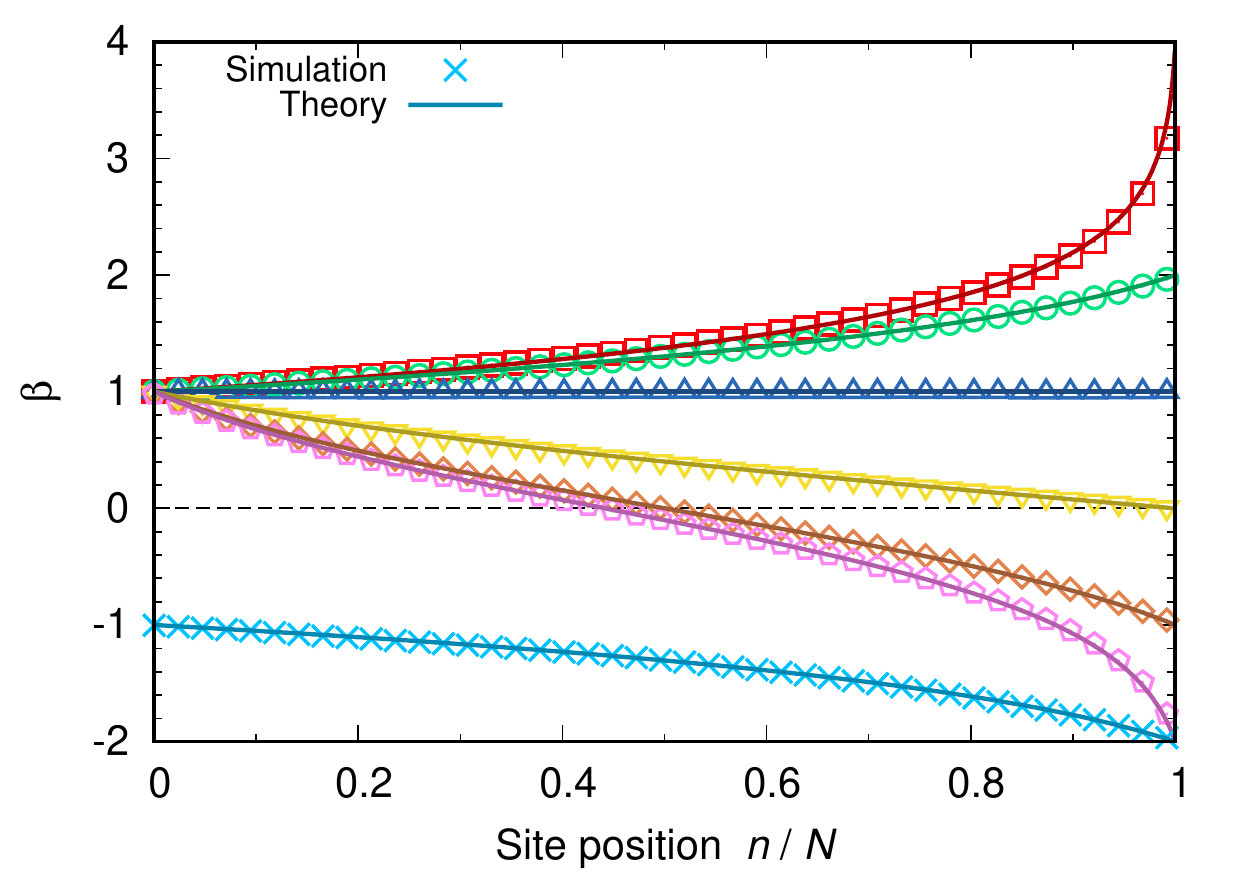}
 \caption{\label{fig:fourierspin1} Inverse temperature profiles for the out-of-equilibrium spin chain.
 Solid lines are computed according to Eq.~\eqref{eq:betaonp}, assuming a linear dependence of $p_n$
 on the site $n$. Points are results of stochastic numerical simulations as described in the main text.
 Here $N=128$, $\varepsilon=1$. Figure adapted from Ref.~\cite{baldiub}.}
\end{figure}

\begin{figure}[h!]
 \centering
 \includegraphics[width=.6\linewidth]{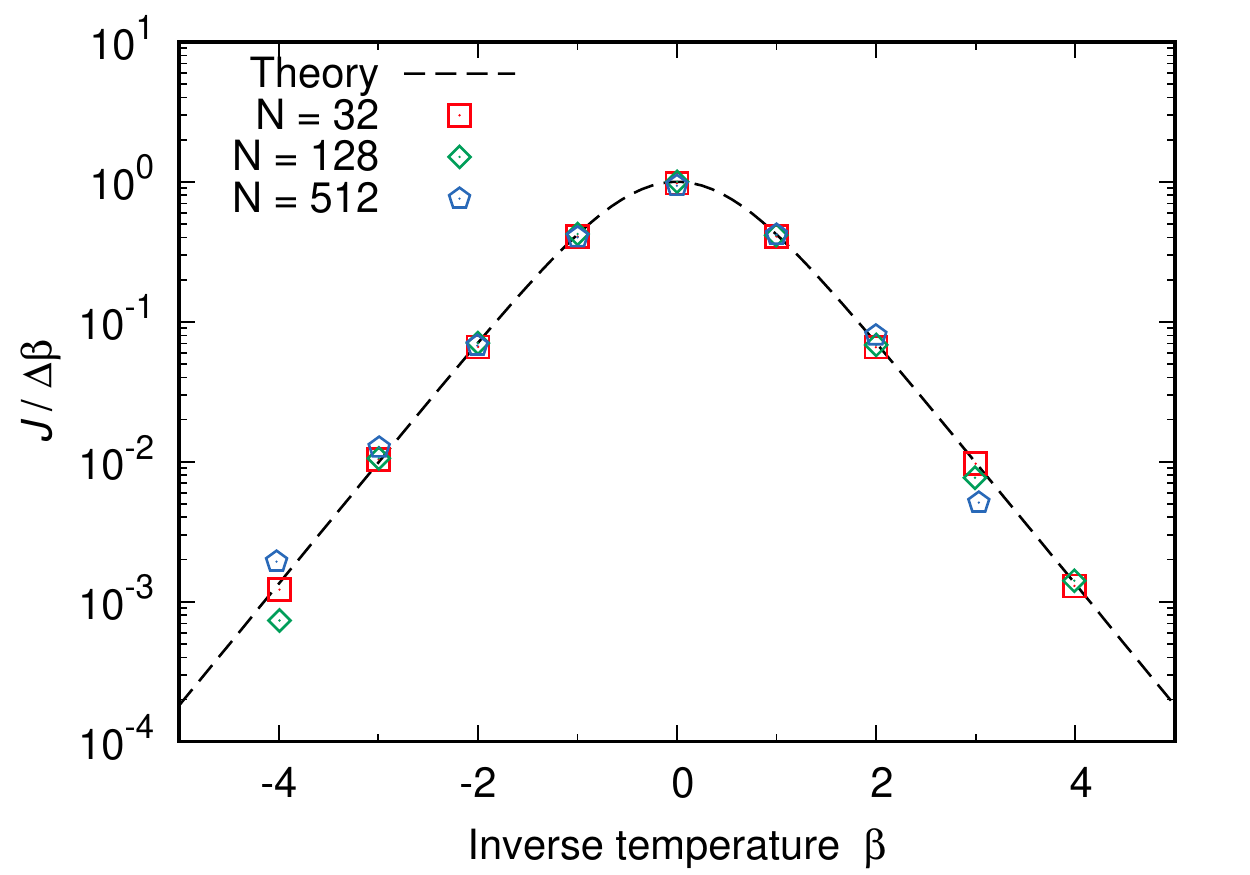}
 \caption{\label{fig:fourierspin2} Ratio between heat flux $J$ and inverse temperature difference between the
 thermal baths. Points are the result of numerical simulations for different values of $N$, with thermal baths at inverse
 temperature $\beta \pm 0.05$; dashed line is obtained
 from Eq.~\eqref{eq:jtheory}. Here $\varepsilon=1$. Figure adapted from Ref.~\cite{baldiub}.}
\end{figure}

In Fig.~\ref{fig:fourierspin1} some of such analytical profiles are compared 
with the results of numerical simulations, showing a nice agreement. It is worth 
noticing that, when the temperatures of the two baths have opposite signs, the 
transition from positive to negative values is realized through a (local) 
$\beta=0$ state, while the $T$ profile shows a singularity. This observation may 
seem to imply that Fourier law Eq.~\eqref{eq:fourier} does not hold in presence 
of such a transition; actually, as briefly discussed in 
Section~\ref{sec:fouriersimple}, a simple change of variable from $T$ to $\beta$
is sufficient to recover the validity of this general principle, as long as close-to-equilibrium
cases are considered.

In this respect, it can be useful to consider a slightly different definition of conductivity, i.e.
the ratio between the total heat flux $J=N \langle j_n \rangle$ (bearing in mind that 
$\langle j_n \rangle$ does not depend on $n$ in the stationary state) and the difference $\Delta \beta$
between the inverse temperatures of the baths. An analytical expression for such ratio
can be found in the limit of small $\Delta \beta$ as  
   \begin{equation}
 \label{eq:jtheory}
 \frac{J}{\Delta \beta} \approx \frac{\partial J}{\partial \beta} = \frac{N}{\tau}\left( \frac{\varepsilon}{\cosh(\beta \varepsilon)}\right)^2\,,
 \end{equation} 
obtained from Eq.~\eqref{eq:jspin} using Eqs.~\ref{eq:wspin} and~\ref{eq:posdensity}.
In figure~\ref{fig:fourierspin2} this theoretical curve is compared with the results of numerical simulations.
In particular, the ``conductivity'' $J/\delta \beta$ displays no singularties on the $\beta=0$
state, which corresponds to the condition of maximal transport for fixed differences $\Delta \beta$.

\subsubsection{Other examples and remarks}
\label{sec:fouriersimple}
In Ref.~\cite{baldiub} two additional one-dimensional systems are considered,
showing a phenomenology quite similar to that encountered in the case of Ising spins.
First, a Hamiltonian chain of the kind~\eqref{eq:minimal}, already discussed in
Section~\ref{sec:measuring}, is studied; the presence of generalized kinetic terms
of the form~\eqref{eq:minimalk} allows negative temperature,
and the local value of $\beta$ at the boundaries can be fixed by mean of stochastic
Langevin-like baths (see Section~\ref{sec:baths}). Then the transport properties of
a discrete linear Schr\"odinger equation model, with Hamiltonian
\begin{equation}
\mathcal{H}=\sum_{n=1}^N   z_{n}^* z_{n+1} + z_n z_{n+1}^*\,,
\end{equation}
are similarly investigated. In both cases, the qualitative scenario already found in the spin chain 
is confirmed. For all these models, it seems possible to state that:
\begin{enumerate}
 \item a stationary state is reached, in which local temperatures are well defined and measurable;
 \item the presence of negative values of temperature does not contradict the basic principle that heat
 flows from hot to cold;
 \item when the two ends of the chain are kept at fixed temperatures with opposite signs, the $\beta$ profile
 is continuous, meaning that a part of the chain in which $\beta=0$ locally can be found; this means that the $T$
 profile shows a singularity. 
\end{enumerate}

The last point may raise some concern. Indeed, it may appear to contradict Fourier's law as stated in Eq.~\eqref{eq:fourier}. However, let us remark that if the same law can be written in terms of $\beta$ as well, yielding
\begin{equation}
\label{eq:fourierbeta}
  \mathbf{J}=\tilde{\kappa}(\beta) \nabla \beta\,,
\end{equation} 
where
\begin{equation}
 \tilde{\kappa}(\beta)=\frac{\kappa(1/\beta)}{\beta^2}
\end{equation} 
for $\beta \ne 0$.

Equation~\eqref{eq:fourierbeta} implies that, when the difference between the inverse temperatures of the two baths is
small and $\tilde{\kappa}(\beta)$ can be reasonably approximated by a constant, the $\beta$ profile should be almost linear. Of course, if such condition is fulfilled and both reservoirs have positive temperature, also the $T$ profile is almost linear, as predicted by the usual formulation of Fourier law~\eqref{eq:fourier}.

In Ref.~\cite{baldiub} a simple argument is provided in order to explain why $\beta$ is the right variable to consider in this case: the basic idea is that, assuming local equilibrium, the probability to find adjacent regions of the chain with very different values of $\beta$ is very small, and this is also true if $\beta\le0$.

\subsection{DNLS equation}
\label{sec:fourierdnls}

When a DNLS chain is put in contact with external reservoirs at its edges, stationary nonequilibrium states
are expected to be controlled by suitable gradients of temperature and chemical potential. A first study
of the nonequilibrium DNLS equation~(\ref{eq:Intro_DNLS})
 was done in~\cite{iubini12} in the positive-temperature region.
The related transport of norm and energy was found to be normal (diffusive), with finite Onsager coefficients
in the limit of large system sizes. In particular, the presence of non-vanishing off-diagonal Onsager
coefficients gives rise to coupled transport with a non-vanishing Seebeck coefficient and related
unconventional  phenomena such as non-monotonic temperature profiles~\cite{iubini12,iubini16}.

\begin{figure}[h!]
\centering
\includegraphics[width=.6\linewidth]{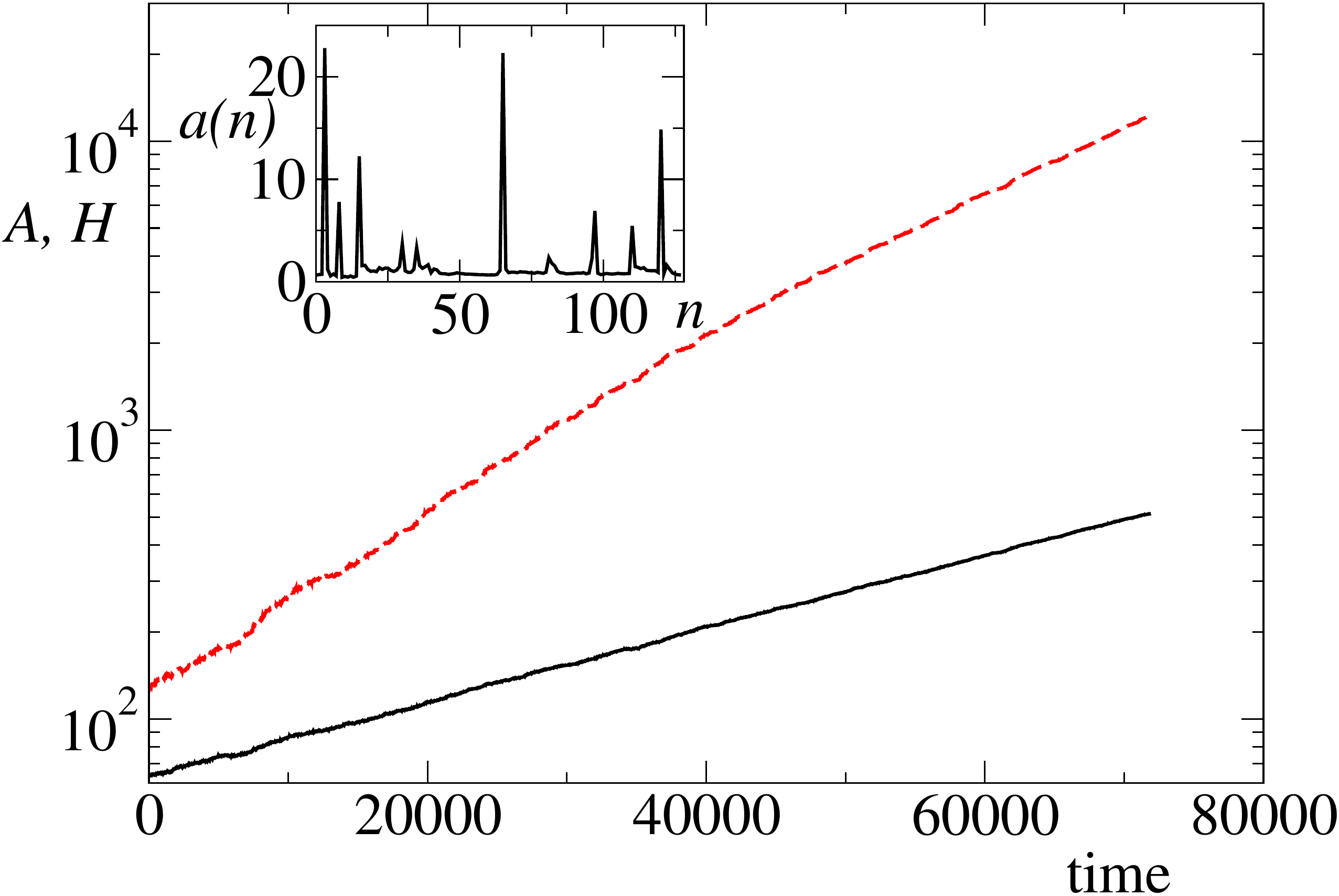}
\caption{Evolution of the total norm $A$ (black solid line) and total energy $H$ (dashed red line)
of a  DNLS chain  Eq.~\ref{eq:Intro_DNLS} with $N=128$ lattice sites and $\Lambda=2$ in contact
with a grand-canonical Monte-Carlo reservoir at inverse temperature $\beta=-1$ and
chemical potential $\mu=0$. Monte-Carlo moves consist in random perturbations of the
complex variable $z_k\rightarrow z_k + \delta$, where $k$ is a randomly selected site of
the chain and $\delta$ is extracted from a uniform
distribution in the square $[-d: d]\times i[-d:d]$, with $d=0.07$. Monte-Carlo updates
occur at random times whose periods $\tau$ are extracted from a Poissonian distribution
$P(\tau)\sim \exp(-\tau)$. The inset shows the final norm profile obtained at $t=7.2\cdot 10^4$
time units. 
}
\label{fig:ch7.2_1}
\end{figure}

The study of the above setup is crucially founded on the possibility to {\it impose}  given values
of temperature and chemical potential locally by means of suitable models of reservoirs. 
In this context, it would be tempting to introduce appropriate reservoirs at negative temperature in order to drive 
the system to the negative-temperature, localized region, as done in Sec.~\ref{sec:fouriersimple}.
 Unfortunately, this program appears to be inconsistent for the DNLS model
because of the nonequivalence of statistical ensembles discussed in Sec.~\ref{sec_6.1}. 
Indeed, a naive attempt of thermalization of the DNLS equation with one of the reservoir schemes discussed in Sec.~\ref{sec:fdr}
would result in an unstable dynamics, as shown in Fig.~\ref{fig:ch7.2_1}. In this example, the interaction
with a reservoir at $\beta=-1$ and $\mu=0$ produces a clear exponential divergence of both the 
total energy $H$ and the total norm $A$, see Eq.~\eqref{eq:Hdnsle} and~\eqref{eq:Adnsle}. In practice, the system localizes a large amount of energy in increasingly high
breather states which grow indefinitely, see the inset for a typical configuration obtained after a 
time of order $10^4$ units. 

\begin{figure}[h!]
\centering
\includegraphics[width=.6\linewidth]{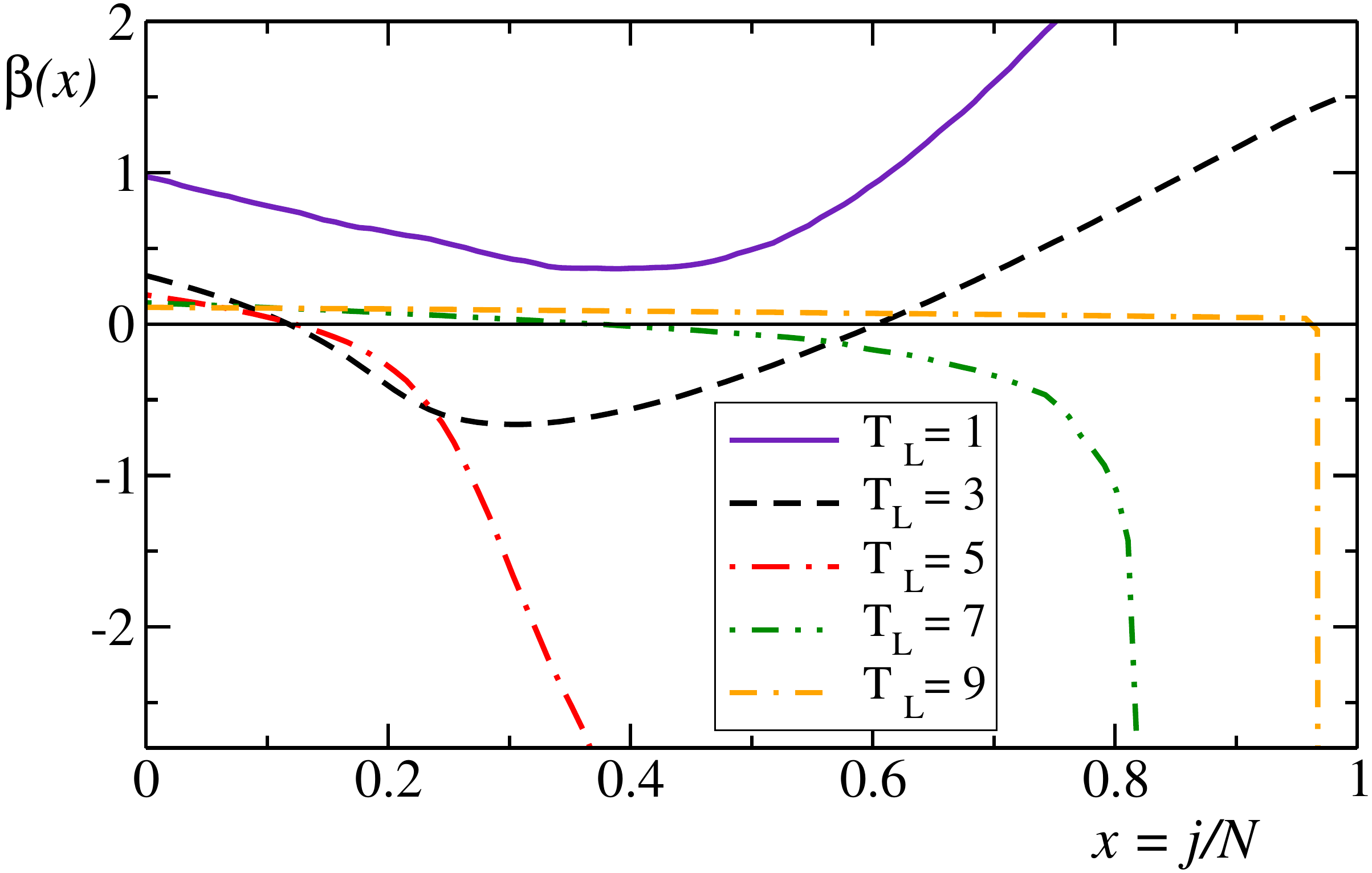}
\caption{
Stationary inverse temperature profiles of a DNLS chain Eq.~(\ref{eq:Intro_DNLS}) with $N=4095$ sites and $\Lambda=2$. The parameter $T_L$ indicates 
the different values of boundary temperature imposed by the reservoir placed at the left edge of the chain, with $\mu=0$.
The rightmost site is attached to a pure norm dissipator with fixed dissipation rate. Figure adapted from Ref.~\cite{iubini17entropy}. 
Details of the numerical setup in~\cite{iubini17entropy}.
}
\label{fig:ch7.2_2}
\end{figure}

These results should clarify that the nonequilibrium setup discussed in Sec.~\ref{sec:fouriersimple} for negative-temperature Fourier transport
can not be applied to the DNLS model in  Eq.~(\ref{eq:Intro_DNLS})~\footnote{ Negative-temperature Fourier transport would instead be accessible for a DNLS equation Eq.~(\ref{eq:Hdnsle}) with $\Lambda<0$. Indeed, in this case NAT states are delocalized, so that normal transport as in~\cite{iubini12} is expected.}
 If this seems to prevent the study of the response of the system to small negative-temperature
unbalances, and thereby the direct measurement of transport coefficients in the negative-temperature region,
it should be  pointed out that nonequilibrium negative-temperature steady states do exist in the DNLS equation 
for sufficiently large unbalances. In this respect, in Ref.~\cite{iubini17entropy} 
 it was studied a setup where a DNLS chain is in contact with a 
positive-temperature reservoir and a pure norm dissipator acting on opposite edges. It was shown that the model can display steady (coupled) transport,
with temperature profiles that enter the region of negative absolute temperatures in the central part of the chain, see 
Fig.~\ref{fig:ch7.2_2}. Such an unusual behaviour occurs when the temperature of the reservoir $T_L$ is large enough ($T_L\gtrsim 3$ for 
the example of Fig.~\ref{fig:ch7.2_2}) and it is accompanied by non-monotonic temperature profiles, which are therefore qualitatively 
different from the ones discussed in Sec.~\ref{sec:fouriersimple}. It should be stressed that the corresponding density profiles obtained in~\cite{iubini17entropy}  are not localized, despite a part of the system is locally above the $\beta=0$ line of the equilibrium
phase diagram (Fig.~\ref{fig:phase_diagDNLS}).
This feature appears to confirm the relevance of delocalized negative-temperature states found in~\cite{GILM1} for an isolated DNLS model, at least for large but finite  system sizes. 
 When $T_L$ is further increased ($T_L\simeq 10$), a different
transport regime sets up and persistent localized excitations (breather states) are observed to
 strongly inhibit transport. 
 Indeed, a sort of ``thermal wall" (see the orange curve) was observed to segregate an extended positive-temperature profile  from an almost empty region close to the dissipator.

Despite the above  phenomenon  requires further investigations and numerical efforts to clarify the role
of finite-size effects and the behaviour in the thermodynamic limit, 
some preliminary considerations can be outlined.
According to Ref.~\cite{iubini17entropy}, there is evidence that negative temperatures can arise locally in steady transport setups
even when the system interacts with  positive-temperature environments.
This genuine nonequilibrium mechanism is  observed when large unbalances and high temperatures are imposed. Accordingly, 
 the related transport regimes appear to be intrinsically far-from-equilibrium, since a description in terms
of linear response coefficients seems to be not feasible.
Moreover, signatures of anomalous transport have been reported  in this setup, a feature that has been previously observed in
 similar setups involving purely dissipative boundary conditions~\cite{iubini14prl}.

\newpage
\section{Conclusions}

In this review we have  analyzed  the statistical properties of systems  which 
admit negative absolute  temperature. Although  the possibility to have  states 
at negative temperature had been established, even at experimental level, many  
decades ago, there is still some confusion about the appropriate 
formalism to use in the  macroscopic description of  this class of systems,  as 
well as  their   real physical relevance and the possibility to have a 
consistent statistical mechanics for them. Going beyond the possible 
difficulties due to terminology,  and some  specious arguments,  our main aim 
was a  detailed physical discussion  of  the general  theoretical  facets of the 
concept of temperature, and overall an illustration of the way to build an appropriate statistical  
 mechanics description   of systems with negative absolute temperature.
 
We have first reviewed the  main phenomenological and theoretical  aspects of  
some  systems  showing negative temperatures, such as nuclear spins, 
two-dimensional vortices, lasers, cold atoms and nonlinear discrete 
Schr\"odinger equation. All these systems are interesting \textit{per se} from a 
physical point of view, and the possibility to observe thermal states which are 
correctly characterized by a negative-temperature description is an experimental 
fact. 

Starting  from the two possible definitions of temperature (Boltzmann vs Gibbs, 
obtained by using the ``surface" or ``volume'' entropy, respectively)  in the 
well-established framework of equilibrium statistical mechanics, we tried to 
clarify some subtle aspects related to entropy and its relations 
with thermodynamics and statistical properties of Hamiltonian systems. In 
particular  we  summarized some different points of view about the role 
of  the two possible definitions of temperature with respect to equipartition theorem, 
adiabatic invariance of  the entropy, Helmholtz's theorem, as well as the problem 
of thermodynamic cycles.
For sure Gibbs formalism shows interesting and relevant properties,
which allow, for instance, to establish a rigorous relation between thermodynamics and statistical
mechanics even in systems with few degrees of freedom; however, the overall picture emerging
from works by many authors indicates that Boltzmann temperature has many advantages
when the problem of equilibrium between systems at contact is considered, and it is
the only meaningful one when dealing with systems with decreasing density of states.

Our journey, then, crossed the realm of  statistical mechanics  description of 
systems with negative absolute temperature. We  began  with a   treatment of the 
equilibrium   case, in particular the problem of the measure of temperature with 
an appropriate thermometer, which allows for  the possibility to establish  a 
zero law even in non standard systems, i.e. systems without the usual 
quadratic kinetic energy. In addition we analyzed the problem of the ensemble 
equivalence (and its  possible violations), both in systems with short-range and
long-range interactions. Then we considered the main topics of the non 
equilibrium statistical mechanics: the possibility to establish in a consistent 
way a Langevin equation for the evolution of slow variables, the problem of
modeling thermal baths (e.g., for numerical simulation purposes) and the validity of
linear response theory. We also discussed the peculiar properties of
the discrete nonlinear Schr\"odinger equation, in which negative temperature states
are related to localization and ensemble inequivalence, and the validity of Fourier
law for heat transport  phenomena in the presence of negative temperature. In all these cases,
the negative-temperature description appears consistent with known results of statistical mechanics
and the outcomes of numerical simulations.

The take-home message from the large body of works considered in this review
appears rather clear and unambiguous: 
systems with decreasing density of states should not be considered pathological. 
They can indeed present some non standard features, but it is nevertheless possible 
to build a consistent macroscopic description both for  equilibrium and out-of-equilibrium properties. Such description
is a natural generalization of the well-established framework of statistical mechanics,
obtained by including also negative values of Boltzmann temperature,
and it is in complete agreement with all known experimental results.

 \newpage
\section*{Acknowledgements}
The authors thankfully acknowledge useful discussions with M.~Campisi, L.~Cerino, S.~Lepri, F.~Miceli, A.~Prados, A.~Puglisi and A.~Sarracino.

 \section*{Funding}
The work of M.~Baldovin, R.~Livi and A.~Vulpiani was supported by the MIUR-PRIN2017 \textit{Coarse-grained description
for non-equilibrium systems and transport phenomena (CO-NEST).}



\newpage
\bibliographystyle{elsarticle-num} 
\bibliography{biblio}


\end{document}